\documentclass[preprint,12pt]{revtex4-1}
\usepackage{bm}%
\usepackage{etex}
\usepackage[all]{xy}
\usepackage[T1]{fontenc}
\usepackage[utf8]{inputenc}
\usepackage[english]{babel}
\usepackage[english]{varioref}
\usepackage{amsthm}
\usepackage{amsmath}
\usepackage[version=4]{mhchem}
\usepackage{graphics} 
\usepackage{subfig}
\usepackage{stmaryrd}
\usepackage{amssymb}
\usepackage{amscd}
\usepackage{tensor}
\usepackage{tabularx}
\newcolumntype{C}[1]{>{\centering\arraybackslash}p{#1}}
\usepackage{hyperref}
\usepackage{mathtools}
\usepackage{amsbsy}
\usepackage{calligra}
\usepackage{dsfont}
\usepackage{bbm}
\usepackage{bbold}
\usepackage{mathtools}
\def\multiset#1#2{\ensuremath{\left(\kern-.3em\left(\genfrac{}{}{0pt}{}{#1}{#2}\right)\kern-.3em\right)}} % multiset coefficients
\DeclareMathAlphabet{\mathpzc}{OT1}{pzc}{m}{it}
\newcommand{\mc}{\multicolumn{1}{c}}
\usepackage{mathrsfs}

\DeclareMathOperator{\diag}{diag}
\newcommand{\vspin}[1]{\pmb{\mathcal{\hat{#1}}}} % spin operator vector with hat
\newcommand{\spin}[1]{\mathcal{\hat{#1}}} % spin operator with hat
\newcommand{\gv}[1]{\ensuremath{\mbox{\boldmath$ #1 $}}}
\newcommand{\op}[1]{\mathscr{\hat{#1}}} % for operators
\newcommand{\sop}[1]{\hat{\hat{\mathfrak{#1}}}} % for superoperators
\newcommand{\ket}[1]{\left| #1 \right>} % for Dirac ket
\newcommand{\bra}[1]{\left< #1 \right|} % for Dirac kets
\newcommand{\braket}[2]{\left< #1 \vphantom{#2} \right|
 \left. #2 \vphantom{#1} \right>} % for Dirac brackets
\newcommand{\matrixel}[3]{\left< #1 \vphantom{#2#3} \right|
 #2 \left| #3 \vphantom{#1#2} \right>} % for Dirac matrix elements
\makeatletter
\renewcommand\paragraph{\@startsection{paragraph}{4}{\z@}%
            {-2.5ex\@plus -1ex \@minus -.25ex}%
            {1.25ex \@plus .25ex}%
            {\normalfont\normalsize\bfseries}}
\makeatother
\usepackage[sort&compress]{natbib}

\begin{document}

\tableofcontents

\pagebreak

\title{An Introduction to the Holstein-Primakoff Transformation,\\with Applications in Magnetic Resonance}

\author{J.A. Gyamfi\footnote{Email: \href{mailto:jerryman.gyamfi@sns.it}{jerryman.gyamfi@sns.it} }}
\affiliation{Scuola Normale Superiore di Pisa, Piazza dei Cavalieri 7, 56126 Pisa, Italy.}

%\date{}

\begin{abstract}
We have witnessed an impressive advancement in computer performance in the last couple of decades. One would therefore expect a trickling down of the benefits of this technological advancement to the borough of computational simulation of multispin magnetic resonance spectra, but that has not been quite the case. Though some significant progress has been made, chiefly by Kuprov and collaborators, one cannot help but observe that there is still much to be done. In our view, the difficulties are not to be entirely ascribed to technology, but, rather, may mostly stem from the inadequacy of the conventional theoretical tools commonly used. We introduce in this paper a set of theoretical tools which can be employed in the description and efficient simulation of multispin magnetic resonance spectra. The so-called Holstein-Primakoff transformation lies at the heart of these, and provides a very close connection to discrete mathematics (from graph theory to number theory). 
\par The aim of this paper is to provide a reasonably comprehensive and easy-to-understand introduction to the Holstein-Primakoff (HP) transformation (and related bosons) to researchers and students working in the field of magnetic resonance. We also focus on how through the use of the HP transformation, we can reformulate many challenging computing problems encountered in multispin systems as enumerative combinatoric problems. This, one could say, is the HP transformation's primary forte. As a matter of illustration, our main concern here will be on the use of the HP bosons to characterize and eigendecompose a class of multispin Hamiltonians often employed in high-resolution magnetic resonance.
\end{abstract}

\maketitle

\section{Introduction}
Consider the deuterated hydroxymethyl radical (\ce{^{.}CH2OD}). It is well understood that the experimental determination of the electron spin resonance (ESR) spectrum for the radical involves the application of a static magnetic field $\mathbf{B}_o$ and a microwave frequency field $\mathbf{B}_1(t)$ (or a succession of such pulses), perpendicular to $\mathbf{B}_o$. How these two magnetic fields are effectively applied defines an experimental method. We could, for example, hold $\mathbf{B}_o$ constant and vary the frequency of the irradiation field, or vice versa. In all cases, though, typical experimental procedures require subjecting the sample first to the static magnetic field $\mathbf{B}_o$, followed then -- after an interval of time $\Delta t$ sufficient enough to allow the system to considerably approach its equilibrium state -- by the irradiation field $\mathbf{B}_1(t)$. 
\par Suppose we want to simulate the ESR spectrum for the \ce{^{.}CH2OD} radical. Like all quantum mechanical calculations, we start with a suitable Hamiltonian. In this case, the Hamiltonian will consist of that for the isolated radical, plus its interaction with the magnetic fields. The full Hamiltonian for the isolated radical alone (which takes into account the core electrons) is way too complicated to deal with. But, fortunately, experience has taught us that the spin Hamiltonian approach, though phenomenological by nature, suffices for the task. With this approach, we ignore the core electrons altogether and take into account only the interaction between: 1) the unpaired electron and the external magnetic fields, 2) the magnetic nuclei (i.e. nuclei with nonzero spin quantum numbers) and the external magnetic fields and 3) the unpaired electron and the magnetic nuclei. Hence, in the case of \ce{^{.}CH2OD} we only need to consider four particles with nonzero spin quantum numbers: the unpaired electron (spin-$1/2$), the two methyl hydrogen nuclei (each of spin-$1/2$) and the deuterium nucleus (spin-$1$) -- where we have obviously assumed that the carbon and oxygen nuclei are of the isotopes \ce{^{12}C} and \ce{^{16}O}, respectively, which are both zero spin nuclei. Let us assume that all couplings are scalar, and that the spin-spin couplings are simply of the type $T_{ii'} \vspin{J}_i \cdot \vspin{J}_{i'}$\citep{art:Corio-1960}, where $T_{ii'}$ is the coupling constant and $\vspin{J}_i(= \spin{J}^x_i \gv{e}_x + \spin{J}^y_i \gv{e}_y + \spin{J}^z_i \gv{e}_z )$ is the spin vector operator for spin $i$. In the absence of the external magnetic fields, the spin Hamiltonian for the isolated radical, $\op{H}_{spin-spin}$, is given by:
				\begin{equation}
				\label{eq:H_spin-spin}
				\op{H}_{spin-spin} = \sum_{i > i'} T_{i,i'} \vspin{J}_i \cdot \vspin{J}_{i'} = \sum_{i > i'} T_{i,i'} \left(\spin{J}^x_i \spin{J}^x_{i'} + \spin{J}^y_i \spin{J}^y_{i'} + \spin{J}^z_i \spin{J}^z_{i'}\right) \ .
				\end{equation}
In the presence of the external static magnetic field $\mathbf{B}_o$, we need to add to the Hamiltonian $\op{H}_{spin-spin}$ an extra term which accounts for the interaction between the spins and $\mathbf{B}_o$. The Hamiltonian for the spins then becomes $\op{H}_o$, where
				\begin{equation}
				\label{eq:H_o}
				\op{H}_o := -\sum_i \gamma_i \vspin{J}_i \cdot \mathbf{B}_o + \op{H}_{spin-spin} 
				\end{equation}							 
where $\gamma_i$ is the gyromagnetic ratio for the $i-$th spin. (We adopt in this article the convention according to which the gyromagnetic ratio carries a sign; for example, it is negative for the electron, and positive for the proton and most atomic nuclei). If we let the system approach its equilibrium state near enough, under the evolution of the Hamiltonian $\op{H}_o$, then the corresponding density matrix operator, $\op{\rho}_o$, of the spin system prior to the application of the irradiation field $\mathbf{B}_1(t)$ could be taken -- without loss of generality -- to be of the form:
				\begin{equation}
				\op{\rho}_o = \frac{e^{-\beta \op{H}_o}}{\mathcal{Z}} 
				\end{equation}		
where $\beta=\frac{1}{k_B T}$ ($k_B$ being the Boltzmann constant and $T$ the absolute temperature), and $\mathcal{Z}$ is the quantum statistical partition function defined as:
				\begin{equation}
				\mathcal{Z}:= \mbox{Tr} \left[ e^{-\beta \op{H}_o} \right] \ . 
				\end{equation}				 		 
The calculation of the expectation value of observables (thus, the magnetic resonance spectrum by extrapolation) related to the spin system after the application of $\mathbf{B}_1(t)$ is highly dependent on $\op{\rho}_o$. But given that the operator $\op{\rho}_o$ commutes with $\op{H}_o$ -- the important implication being that they both share the same set of eigenvectors -- the eigenvectors and eigenvalues of $\op{H}_o$ thus assume a fundamental role in the computational endeavor to simulate the magnetic resonance spectrum of a spin system -- because knowing them greatly simplifies the computation. 
\par One simple way to determine the eigendecomposition of the operator $\op{H}_o$ is to take its matrix representation in some spin basis (the uncoupled spin representation being the most common one) and diagonalize the whole matrix tout court (let us call this the "tout court approach"). But there is a catch here: as it is well known, the dimension of the Hilbert space for a system of nonzero spins grows exponentially with the number of these spins. Thus, the dimension of the matrix representation of $\op{H}_o$ also inflates exponentially with the number of nonzero spins. (For example, a system which consists of $N$ spins-$1/2$ has a spin Hilbert space of dimension $2^N$.) The practical consequence of this -- as far as the art of simulation is concerned -- is that the computational cost of the eigendecomposition also scales exponentially with the number of spins, which in turn translates into far longer computing time. There is also the problem of computing memory to mention: the finite memory available for use on computing devices means that the dimension of $\op{H}_o$ cannot exceed a given threshold inherent to the device. All these fall under what has been dubbed as the "curse of dimensionality". The mentioned inconveniences, together with the numerical inaccuracies one easily encounters when numerically diagonalizing very large matrices, cast the tout court approach as practically unsuitable for normal systems with more than about fifteen nonzero spins.
\par Symmetry and arguments from the theory of angular momentum can also be used to reduce the computational complexity of eigendecomposing spin Hamiltonians of the form of $\op{H}_o$. These arguments certainly depend on the symmetry properties of the spin Hamiltonian and may not be enough to alleviate the curse of dimensionality as we shall see later.
\par In this article we show how one can further reduce the eigendecomposition computational cost using the so-called Holstein-Primakoff (HP) representation, first introduced by Theodore Holstein and Henry Primakoff in 1940 to study spin waves in ferromagnets\citep{art:Holstein-1940}. Simply put, the HP representation is just another way of representing spin states and operators. So far, its application has been mainly confined to the spin wave theory in solid state physics but it is becoming increasingly clear that it has far reaching applications in the theory of angular momentum and group theory than previously thought\citep{art:Gyamfi-2018}. The fundamental role played by angular momentum theory in magnetic resonance theory cannot be overemphasized; thus, it is not surprising that the HP representation also finds important applications in the latter. Our scope here is to illustrate some of these applications, namely, in relation to the eigendecomposition of spin Hamiltonians of the type $\op{H}_o$, Eq. \eqref{eq:H_o}. If we take the \ce{^{.}CH2OD} radical for example, the spin Hilbert space is of dimension 24. The tout court approach will have us diagonalize a matrix of dimension 24 to find the eigenvectors and eigenvalues of the radical, but in the HP representation the problem reduces to effectively diagonalizing just two matrices: one of dimension 4 and the other of dimension 7. As we shall see, besides enabling us to predict how many smaller matrices we will have to diagonalize and their respective dimensions, the HP transformation approach can also be used to infer the best strategy to construct these matrices in a computationally efficient manner.
\par But more importantly, our scope here is to introduce this type of spin representation to researchers and students -- be they experimentalists or theorists -- working in the field of magnetic resonance. We therefore give here a very comprehensive introduction to the subject before applying it. No advanced knowledge of quantum mechanics, group theory or even some knowledge of number theory is required of the Reader. The scope of the paper is therefore twofold: 1) serve as a hard-to-come-by simple and comprehensive introduction to the HP transformation, and 2) show some novel applications of the latter in magnetic resonance, with particular reference to multispin systems. 
\par What we have discussed above is the diagonalization approach to the simulation of magnetic resonance spectra. An alternative is the propagation approach, where one propagates the density matrix $\op{\rho}_o$, usually in Liouville space. One notable computational framework based on this approach is the ingenious state space restriction (SSR) method proposed by Kuprov and collaborators\citep{art:Kuprov-2007,art:Kuprov-2008}. The density matrix propagation approach is more appropriate than the diagonalization approach when dealing with a large number of spins but we limit ourselves in this article to the latter because it makes the exposition of the HP transformation and the related mathematical apparatuses to be discussed more intelligible. Nevertheless, in the framework of HP transformation, the diagonalization approach in the case of isotropic Hamiltonians can be pursued at a computational cost far less than the tout court approach, even though the scaling may still remain exponential with respect to the number of spins in some cases. The propagation approaches can all be reformulated in the framework of HP transformation, allowing the simulation of the resonance spectra at an even lower computational cost.
\par The paper is organized as follows: In \S \ref{sec:Notations} we clarify the notations to be employed in the paper, together with some definitions. The Holstein-Primakoff transformation is introduced in \S \ref{sec:intro_HP} after we have revisited the quantum harmonic oscillator. This is to highlight some of the important features between the two. Then follows \S \ref{sec:HP_trans_and_multispin}, where we use the HP transformation and other theoretical tools to be introduced there to, first, study the eigendecomposition problem in relation to the isotropic spin Hamiltonian for an isolated multispin system; that is then followed by a similar study on the same system but, this time, in the presence of an external static magnetic field. In \S \ref{sec:Schwinger_bosons} we introduce the Schwinger bosons, which is also another type of spin representation but closely related to the HP transformation. 

\section{Notations and definitions}\label{sec:Notations}
\par To make the concepts we will discuss later on more intelligible, it is important we get our terminologies and notations in order now. For example, we shall often speak of \emph{multisets}\citep{misc:Stanley-2011}. Unlike ordinary sets where each distinct element must appear only once, in a multiset distinct elements can repeat any number of times. A set therefore can only tell us how many distinct elements we have and their identities. For instance, say $P_4$ the set of the first four prime numbers; then, $P_4 = \{2,3,5,7\}$. Consider all the possible three digit (positive) integers we can create from the elements of $P_4$. The number "537" for example is a valid one. By representing the digits composing these numbers as elements of a collection, we may write "537" as "$\{5,3,7\}$",  using the set notation. In the same spirit, $\{2,2,2\}, \{3,7,7\}$ and $\{5,7,5\}$ are all valid. While $\{5,3,7\}$ is a set, $\{2,2,2\},\{3,7,7\}$ and $\{5,7,5\}$ are not, but are multisets. According to the criterion chosen to represent all possible three digit numbers from the set $P_4$, it is clear that the multiset $\{5,3,7\}$ is different from $\{3,7,5\}$. Order is therefore important here. Like a set, a multiset can also be ordered. If we are only interested in the number of times a number repeats as a digit, then order is no longer important and so we may indicate "377" and "737" as $\{3,7,7\}$ or $\{7,3,7\}$ or $\{7,7,3\}$, or -- even more succinctly, using a customary notation in multiset theory -- $\{3,7^2\}$, the exponent (or \emph{multiplicity}) here indicates the number of times an element repeats itself. 
\par Consider now a multiset $\mathpzc{A}=\{j_1,j_2, \ldots ,j_i, \ldots , j_N\}$ of $N$ spins, where $j_i$ is the spin quantum number of the $i-$th spin. In the following, we shall reserve the Latin letter $i$ to lower index $\mathpzc{A}$'s elements when the latter is expressed in this laborious manner. A more efficient way to indicate the same multiset (if we are not interested in order as we are now) is to specify all the \emph{distinct} elements and their multiplicities. We shall use the Greek letter $\alpha$ to index distinct elements of the multiset. Thus, if the spin multiset $\mathpzc{A}$ has $\sigma$ distinct elements, then we can express $\mathpzc{A}$ also as $\mathpzc{A}=\{j_\alpha^{N_\alpha}\}, \ \alpha = 1, 2, \ldots , \sigma$, where $N_\alpha$ is the multiplicity of the $\alpha-$th distinct spin quantum number. Naturally, $N = \sum^\sigma_{\alpha=1} N_\alpha $. It is crucial to point out that two elements $j_i$ and $j_{i'}$ of $\mathpzc{A}$ are distinct only if they correspond to different spin quantum numbers -- other characteristics of the spins (like charge, magnetic moment etc.) are not of any merit whatsoever here. For example, the multiset $\mathpzc{A}=\{\frac{1}{2}^9, 1^3\}$ indicates \emph{any} aggregate of nine spin-$1/2$ and three spin-$1$; the actual composition could consist of, for instance, 1) nine electrons and three protons, or 2) five \ce{^{15}_{7}N_8} plus four muons -- both spin-$1/2$s -- and three \ce{^{14}_{7}N_7} nuclei (which are spin-$1$)\cite{art:Fuller-1976}, etc.
\par A very important distinction is due here. A system of $N$ spins whereby $j_1=j_2 = \ldots = j_N=j$ is termed as \emph{univariate spin system} (USS)\citep{art:Gyamfi-2018}. The multiset representation of the system is then $\mathpzc{A}=\{j^N\}$. In a USS, the $N$ spins could be mutually different in regards to mass, charge, magnetic moment, etc. On the other hand, a system of \emph{identical spins} (IS)\citep{art:Gyamfi-2018} is a univariate spin system whereby all the $N$ spins share exactly the same intrinsic fundamental properties like mass and charge, and are indistinguishable from one another when placed under the same external conditions. Thus, an identical spin system is also necessarily USS, but a USS is in general not an IS.
\par In addition to the concept of univariate and identical spin systems, we also have \emph{equivalent} spins (ES). Consider a given multiset $\mathpzc{A}=\{j_1, j_2 , \ldots , j_N\}$ of spins. Say $\mathpzc{A}'$ a submultiset of $\mathpzc{A}$, i.e. every element of $\mathpzc{A}'$ is also an element of $\mathpzc{A}$. $\mathpzc{A}'$ is said to be a system of equivalent spins if every element of $\mathpzc{A}'$ couples to all other elements of $\mathpzc{A}$ and any external field in the same manner. Specifically, the elements of $\mathpzc{A}'$ are said to be \emph{equivalent} if one cannot distinguish between them on the basis of their coupling tensors with other spins and external fields. It mostly happens that equivalent spins are also identical, but in principle they do not need to be. The concept of equivalent nuclei in NMR, for example, is just a limit case of equivalent spins. If we take the methyl radical \ce{^{.}CH3} for example (assuming all three hydrogen nuclei are \ce{^1H} and the carbon atom is \ce{^{12}C}), the spin system $\mathpzc{A}=\{j_1,j_2,j_3,j_4\}=\{\frac{1}{2}^4\}$ is clearly univariate; the three hydrogen nuclei are identical spins. The same trio of spins also constitute a collection of equivalent spins when the system's Hamiltonian is invariant under the operation of the point group $C_3$. 
\par Operators will be indicated with their usual hats while their matrix representations will bear none: for example, the matrix representation of the operator $\op{A}$ will be simply indicated as $\mathscr{A}$.
\par If $j_i$ is the spin quantum number of the $i-$th spin, therefore a scalar ($j_i=\frac{1}{2},1,\frac{3}{2},2,\frac{5}{2},\ldots $), then its corresponding spin vector operator will be indicated as $\vspin{J}_i$, which, as we saw above, corresponds to the sum: $\vspin{J}_i = \spin{J}^x_i \gv{e}_x + \spin{J}^y_i \gv{e}_y + \spin{J}^z_i \gv{e}_z$, where $\spin{J}^x_i, \ \spin{J}^y_i, \ \spin{J}^z_i$ are the spin angular operators along the respective axis and defined on the spin Hilbert space of spin $i$. In this article, we will not try to indicate spin vector operators of electrons and nuclei with different symbols: they will all be indicated simply as $\vspin{J}_i$, and the index $i$ will serve the purpose of recording the specific identity of the spins based on our chosen choice of integer-labelling of the latter.

\section{The Holstein-Primakoff Transformation}\label{sec:intro_HP}
Before going on to discuss the Holstein-Primakoff transformation, it is useful we briefly recall here some salient features of the quantum harmonic oscillator. The reason why we do so is that the Holstein-Primakoff transformation, like many other techniques developed in many-body quantum theory, can be traced back to the quantum harmonic oscillator. The short review we give below on the latter looks at the subject from a slightly different point of view with respect to the traditional treatments, and focuses on those important features it shares with the HP transformation. %Almost every Reader is already quite familiar with the subject so we apologize for having to bring it up again. But as one very prominent physicist is attributed as saying, "The career of a young theoretical physicist consists of treating the harmonic oscillator at ever-increasing levels of abstraction."\citep{book:Kaiser-2018} This is not far from the truth for theoretical chemists as well.
\subsection{The quantum harmonic oscillator}
\par The harmonic oscillator is perhaps the most important model in quantum mechanics. We are not going to belabor it here as it is extensively treated in almost every textbook on quantum mechanics. We only recall here that the Hamiltonian of a body of mass $m$ tied to a spring able to move only in one direction (let us call it the $x$ axis) is\citep{book:Atkins-2011}: 
			\begin{equation}
			\label{eq:Hamil_harmonic_oscil}
			\op{H} = \frac{\op{P}_x^2}{2m} + \frac{1}{2}k \op{X}^2
			\end{equation}
where $\op{P}_x$ is the linear momentum operator along the $x-$axis, $\op{X}$ is the spatial displacement operator from the equilibrium position, and $k$ is the force constant.
\par Since we shall be working in the position space, it is important to recall that the position eigenkets $\{\ket{x}\}$ are such that:
			\begin{equation}
			\op{X}\ket{x} = x \ket{x}
			\end{equation}
and they obey the orthogonality condition:
			\begin{equation}
			\label{eq:x'_x}
			\braket{x'}{x} = \delta(x'-x)
			\end{equation}
where $\delta(\bullet)$ is the Dirac delta function. If we multiply Eq. \eqref{eq:x'_x} by $\ket{x'}$ and integrate over the entire range of $x'$ we get:
			\begin{equation}
			\int dx' \ket{x'}\braket{x'}{x} = \int dx' \ket{x'} \delta(x'-x) = \ket{x} 
			\end{equation}
from which we deduce that:
			\begin{equation}
			\label{eq:completeness_space}
			\int dx' \ket{x'}\bra{x'} = \hat{\mathbb{I}} 
			\end{equation}
where $\hat{\mathbb{I}}$ is the identity operator. Eq. \eqref{eq:completeness_space} is the completeness relation for the position eigenkets.
\par  To study the behavior of the body quantum mechanically, what we can do is to solve the time-independent Schr\"odinger's equation for the body's energy eigenkets, $\{\ket{E}\}$:
			\begin{equation}
			\label{eq:SE_harmonic}
			\begin{split}
			\op{H} \ket{E} & = E \ket{E} \\
			\left(\frac{\op{P}_x^2}{2m} + \frac{1}{2}k \op{X}^2 \right) \ket{E} & = E \ket{E}
			\end{split}
			\end{equation}			 
where $E$ is the energy of the body. Note that we may expand $\ket{E}$ in terms of the position eigenkets:
			\begin{equation}
			\label{eq:Harmonic_energy_eigenstates}
			\ket{E} = \int dx \ket{x} \braket{x}{E} 
			\end{equation}
where $(\lVert \braket{x}{E}\rVert^2 \ dx)$ can be interpreted as the probability to find the body in any of the points between $x$ and $x+dx$ given that its energy is $E$. The coefficient $\braket{x}{E}=:\psi_E(x)$ is therefore the probability amplitude, or wavefunction, in position space.
\par If we multiply both sides of Eq. \eqref{eq:SE_harmonic} from the left by the identity operator $\int dx \ket{x}\bra{x}$, we obtain:
			\begin{equation}
			\label{eq:harmonic_wave_1}
			\int dx \ket{x} \left(\frac{1}{2m}\matrixel{x}{\op{P}_x^2}{E} + \frac{1}{2}k x^2 \braket{x}{E}\right)= \int dx \ket{x} E \braket{x}{E} \ . 
			\end{equation}
Making use of the fact that $\matrixel{x}{\op{P}_x^2}{E}=-\hbar^2 \frac{d^2}{dx^2} \braket{x}{E}$\citep{book:Sakurai-2011}, Eq. \eqref{eq:harmonic_wave_1} becomes:
			\begin{equation}
			\int dx \ket{x} \left(\frac{-\hbar^2}{2m}\frac{d^2}{dx^2} + \frac{1}{2}k x^2 \right)\psi_E(x)= \int dx \ket{x} E \ \psi_E(x) 
			\end{equation}
that is,
			\begin{equation}
			\label{eq:harmonic_osci_wave_equation}
			\left(\frac{-\hbar^2}{2m}\frac{d^2}{dx^2} + \frac{1}{2}k x^2 \right)\psi_E(x)=  E \ \psi_E(x) \ .
			\end{equation}
The solution to this eigenspectrum problem is well known. The energy eigenvalues are found to be parameterized by $n$\citep{book:Atkins-2011,book:Lancaster-2014}:
			\begin{equation}
			\label{eq:Energy_harmonic_osci}
			E_n= \left(n + \frac{1}{2} \right) \hbar \omega \ , \qquad n=0,1,2,3, \ldots 
			\end{equation}
Thus, $\psi_E(x)=\psi_{\left(n + \frac{1}{2} \right) \hbar \omega} (x)$, which may be simply written as $\psi_n(x)$. Since there is a one-to-one correspondence between $E$ and $n$ (for fixed mass and force constant), we may henceforth simply indicate $\ket{E}$ as $\ket{n}$. Moreover, it is found that\citep{book:Atkins-2011,book:Lancaster-2014}:
			\begin{equation}
			\label{eq:psi_n_x_wavefunction}
			\psi_n(x) = \frac{1}{\sqrt{2^n n!}} \left( \frac{m\omega}{\pi \hbar}\right)^{1/4} H_n\left(\zeta \right) e^{-\zeta^2 / 2} 
			\end{equation}
where $\omega \equiv \sqrt{k/m }$, $\zeta \equiv x \sqrt{m \omega/\hbar}$ and $H_n(z)$ is the Hermite polynomial of order $n$ in $z$.
\par At this point we may rewrite Eq. \eqref{eq:Harmonic_energy_eigenstates} as:
			\begin{equation}
			\label{eq:Harmonic_ket_n_x-space}
			\ket{n} = \int dx \ket{x} \psi_n(x)
			\end{equation}
where we have effected the substitutions $\ket{E} \to \ket{n}$ and $\braket{x}{E} \to \psi_n(x)$. It is evident from Eq. \eqref{eq:Harmonic_ket_n_x-space} that the wavefunction $\psi_n(x)$ is just the expansion coefficient when we expand $\ket{n}$ in terms of position eigenstates. Naturally, Eq. \eqref{eq:Harmonic_ket_n_x-space} is not the only possible expansion we could think of. We could have equally expanded $\ket{n}$ in terms of the momentum eigenstates (along the $x-$ axis), i.e. $\{\ket{p_x}\}$, where:
			\begin{equation}
			\op{P}_x \ket{p_x} = p_x \ket{p_x} 
			\end{equation}
and, 
			\begin{equation}
			\braket{p'_x}{p_x} = \delta(p'_x - p_x) 
			\end{equation}
from which follows the completeness relation:
			\begin{equation}
			\label{eq:completeness_momentum}
			\int dp_x' \ket{p_x'}\bra{p_x'} = \hat{\mathbb{I}}\ .		
			\end{equation}
In fact, introducing the momentum space completeness relation into Eq. \eqref{eq:Harmonic_ket_n_x-space} we get:
			\begin{equation}
			\ket{n} = \int dp_x \ket{p_x} \left( \int dx \braket{p_x}{x} \psi_n(x) \right) = \int dp_x \ket{p_x}  \phi_n(p_x) 
			\end{equation}
where $\phi_n(p_x)$ is the wavefunction in momentum space. In complete analogy to $\psi_n(x)$, $\left( \lVert \phi_n(p_x) \rVert^2 dp_x \right)$ gives the probability of measuring the body's momentum to be between $p_x$ and $p_x + dp_x$ if its energy is fixed at $E_n$. We briefly mention that because $\braket{p_x}{x}=\frac{1}{\sqrt{2\pi \hbar}} \exp\left( - \frac{ip_x x}{\hbar}\right)$\citep{book:Sakurai-2011}, it follows that
			\begin{equation}
			\phi_n(p_x) = \frac{1}{\sqrt{2\pi \hbar}} \int dx \ \exp\left( - \frac{ip_x x}{\hbar}\right) \psi_n(x) 
			\end{equation}
which means $\phi_n(p_x)$ and $\psi_n(x)$ are related through the Fourier transform, as one would expect.
\par The gist of the above arguments is that depending on the basis in which we expand $\ket{n}$, we get different wavefunctions. The appropriate wavefunction to talk about depends on how we intend to probe the quantum system: for example, if we intend to measure position given a fixed energy of the system, then $\{\psi_n(x)\}$ are the wavefunctions to go with; if, instead, we want to measure momentum then we need the $\{\phi_n(p_x)\}$. But in all these discussions, the nature of the ket $\ket{n}$ remains the same independent of whether we work in momentum or position space. As a matter of course, one might then ask why don't we just deal solely with the energy eigenkets $\ket{n}$ without resorting to wavefunctions -- and that is precisely the quintessence of the so-called \emph{second quantization} scheme (also known as the \emph{occupation number representation}). When we expand $\ket{n}$ in some basis and then solve for the related coefficients, or wavefunctions, as we did above, Eq. \eqref{eq:harmonic_osci_wave_equation}, we speak then of \emph{first quantization}. Thus, in first quantization, it is important to specify in which basis the expansion of the ket $\ket{n}$ is being done, normally in position or momentum space (but, in general, in the space of either one of a canonically conjugated pair of operators).
\par How then do we go about doing quantum mechanics without explicitly talking of wavefunctions? In other words, how does second quantization work? To see how it works, let us go back to Eq. \eqref{eq:SE_harmonic}, which at this point, we may conveniently write as: 
				\begin{equation}
				\label{eq:SE_harmonic_n}
				\op{H} \ket{n} = E_n \ket{n} \ .
				\end{equation}
Instead of trying to find an expression for $\ket{n}$, we take it as it is -- \emph{viz.} $\{\ket{n}\}$ are eigenstates of $\op{H}$, period. What we rather do is to rewrite the Hamiltonian $\op{H}$ in terms of a set of mutually commuting operators which have $\{\ket{n}\}$ as their eigenstates. It is worth noting that the eigenenergy $E_n$,   
Eq. \eqref{eq:Energy_harmonic_osci}, depends only on $n$ and not on neither the position $(x)$ nor the momentum $(p_x)$ of the body. Indeed, the mathematical expression for $E_n$ in terms of $n$ gives important clues on how we may rewrite $\op{H}$ in such a way that the new operators in the latter do really have $\{\ket{n}\}$ as their eigenkets. If we combine Eqs. \eqref{eq:SE_harmonic_n} and \eqref{eq:Energy_harmonic_osci}, we find that,
				\begin{equation}
				\label{eq:SE_harmonic_n_2}
				\op{H} \ket{n} = \left(n + \frac{1}{2} \right)\hbar \omega \ket{n}
				\end{equation}
which means that,
				\begin{equation}
				\label{eq:Hamornic_hamil_2}
				\op{H} = \left( \op{A} +  \frac{1}{2} \right)\hbar \omega
				\end{equation}
where $\op{A}$ is, for now, an unknown operator with the following property:
				\begin{equation}
				\op{A} \ket{n} = n \ket{n} \ .
				\end{equation}
Comparing Eq. \eqref{eq:SE_harmonic} with Eq. \eqref{eq:Hamornic_hamil_2}, we find that:
				\begin{equation}
				\op{A} = \frac{\omega m}{2\hbar} \left( \op{X}^2 + \frac{\op{P}^2}{\omega^2 m^2} - \frac{\hbar}{\omega m}\right) \ .
				\end{equation}
Traditionally, this form of $\op{A}$ is less preferred because it obscures some interesting insights. Rather, it is the factorized form:
				\begin{equation}
				\op{A} = \hat{a}^\dagger \hat{a}
				\end{equation}
where\citep{book:Atkins-2011},
			\begin{subequations}
			\label{eq:def_a_operators}
			\begin{align}
			\hat{a}^\dagger & = \sqrt{\frac{m\omega}{2\hbar}} \left(\op{X} - \frac{i}{m\omega} \op{P}_x \right)\\
			\hat{a} & = \sqrt{\frac{m\omega}{2\hbar}} \left(\op{X} + \frac{i}{m\omega} \op{P}_x \right) 
			\end{align}
			\end{subequations}
that we prefer and use. The reason is that, despite the fact that the operators $\hat{a}^\dagger$ and $\hat{a}$ are not Hermitian, they allow for transition between the eigenstates -- so they are extremely useful when discussing emission and absorption processes. As a matter of fact, one can derive that:
			\begin{equation}
			\label{eq:a_dagger_a}
			\hat{a} \ket{n} = \sqrt{n} \ket{n-1} \ , \quad \  \hat{a}^\dagger \ket{n} = \sqrt{n+1} \ket{n+1} \ .
			\end{equation}
\par It is truly remarkable that the form $\op{H}$ assumes in light of the operators $\hat{a}$ and $\hat{a}^\dagger$, namely, 
			\begin{equation}
			\label{eq:harmonic_hamil_adagger_a}
			\op{H} = \left( \hat{a}^\dagger \hat{a} +  \frac{1}{2} \right)\hbar \omega 
			\end{equation}
appears in many problems in quantum mechanics. Perhaps the most important is the quantization of the electromagnetic field, which was first done in the early years of quantum theory. There, one sees that the Hamiltonian of the field becomes a sum over an infinite number of harmonic oscillators, each representing a specific mode of the field. And the operators $\hat{a}^\dagger$ and $\hat{a}$ are tasked with increasing and decreasing by a quanta of energy their respective modes. Einstein had already discovered the photoelectric effect and the notion of the electromagnetic field as being composed of particles called photons was well established by then, so interpreting the operators $\hat{a}^\dagger$ and $\hat{a}$ as creating and annihilating photons of certain momentum was leapt at very easily -- thus their eponymous current names. Soon after, the language introduced by the quantization of the electromagnetic field crept into all physical problems where the Hamiltonian could be recast into the likeness of Eq. \eqref{eq:harmonic_hamil_adagger_a}. In every instance (with the exception of the harmonic oscillator), the particles which the creation and annihilation operators were meant to create or annihilate were given a specific name. In the case of lattice vibrations, for example, the particles are called phonons\citep{book:Sakurai-2011, book:Lancaster-2014}. In this narrative, the operator $\hat{a}^\dagger \hat{a}$ is interpreted as counting the number of particles occupying a certain state, hence its name \emph{occupation number operator}, usually indicated as $\hat{n}$. In this view, the state $\ket{n=0}$ contains no particle, so we call it the \emph{vacuum state}. The absence of particles in the vacuum state though does not necessarily imply a state of zero energy. As it mostly happens, it is characterized by a specific energy. For the harmonic oscillator, Eq. \eqref{eq:Energy_harmonic_osci}, this vacuum energy corresponds to $E_0 = \frac{1}{2}\hbar \omega$. And anytime a particle is added to the system, the energy of the latter increases by $\Delta E = \hbar \omega$ -- which is commonly referred to as a \emph{quanta of energy}. In addition, we can imagine creating any state $\ket{n}$ from the vacuum state $\ket{0}$. It is easy to prove from Eq. \eqref{eq:a_dagger_a} that:
				\begin{equation}
				\label{eq:n_from_vacuum_harmonic}
				\ket{n} = \frac{\left(\hat{a}^\dagger\right)^n}{\sqrt{n!}} \ket{0} \ .
				\end{equation}				  
The vacuum state of the harmonic oscillator can thus accommodate any finite number of particles, hence the particles must be bosons. The normalization factor $\frac{1}{\sqrt{n!}}$ can be interpreted as accounting for the indistinguishability of these bosons. (It is interesting to note the analogy here with the solution to the Gibbs paradox.). The particle states $\{\ket{n}\}$ are the orthonormal basis elements of a vector space called the \emph{Fock space}\citep{book:Kaiser-2018, book:Sakurai-2011, book:Lancaster-2014}. In this space, two basis states $\ket{n}$ and $\ket{n'}$ are orthogonal to each other if they differ in their occupation numbers. The Fock space for the harmonic oscillator is \emph{bosonic} because, as stated above, we can fill the vacuum state with any number of bosons. There are Fock spaces where one cannot fill the vacuum state with more than one particle. These are called \emph{fermionic} Fock spaces, and the particles in question are fermions. Finally, we mention that from the properties of the operators $\hat{a}^\dagger$ and $\hat{a}$, Eq. \eqref{eq:a_dagger_a}, one derives the commutation relation:
				\begin{equation}
				\left[ \hat{a},\hat{a}^\dagger\right] = \hat{\mathbb{I}} 
				\end{equation}
which can also be easily derived applying the definition of $\hat{a}^\dagger$ and $\hat{a}$ given at Eq. \eqref{eq:def_a_operators}. 
\par In discussing the harmonic oscillator above, we made a very deep conceptual leap when we transitioned from first to second quantization. This has to do with the interpretation of the integer $n$. Under the first quantization scheme, $n$ just indexed eigenfunctions like $\psi_n(x)$ and their respective energies (Eq. \eqref{eq:psi_n_x_wavefunction}). But according to the second quantization scheme, we came to see the same integer $n$ as being an eigenvalue of the operator $\hat{a}^\dagger \hat{a}$ and also indicates the number of bosons occupying a Fock space ket. Both interpretations are correct and can be used interchangeably. However, some caution is needed in how far we drag the meaning of $n$ together with the creation and annihilation operators in second quantization. In general, the particles created or annihilated according to second quantization represent excitations in the system under study. This is quite clear from the relation between $n$ and the quantum harmonic oscillator's eigenenergy, Eq. \eqref{eq:Energy_harmonic_osci}: $n$ indicates how energetically excited the state of the oscillator is. And the notion of $\hat{a}^\dagger$ and $\hat{a}$ creating and annihilating some bosons, respectively, according to the second quantization scheme,  is just a mathematical construct which provide an alternative way of talking about these same harmonic excitations. The same applies to lattice vibrations: phonons, like the bosons for the harmonic oscillator, are just mathematically constructed particles which occupation numbers represent excitations in lattice vibrations. 
\par However, there are also many instances whereby these particles one get from second quantizing a system are not just the fruits of some mathetmatical trickery, but are real particles to reckon with in Nature. For example, photons represent the excitations in the electromagnetic field according to second quantization, and are real. The Higgs boson represent excitations in the Higgs field, and it has recently been detected experimentally. The electromagnetic field and the Higgs field are examples of what we call \emph{quantum fields}. In fact, all the elementary particles in physics (including the electron) are excitations of a particular quantum field. The study of quantum fields and their excitations is the subject of \emph{quantum field theory} (QFT)\citep{book:Lancaster-2014}. We are not going to need QFT in the discussions that follow, but it is important we bear in mind the episteme related to these particles which transpire through second quantization.
\par To conclude this brief discussion on the harmonic oscillator and second quantization, we consider a collection of noninteracting harmonic oscillators, say $N$ in total. If we work in the occupation number representation, the Hamiltonian here is a direct generalization of Eq. \eqref{eq:Hamornic_hamil_2}:
				\begin{equation}
				\op{H} = \sum^N_{i=1} \left(\hat{a}^\dagger_i \hat{a}_i + \frac{1}{2} \right) \hbar \omega_i 
				\end{equation}
It is then very easy to see that  a generic state of the system can be expressed as $\ket{n_1, n_2, \ldots, n_N}$ where $n_i$ indicates the number of bosons present in the $i-$th harmonic oscillator. The (overall) vacuum state of the system is the state in which $n_i=0$ for any given oscillator $i$, i.e. $\ket{0,0,\ldots, 0}$, which we shall simply indicate as $\ket{\mathbb{0}}$. The eigenenergies are also easily found to be:
				\begin{equation}
				E_{n_1,n_2, \ldots,n_N} = \sum^N_{i=1} \left(n_i + \frac{1}{2} \right) \hbar \omega_i \ .
				\end{equation}
The operators $\hat{a}^\dagger_i$ and $\hat{a}_i$ operate only on the Fock space of the $i-$th oscillator, therefore such operators with different indexes commute:
				\begin{equation}
				\left[\hat{a}_i, \hat{a}_{i'} \right] = \left[\hat{a}^\dagger_i, \hat{a}^\dagger_{i'} \right]=\hat{0} \ , \quad \left[\hat{a}_i, \hat{a}^\dagger_{i'} \right] = \delta_{i,i'} \hat{\mathbb{I}} \ .
				\end{equation}
As in the case of a single harmonic oscillator, the generic state $\ket{n_1,n_2,\ldots,n_N}$ can be generated from the vacuum state $\ket{\mathbb{0}}$:
				\begin{equation}
				\ket{n_1,n_2,\ldots,n_N} = \prod^N_{i=1} \frac{(\hat{a}_i^\dagger)^{n_i}}{\sqrt{n_i!}} \ket{\mathbb{0}} \ .
				\end{equation}
The generic multi-harmonic oscillator state $\ket{n_1, n_2, \ldots, n_N}$ is simply the direct product $\ket{n_1} \otimes \ket{n_2} \otimes \ldots \otimes \ket{n_N}$. The Fock space for this collection of oscillators is thus the vector space tensor product of the Fock spaces of the separated oscillators. The collection of integers $\{n_1,n_2, \ldots , n_N\}$ is an ordered multiset: the first element denotes the number of bosons in the first oscillator, the second element is the number of bosons in the second oscillator and so on. Each individual oscillator has its own vacuum state, and each of these has its own energy content. The energy ($E_\mathbb{0}$) of the overall vacuum state, $\ket{\mathbb{0}}$, is the sum of the energy of the various local vacuum states: $E_\mathbb{0}=\frac{1}{2} \sum^N_{i=1} \hbar \omega_i$. 
\subsection{The Holstein-Primakoff Transformation}\label{subsec:HP_transformation}
Having discussed the harmonic oscillator, we are now ready to discuss the Holstein-Primakoff (HP) transformation.
\par Given an arbitrary particle of spin-$j$ ($j=\frac{1}{2},1,\frac{3}{2},2,\frac{5}{2},\ldots $), we commonly represent its spin quantum states through a set of orthonormal states $\{\ket{j,m}\}$ which are each a simultaneous eigenstate of the operators $\vspin{J}^2$ and $\spin{J}^z$:
				\begin{subequations}
				\label{eq:J^2_J_z_eigenspectrum}
				\begin{align}
				\vspin{J}^2 \ket{j,m} & = j(j+1) \hbar^2 \ket{j,m} \\
				\spin{J}^z \ket{j,m} & = m \hbar \ket{j,m}
				\end{align}
				\end{subequations}
As we know, the possible values of the magnetic spin quantum number, $m$, depends on $j$ and the \emph{nature} of $j$. If $j$ is an integral integer, then $m= \pm j,\pm (j-1), \ldots , 0$, while for half-integral $j$, $m= \pm j , \pm (j-1), \ldots , \pm 1/2$. The fact that the values of $m$ can be either: 1) all integral or, 2) all half-integral integers, creates some discomfort when writing algorithms for doing computations on systems which may potentially involve spin quantum numbers of different types. The ideal way to go about this would be to have a simple way of representing the possible $m$ values as function of some parameter  which is independent of whether $j$ is integral or half-integral. In fact, for a given $j$, it is easy to verify that the possible values of $m$ can be simply expressed as:
				\begin{equation}
				\label{eq:m_n_relation_a}
				m = j - n \  , \ \qquad \mbox{where } n = 0, 1, 2 , \ldots, 2j \ .
				\end{equation}
Note that, alternatively, we could also have chosen the form:
				\begin{equation}
				\label{eq:m_n_relation_b}
				m = j + n' \  ,  \qquad \mbox{where } n' = 0, -1, -2 , \ldots, -2j \ .
				\end{equation}
However, the form given in \eqref{eq:m_n_relation_a} is preferred because the parameter $n$ admits only nonnegative integers. But another reason why we choose $n$ (Eq. \ref{eq:m_n_relation_a}) over $n'$ (Eq. \ref{eq:m_n_relation_b}) is that the former allows a very natural transition to the second quantization scheme, and, as we shall see shortly, this is what leads to the Holstein-Primakoff transformation.
\par Indeed, the relation \eqref{eq:m_n_relation_a} implies that there is one-to-one correspondence between $m$ and $n$ for fixed $j$. So instead of the eigenstates $\{\ket{j,m}\}$, we can equally make use of the states $\{\ket{j,n}\}$ -- which are also simultaneous eigenstates of $\vspin{J}^2$  and $\spin{J}^z$. Just as the states $\{\ket{j,m}\}$ are often simply indicated as $\{\ket{m}\}$, we will often indicate the states $\{\ket{j,n}\}$ as $\{\ket{n}\}$. For example, for a spin-$1/2$, the states $\ket{m=+1/2}$ and $\ket{m=-1/2}$ in the $\ket{m}$ representation become $\ket{n=0}$ and $\ket{n=1}$ in the $\ket{n}$ representation, respectively. If the new states $\{\ket{n}\}$ bring to mind the simple quantum harmonic oscillator, then you already get where this is going. In particular, if we combine Eq. \eqref{eq:J^2_J_z_eigenspectrum} and Eq. \eqref{eq:m_n_relation_a}, we get:
				\begin{equation}
				\spin{J}^z \ket{n} = \hbar (j-n) \ket{n} 
				\end{equation}
where we have effected the transformation $\ket{j,m} \to \ket{n}$. Repeating the reasoning which led to Eq. \eqref{eq:Hamornic_hamil_2}, we see that we may express $\spin{J}^z$ as
				\begin{equation}
				\label{eq:J_z_in_b_b}
				\spin{J}^z = \hbar\left(j - \hat{b}^\dagger \hat{b}\right)
				\end{equation}
where 
				\begin{equation}
				\label{eq:b_occupation_no}
				\hat{b}^\dagger \hat{b} \ket{n} = n \ket{n} \ .
				\end{equation}
In the language of occupation number representation, the operator $\hat{b}^\dagger \hat{b}$ is the occupation number operator for some particles. The vacuum state of these particles, i.e. $\ket{n=0}$, is seen to correspond to the state $\ket{j,j}$; and as we increase $n$, $m$ decreases by the same degree.
\par To delineate further the parallelism between this way of talking about spin states and the harmonic oscillator, let us consider a single spin$-j$ interacting with a static magnetic field $\mathbf{B}_o = B_o \gv{e}_z$. The spin Hamiltonian, as we know, is:
				\begin{equation}
				\op{H} = -\gamma B_o \spin{J}^z 
				\end{equation}
where $\gamma$ is the spin's gyromagnetic ratio. According to Eq. \eqref{eq:J_z_in_b_b}, we can rewrite this Hamiltonian also as:
				\begin{equation}
				\label{eq:spin_hamil_in_n_single}
				\op{H} = \left(j-\hat{b}^\dagger \hat{b} \right) \hbar \omega
				\end{equation}
where $\omega := -\gamma B_o$ is the Larmor frequency. And the eigenvalues of $\op{H}$ are easily seen to be:
				\begin{equation}
				\label{eq:spin_eigenenergy_n_single}
				E_n = (j-n)\hbar \omega \ .
				\end{equation}
The quanta of energy is still $\hbar \omega$ and the vacuum energy here is thus $E_0 = j \hbar \omega$. We also observe from Eq. \eqref{eq:spin_eigenenergy_n_single} that the occupation number for the particles which $\hat{b}^\dagger \hat{b}$ counts indicates excitations in the spin, just as we saw for the quantum harmonic oscillator (Eq. \eqref{eq:Energy_harmonic_osci}). It is worthwhile to point out that while the vacuum state of the harmonic oscillator is a true ground state (that is, the state with the lowest energy), in the present case, the vacuum state $\ket{0}$ is not always the ground state. Here, $\ket{0}$ is the ground state only when $\gamma > 0$, while for $\gamma <0$ the $\ket{0}$ corresponds to the highest excited state.
\par Eqs. \eqref{eq:spin_eigenenergy_n_single} and \eqref{eq:spin_hamil_in_n_single} rightly reminds us of Eqs. \eqref{eq:harmonic_hamil_adagger_a} and \eqref{eq:Energy_harmonic_osci} from the harmonic oscillator problem, respectively. In addition, the operators $\hat{b}^\dagger$ and $\hat{b}$ have the same properties as the $\hat{a}^\dagger$ and $\hat{a}$ encountered when we discussed the harmonic oscillator, respectively, Eq. \eqref{eq:a_dagger_a}; namely:
			\begin{subequations}
			\begin{align}
			\label{eq:b_dagger_b}
			\hat{b} \ket{n} & = \sqrt{n} \ket{n-1} \ , \quad \quad \ \ \  1 \leq n \leq 2j \\
			\hat{b}^\dagger \ket{n} & = \sqrt{n+1} \ket{n+1} \ , \quad \  0 \leq n \leq 2j-1 
			\end{align}
			\end{subequations}
where the ranges on $n$ have been set so as to remain consistent with Eq. \eqref{eq:m_n_relation_a}. The above bounds imposed on $n$ mark a very important difference between the new states $\{\ket{n}\}$ for the spin and those for the quantum harmonic oscillator. We shall come back to this point very soon.
\par Eq. \eqref{eq:J_z_in_b_b} gives a second quantization representation of the spin operator $\spin{J}^z$, but this is not enough to allow us to fully do spin dynamics in second quantization. We also need to express the operators $\spin{J}^x$ and $\spin{J}^y$ in terms of the operators $\hat{b}^\dagger$ and $\hat{b}$. To achieve this, it is rather convenient to deal with their linear combinations $\spin{J}^\pm= \spin{J}^x \pm i \spin{J}^y $. From the theory of angular momentum, we know that, for example,
			\begin{equation}
			\label{eq:J_+_1}
			\spin{J}^+ \ket{m} = \hbar \sqrt{(j-m)(j+m+1)} \ket{m+1} \ .
			\end{equation}
From Eq. \eqref{eq:m_n_relation_a}, we have that if $\ket{m} \to \ket{n}$, then $\ket{m+1} \to \ket{n-1}$, therefore Eq. \eqref{eq:J_+_1} in the occupation number representation becomes:
			\begin{equation}
			\label{eq:J_+_2}
			\spin{J}^+ \ket{n} \equiv f^+(\hat{b}^\dagger, \hat{b}) \ket{n}= \hbar \sqrt{2j-(n-1)}\sqrt{n} \ket{n-1} 
			\end{equation}
where $f^+(\hat{b}^\dagger, \hat{b})$ is simply the operator $\spin{J}^+$ written in terms of $\hat{b}^\dagger$ and $\hat{b}$. Our objective is to find $f^+(\hat{b}^\dagger, \hat{b})$. From the first equation of Eq. \eqref{eq:b_dagger_b}, we note that:
			\begin{equation}
			\label{eq:J_+_in_b_2}
			f^+(\hat{b}^\dagger, \hat{b}) \ket{n}= \hbar \sqrt{2j-(n-1)} \ \hat{b}\ket{n} \ .
			\end{equation}
Given that $\hat{b}\ket{n} \propto \ket{n-1}$, and the final state must remain $\ket{n-1}$, the operator which generates the coefficient $\sqrt{2j-(n-1)}$ must be in function of the occupation number operator, $\hat{b}^\dagger \hat{b}$, Eq. \eqref{eq:b_occupation_no}. We are then lead to the conclusion that:
			\begin{equation}
			\label{eq:J_+_in_b}
			\spin{J}^+ = \hbar\sqrt{2j-\hat{b}^\dagger \hat{b}} \ \hat{b}\ .
			\end{equation}			  
Since $\spin{J}^-$ and $\spin{J}^+$ are Hermitian conjugate of each other, it follows immediately from Eq. \eqref{eq:J_+_in_b} that:
			\begin{equation}
			\label{eq:J_-_in_b}
			\spin{J}^- = \hbar \ \hat{b}^\dagger \ \sqrt{2j-\hat{b}^\dagger \hat{b}} \ .
			\end{equation}	
Eqs. \eqref{eq:J_z_in_b_b}, \eqref{eq:J_+_in_b} and \eqref{eq:J_-_in_b} constitute the \emph{Holstein-Primakoff transformation}. In the HP transformation the usual spin operators $\spin{J}^x, \spin{J}^y, \spin{J}^z$ are all written in function of a single operator $\hat{b}^\dagger$ and its Hermitian conjugate, $\hat{b}$. The particles which $\hat{b}^\dagger$ ($\hat{b}$) creates (annihilates) are called the \emph{Holstein-Primakoff bosons}.	We emphasize that the HP bosons simply represent spin excitations, and we should not go beyond this interpretation. Interestingly, while $\spin{J}^+ \ket{m} \propto \ket{m+1}$ and $\spin{J}^-\ket{m} \propto \ket{m-1}$, we observe from Eqs. \eqref{eq:J_+_in_b} and \eqref{eq:J_-_in_b} that  $\spin{J}^+ \ket{n} \propto \ket{n-1}$ and $\spin{J}^-\ket{n} \propto \ket{n+1}$, which is consistent with the fact that an increase in $m$ implies a decrease in $n$ and vice versa (because the sum $m+n$ is conserved, Eq. \eqref{eq:m_n_relation_a}). Therefore, the operator $\spin{J}^+$ (unlike $\hat{b}^\dagger$) annihilates HP bosons, but $\spin{J}^-$ creates them. In complete analogy to the $m-$representation where we know $\spin{J}^+$ cannot increase the magnetic spin quantum number $m$ of the state $\ket{m}$ indefinitely, in the occupation number representation $\spin{J}^+$ cannot annihilate the HP bosons indefinitely but ends when $n=0$, which corresponds to the HP vacuum state. Analogously, $\spin{J}^-$ can fill the vacuum state with a maximum number of HP bosons, namely, $n=2j$. This limit is set by the operator $\sqrt{2j-\hat{b}^\dagger \hat{b}}$ in the definition of $\spin{J}^-$, Eq. \eqref{eq:J_-_in_b}. This is all consistent with Eq. \eqref{eq:m_n_relation_a} where it was evident that $n$ is a nonnegative integer, and whose range is bounded by the spin quantum number $j$: $0 \leq n \leq 2j$. 

From Eqs. \eqref{eq:J_z_in_b_b}, \eqref{eq:J_+_in_b} and \eqref{eq:J_-_in_b} we also derive the following:
			\begin{subequations}
			\begin{align}
			\spin{J}^x & = \hbar\frac{\sqrt{2j-\hat{b}^\dagger \hat{b}} \ \hat{b} + \hat{b}^\dagger \ \sqrt{2j-\hat{b}^\dagger \hat{b}}}{2} \\
			\spin{J}^y & = \hbar\frac{\sqrt{2j-\hat{b}^\dagger \hat{b}} \ \hat{b} - \hat{b}^\dagger \ \sqrt{2j-\hat{b}^\dagger \hat{b}}}{2i} \\
			 \vspin{J}^2 & = \hbar^{2}j(j+1) \hat{\mathbb{I}} \label{eq:J^2}\ . 			
			\end{align}
			\end{subequations}
Most importantly, one can verify that the Holstein-Primakoff transformation preserves the commutation relations:
			\begin{equation}
			\left[\spin{J}^\alpha, \spin{J}^\beta \right] = i \hbar \epsilon^{\alpha \beta \gamma} \spin{J}^\gamma 
			\end{equation}
or their equivalent:
			\begin{subequations}
			\label{eq:commutations_J_z_J_pm}
			\begin{align}
			\left[\spin{J}^z, \spin{J}^\pm \right] & = \pm \hbar \spin{J}^\pm \\ 
			\left[\spin{J}^+, \spin{J}^- \right] & = 2\hbar \spin{J}^z 
			\end{align}
			\end{subequations}
where $\alpha,\beta$ and $\gamma$ represent any of the three directions $x,y,z$ and $\epsilon^{\alpha \beta \gamma}$ is the three-dimensional Levi-Civita symbol. Hence, working in the HP representation is just the same as in the $m-$representation, in the sense that the physics does not change.
\par In analogy to the quantum harmonic oscillator, the HP spin states $\{\ket{n}\}$ can all be generated from the vacuum state $\ket{0}$:
			\begin{equation}
			\ket{n} = \frac{( \hat{b}^\dagger)^n}{\sqrt{n!}} \ket{0} \ , \qquad 0\leq n \leq 2j \ .
			\end{equation}
We could also obtain $\ket{n}$ from $\ket{0}$ by applying repeatedly $\spin{J}^-$. This has the advantage of incorporating inevitably the bound on $n$. Certainly, 
				\begin{equation}
				\ket{n} \propto \left(\frac{\spin{J}^-}{\hbar}\right)^n \ket{0} \ .
				\end{equation}
Indeed, the following identities can be easily proved:
				\begin{equation}
				\label{eq:n_def_1}
				\ket{n} = \frac{\left(\frac{\spin{J}^-}{\hbar}\right)^n}{n! \sqrt{\binom{2j}{n}}} \ket{0}
				\end{equation}
				
				\begin{equation}
				\label{eq:n_def_2}
				\ket{n} = \frac{\prod^n_{k=1}(2j+k-\hat{b}^\dagger \hat{b})^{1/2}   }{n! \sqrt{\binom{2j}{n}}}\ (\hat{b}^\dagger)^n \ket{0}
				\end{equation}								
				
				\begin{equation}
				\label{eq:n_def_3}
				\ket{n} = \hat{\Lambda}(j,n) \  \frac{(\hat{b}^\dagger)^n}{\sqrt{n!}} \ket{0} 
				\end{equation}
where, the operator $\hat{\Lambda}(j,n)$ is defined as:
				\begin{equation}
				\label{eq:controller_op}
				\hat{\Lambda}(j,n) := \frac{\sqrt{\multiset{2j+1-\hat{b}^\dagger \hat{b}}{n}}}{\sqrt{\binom{2j}{n}}} 
				\end{equation}
and where $\multiset{x}{n}$ is defined as:
				\begin{equation}
				\label{eq:rising_factorial}
				\multiset{x}{n} := \frac{x(x+1)(x+2) \ldots (x+n-1) }{n!} \ .
				\end{equation}
$x$ in Eq. \eqref{eq:rising_factorial} can be a number or an operator. For a nonnegatve integer $x$, $\multiset{x}{n}$ is what we call the \emph{multiset coefficient} (read as "\emph{x multichoose n}"), and it gives the number of combinations of length $n$ we can get from a set of $x$ elements, if we allow repetition of elements and disregard order\citep{misc:Stanley-2011}. For example, the multisets of cardinality $2$ we can get from a set of cardinality $3$ like $\{i,j,k\}$ are: $\{i,i\}, \{i,j\}, \{i,k\}, \{j,j\}, \{j,k\}, \{k,k\}$; so $\multiset{3}{2}=6$, which can be checked from Eq. \eqref{eq:rising_factorial}. By definition, $\multiset{x}{n}=1$ when $n=0$ and $x$ is a scalar; while $\multiset{x}{n}=\hat{\mathbb{I}}$ when $n=0$ and $x$ is an operator. To prove Eqs. \eqref{eq:n_def_2} and \eqref{eq:n_def_3} from \eqref{eq:n_def_1} the following identity is very useful:
				\begin{equation}
				(\hat{b}^\dagger)^k \sqrt{2j-\hat{b}^\dagger \hat{b}} = \sqrt{2j+k-\hat{b}^\dagger \hat{b}} \ (\hat{b}^\dagger)^k 
				\end{equation}
for nonnegative integer $k$. The identity can be easily proved using the properties of the operators $\hat{b}^\dagger$ and $\hat{b}$, Eqs. \eqref{eq:b_dagger_b}.
\par The operator $\hat{\Lambda}(j,n)$ defined in Eq. \eqref{eq:controller_op} is a very interesting one. It can be easily shown that:
				\begin{equation}
				\label{eq:cut-off_2}
				\hat{\Lambda}(j,n) \ket{n} = \begin{cases}
				\ket{n}  \ , & \mbox{if } 0 \leq n \leq 2j \\
				\mbox{nonexistent}  \ , & \mbox{otherwise}
				\end{cases} \ .
				\end{equation}
  Eq. \eqref{eq:n_def_3} is very telling: $\frac{(\hat{b}^\dagger)^n}{\sqrt{n!}} \ket{0}$ certainly yields the quantum harmonic oscillator state $\ket{n}$ -- Eq. \eqref{eq:n_from_vacuum_harmonic}. So, for now, $n$ can be any nonnegative integer, just like with the quantum harmonic oscillator. But by multiplying $\frac{(\hat{b}^\dagger)^n}{\sqrt{n!}} \ket{0}$ from the left by the operator $\hat{\Lambda}(j,n)$, we restrict the range of values for $n$ to $0 \leq n \leq 2j$. Evidently, $\hat{\Lambda}(j,n)$ plays the role of a "cut-off" operator: while it forbids $n$ from being outside the range $0\leq n \leq 2j$, it reduces to the identity operator when $n$ is within the same range. Interestingly, this observation on the properties of $\hat{\Lambda}(j,n)$ entails a very peculiar implication. It begins with the observation that,
  				\begin{equation}
  				\lim_{j \to \infty}  \hat{\Lambda}(j,n) = \hat{\mathbb{I}}
  				\end{equation}
according to Eq. \eqref{eq:controller_op}. But if $\hat{\Lambda}(j,n)$ becomes the identity operator for any occupation number $n$, then there is a one-to-one correspondence between the states $\{\ket{n}\}$ in Eq. \eqref{eq:n_def_3} and the quantum harmonic oscillator states $\{\ket{n}\}$ we saw in Eq. \eqref{eq:n_from_vacuum_harmonic}. This means that the eigenvectors of the quantum harmonic oscillator are the same as those of a spin-$\infty$ particle in Fock space\citep{art:Gyamfi-2018}. Conversely, we can interpret the Fock space in which the spin states $\{ \ket{n}\}$ are defined as a truncated version of the quantum harmonic's Fock space by means of the the operator $\hat{\Lambda}(j,n)$.
\par The next sections dedicated to multispin systems notwithstanding, we briefly introduce the topic here. Given an arbitrary finite collection of spins, their spin quantum numbers form a multiset, which we shall indicate as $\mathpzc{A}$, i.e. $\mathpzc{A}=\{j_1,j_2,\ldots, j_N\}$. Multispin states are normally indicated in the $m-$representation either using the coupled or uncoupled representations\citep{book:Zare-2018}. In the latter, the states, $\{\ket{j_1m_1,j_2m_2, \ldots , j_N m_N}\}$, are simultaneous eigenstates of the individual $\vspin{J}^2_i$ and $\spin{J}^z_i$ operators. In the HP occupation number representation, $\ket{j_1m_1,j_2m_2, \ldots , j_N m_N} \mapsto \ket{n_1, n_2, \ldots , n_N}$, where $n_i$, $0\leq n_i \leq 2j_i$, is the number of HP bosons introduced into the HP vacuum state of the $i-$th spin, and $N$ is the total number of spins. Each possible combination of occupation numbers $\{n_1,n_2, \ldots ,n_N\}$ also forms an ordered multiset and corresponds to a specific multispin state. The overall HP vacuum state is obviously the $\ket{n_1, n_2, \ldots , n_N}$ state with $n_i=0$ for every $i$. If we indicate such state as $\ket{\mathbb{0}}$, it trivially follows from Eq. \eqref{eq:n_def_3} that:
				\begin{equation}
				\ket{n_1, n_2, \ldots , n_N} = \hat{\Lambda}(j_1,n_1;\ldots;j_N,n_N) \ \prod^N_{i=1}\frac{(\hat{b}_i^\dagger)^{n_i}}{\sqrt{n_i!}} \ket{\mathbb{0}}  
				\end{equation}
where,
				\begin{equation}
				\hat{\Lambda}(j_1,n_1;\ldots;j_N,n_N) := \prod^N_{i=1} \frac{\sqrt{\multiset{2j_i+1-\hat{b}_i^\dagger \hat{b}_i}{n_i}}}{\sqrt{\binom{2j_i}{n_i}}}
				\end{equation}
If we consider a multiset of noninteracting spins $\mathpzc{A}=\{j_1,j_2,\ldots, j_N\}$ placed in a region with static magnetic field $\mathbf{B}_o=B_o \gv{e}_z$, the multispin Hamiltonian of the system is a facile generalization of Eq. \eqref{eq:spin_hamil_in_n_single}:
				\begin{equation}
				\op{H} = \sum^N_{i=1} \left(j_i - \hat{b}^\dagger_i \hat{b}_i \right) \hbar \omega_i
				\end{equation}
and the eigenenergies are simply:
				\begin{equation}
				E_{n_1,n_2, \ldots , n_N} = \sum^N_{i=1} \left(j_i - n_i \right) \hbar \omega_i \ .
				\end{equation}
and the total vacuum energy is $E_\mathbb{0} = \sum_i j_i \hbar \omega_i$. Like the single spin HP vacuum state, the multispin analogue is not necessarily the ground state. To get the ground state from $\ket{\mathbb{0}}$, one needs to fill all localized vacuum states characterized by nonnegative $\omega_i$ (i.e. negative $\gamma_i$) with HP bosons to their maximum capacity and leave those with negative $\omega_i$ empty. 

\section{Multispin magnetic resonance with the Holstein-Primakoff Transformation}\label{sec:HP_trans_and_multispin}
\par We have seen that the uncoupled representation of the multispin states associated with $\mathpzc{A}=\{j_1,j_2,\ldots, j_N\}$ are $\{\ket{n_1,n_2,\ldots,n_N}\}$ in the HP occupation number representation, and the possible occupation numbers $(n_1,n_2,\ldots,n_N)$ are multisets of integers. To effectively discuss multispin dynamics, it is convenient to have a reasonable shorthand for indicating a generic state $\ket{n_1,n_2,\ldots,n_N}$. The so-called \emph{index compression maps} are devised just for this purpose and they will be our first order of business in this section. In the second part, we will analyze the eigendecomposition problem for some general forms of isotropic multispin Hamiltonians using the HP representation.

\subsection{Index compression maps}\label{subsec:Index_comp_map}
A detailed discussion on index compression maps is beyond the scope of this article, so we are going to keep the mathematical details to the minimum. We will concern ourselves here with only one special kind of index compression maps, denoted $\eta_0$. For more on index compression maps see \citep{misc:Gyamfi-2018}. 
Given a spin multiset $\mathpzc{A}=\{j_1,j_2,\ldots, j_N\}$, the index compression map,  $\eta_0$, maps each possible combination of occupation numbers $(n_1,n_2,\ldots,n_N)$ to a unique integer of the set $\{\mathbb{n}\}$. The mapping is one-to-one. $\eta_0$ is simply a function which has the HP occupation numbers $n_1,n_2, \ldots , n_N$ as variables. To be more specific:
			\begin{equation}
			\label{eq:eta_0}
			\mathbb{n} = \eta_0(n_1,n_2, \ldots , n_N) = \sum^N_{i=1} \mathcal{W}_{R,i} \cdot n_i \ .
			\end{equation}
where,
			\begin{equation}
			\label{eq:W_R_i}
			\mathcal{W}_{R,i} := \delta_{N,i} + (1-\delta_{N,i}) \prod^{N-i}_{k=1} d_{i+k}
			\end{equation}
where $\delta_{i,i'}$ is the Kronecker delta and $d_i$ is the dimension of the spin-$i$'s spin Hilbert space, i.e. $d_i = 2j_i + 1$. In more comprehensible terms, for $i \neq N$, $\mathcal{W}_{R,i}$ is the product of all the $d_{i'}$ with $i' > i$; while for $i=N$,  $\mathcal{W}_{R,i}=1$. For instance, $\mathcal{W}_{R,2}=d_3 \cdot d_4 \cdots d_N $. In the same spirit, $\mathcal{W}_{R,0}$ then corresponds to the dimension $D_{\mathcal{H}}$ of the entire spin Hilbert space of the multispin system. We thus infer from Eq. \eqref{eq:eta_0} that the possible values of $\mathbb{n}$ are: $0,1,2, \ldots , (D_{\mathcal{H}} - 1)$\citep{misc:Gyamfi-2018}. Note that since each of the states $\ket{n_1,n_2, \ldots, n_N}$ constitute a basis for the multispin Hilbert space, the dimension of the latter must coincide with the number of all possible combinations of the HP occupation numbers  $\{n_1,n_2, \ldots, n_N\}$, which is precisely $D_{\mathcal{H}}$. These combinations are ordered and restricted: restricted in the sense that for each $n_i$, $0 \leq n_i \leq 2j_i$, as we mentioned above. The range of $\mathbb{n}$ is thus consistent with the dimension of the multispin Hilbert space.
\par We adopt the following notation: if $\mathbb{n}$ is related to the multiset $\{n_1,n_2,\ldots,n_N\}$ through the index compression map $\eta_0$, Eq. \eqref{eq:eta_0}, then we shall write the ket $\ket{n_1,n_2,\ldots,n_N}$ as $\ket{\mathbb{n}}$. The integers $\mathbb{n}$ will be indicated with the font $\mathbb{O}, \mathbb{1}, \mathbb{2}, \mathbb{3}, \ldots$ to differentiate them from the single spin states. The complete HP vacuum state of the system always has $\mathbb{n}=\mathbb{0}$, and so $\ket{0,0,\ldots, 0} \mapsto \ket{\mathbb{0}}$, consistent with our previous notation for this particular state.
\par Let us return to the deuterated hydroxymethyl radical (\ce{^{.}CH2OD}). There are only four nonzero spins we need to take into consideration for the simulation of its ESR spectrum: 1) the radical electron, 2) the first hydrogen nuclear, 3) the second hydrogen nuclear, and 4) the deuterium nuclear. If we number the spins according to the same order, then the spin multiset in this case is $\mathpzc{A}=\{j_1 , j_2 , j_3 , j_4\}=\{\frac{1}{2} , \frac{1}{2} , \frac{1}{2} , 1\}$. And from Eq \eqref{eq:eta_0} we can write:
				\begin{equation}
				\label{eq:DE_CH2OD_rad}
				\mathbb{n} = 12 n_1 + 6 n_2 + 3 n_3 + n_4 \ .
				\end{equation}
In table \ref{tab:CH2OD_table_1} the elements of \ce{^{.}CH2OD}'s spin Hilbert space basis  in the uncoupled representation ($\ket{m_1,m_2,m_3,m_4}$) are translated into the HP occupation number representation ($\ket{n_1,n_2,n_3,n_4}$), and their respective shorthand notation $\ket{\mathbb{n}}$ according to the index compression map $\eta_0$ is also given. As one can easily observe, the relationship between the integers $\mathbb{n}$ and the occupation numbers $(n_1,n_2,n_3,n_4)$ according to the mapping $\eta_0$ strictly depends on how we label the spins.
	\begin{table}[htb!]
	\centering
	\begin{tabular}{| c | c | c |}
	\hline	
	$ \ket{m_1,m_2,m_3,m_4}$ & $ \ket{n_1,n_2,n_3,n_4}$ & $\ket{\mathbb{n}}$ \\
	\hline
	$ \ket{+\frac{1}{2},+\frac{1}{2},+\frac{1}{2},+1}$ & $ \ket{0,0,0,0}$ & $\ket{\mathbb{0}}$ \\
	$ \ket{+\frac{1}{2},+\frac{1}{2},+\frac{1}{2},0}$ & $ \ket{0,0,0,1}$ & $\ket{\mathbb{1}}$ \\
	$ \ket{+\frac{1}{2},+\frac{1}{2},+\frac{1}{2},-1}$ & $ \ket{0,0,0,2}$ & $\ket{\mathbb{2}}$ \\
	$ \ket{+\frac{1}{2},+\frac{1}{2},-\frac{1}{2},+1}$ & $ \ket{0,0,1,0}$ & $\ket{\mathbb{3}}$ \\
	$ \ket{+\frac{1}{2},+\frac{1}{2},-\frac{1}{2},0}$ & $ \ket{0,0,1,1}$ & $\ket{\mathbb{4}}$ \\
	$ \ket{+\frac{1}{2},+\frac{1}{2},-\frac{1}{2},-1}$ & $ \ket{0,0,1,2}$ & $\ket{\mathbb{5}}$ \\
	$ \ket{+\frac{1}{2},-\frac{1}{2},+\frac{1}{2},+1}$ & $ \ket{0,1,0,0}$ & $\ket{\mathbb{6}}$ \\
	$ \ket{+\frac{1}{2},-\frac{1}{2},+\frac{1}{2},0}$ & $ \ket{0,1,0,1}$ & $\ket{\mathbb{7}}$ \\
	$ \ket{+\frac{1}{2},-\frac{1}{2},+\frac{1}{2},-1}$ & $ \ket{0,1,0,2}$ & $\ket{\mathbb{8}}$ \\
	$ \ket{+\frac{1}{2},-\frac{1}{2},-\frac{1}{2},+1}$ & $ \ket{0,1,1,0}$ & $\ket{\mathbb{9}}$ \\
	$ \ket{+\frac{1}{2},-\frac{1}{2},-\frac{1}{2},0}$ & $ \ket{0,1,1,1}$ & $\ket{\mathbb{10}}$ \\
	$ \ket{+\frac{1}{2},-\frac{1}{2},-\frac{1}{2},-1}$ & $ \ket{0,1,1,2}$ & $\ket{\mathbb{11}}$ \\
	$ \ket{-\frac{1}{2},+\frac{1}{2},+\frac{1}{2},+1}$ & $ \ket{1,0,0,0}$ & $\ket{\mathbb{12}}$ \\
	$ \ket{-\frac{1}{2},+\frac{1}{2},+\frac{1}{2},0}$ & $ \ket{1,0,0,1}$ & $\ket{\mathbb{13}}$ \\
	$ \ket{-\frac{1}{2},+\frac{1}{2},+\frac{1}{2},-1}$ & $ \ket{1,0,0,2}$ & $\ket{\mathbb{14}}$ \\
	$ \ket{-\frac{1}{2},+\frac{1}{2},-\frac{1}{2},+1}$ & $ \ket{1,0,1,0}$ & $\ket{\mathbb{15}}$ \\
	$ \ket{-\frac{1}{2},+\frac{1}{2},-\frac{1}{2},0}$ & $ \ket{1,0,1,1}$ & $\ket{\mathbb{16}}$ \\
	$ \ket{-\frac{1}{2},+\frac{1}{2},-\frac{1}{2},-1}$ & $ \ket{1,0,1,2}$ & $\ket{\mathbb{17}}$ \\
	$ \ket{-\frac{1}{2},-\frac{1}{2},+\frac{1}{2},+1}$ & $ \ket{1,1,0,0}$ & $\ket{\mathbb{18}}$ \\
	$ \ket{-\frac{1}{2},-\frac{1}{2},+\frac{1}{2},0}$ & $ \ket{1,1,0,1}$ & $\ket{\mathbb{19}}$ \\
	$ \ket{-\frac{1}{2},-\frac{1}{2},+\frac{1}{2},-1}$ & $ \ket{1,1,0,2}$ & $\ket{\mathbb{20}}$ \\
	$ \ket{-\frac{1}{2},-\frac{1}{2},-\frac{1}{2},+1}$ & $ \ket{1,1,1,0}$ & $\ket{\mathbb{21}}$ \\
	$ \ket{-\frac{1}{2},-\frac{1}{2},-\frac{1}{2},0}$ & $ \ket{1,1,1,1}$ & $\ket{\mathbb{22}}$ \\
	$ \ket{-\frac{1}{2},-\frac{1}{2},-\frac{1}{2},-1}$ & $ \ket{1,1,1,2}$ & $\ket{\mathbb{23}}$ \\
	\hline
	\end{tabular}
	\caption{Basis kets of \ce{^{.}CH2OD}'s spin Hilbert space in the uncoupled representation ($\ket{m_1,m_2,m_3,m_4}$), the HP representation ($\ket{n_1,n_2,n_3,n_4}$) and their shorthand notation $\ket{\mathbb{n}}$ according to the index compression map $\eta_0$.}
	\label{tab:CH2OD_table_1}
	\end{table}
\par It is interesting to note that once we know the labels on the spins from the multiset $\mathpzc{A}=\{j_1,j_2, \ldots,j_N\}$, we can determine the corresponding HP occupation numbers $n_1,n_2,\ldots,n_N$ for any given $\mathbb{n}$. This is because the map $\eta_0$ is invertible and has a unique inverse: $\eta^{-1}_0(\mathbb{n})=\{n_1,n_2,\ldots, n_N\}$. In this sense, Eq. \eqref{eq:eta_0} can be viewed as a linear multivariable Diophantine equation in the variables $n_1,n_2,\ldots, n_N$. The problem is somehow complicated by the fact that each variable $n_i$ is restricted to a specific range. In any event, there is a simple algorithm one can follow to solve Eq. \eqref{eq:eta_0} for $n_1,n_2,\ldots, n_N$ when $\mathbb{n}$ is fixed. It goes as follows: Find the integer $n_1$, $0 \leq n_1 \leq 2j_1$, which maximizes the product $\mathcal{W}_{R,1} n_1$ but still keeps the latter less or equal to the given $\mathbb{n}$; find then $n_2$ in the range $0 \leq n_2 \leq 2j_2$ which maximizes the sum $\mathcal{W}_{R,1} \cdot n_1 + \mathcal{W}_{R,2} \cdot n_2$, still keeping the sum less or equal to $\mathbb{n}$. Continue with the procedure till you get to $n_N$. In the particular case where along the way you get to a certain occupation number $n_k$, where $n_1 \leq n_k < n_N$, such that $\mathbb{n}= \mathcal{W}_{R,1} \cdot n_1 + \mathcal{W}_{R,2} \cdot n_2 + \ldots + \mathcal{W}_{R,k} \cdot n_k$, then $n_{k+1}=n_{k+2} = \ldots = n_{N}=0$.
\par For instance, say we want to find the HP occupation numbers $n_1,n_2,n_3,n_4$ corresponding to $\mathbb{n}=\mathbb{11}$ for the \ce{^{.}CH2OD} radical (assuming we maintain the same choice of spin labelling as done above). This amounts to solving the Diophantine equation:
				\begin{equation}
				\label{eq:DE_CH2OD_rad_2}
				\mathbb{11} = 12 n_1 + 6 n_2 + 3 n_3 + n_4 
				\end{equation}
in accordance with Eq. \eqref{eq:DE_CH2OD_rad}, knowing that $n_1,n_2,n_3$ can be or $0$ or $1$, while $n_4$ can be any nonnegative integer not greater than $2$. According to the algorithm discussed above, we see that $n_1$ must necessarily be $0$, $n_2$ and $n_3$ must be $1$ and, finally, $n_4=2$. Thus, $\ket{\mathbb{11}}=\ket{0,1,1,2}$, which is in agreement with Table \eqref{tab:CH2OD_table_1}. It therefore goes without saying that by means of the map $\eta_0$, we can easily encode multispin states in the form of integers, and the decoding process is as easy as that of encoding.
\par Before moving on, we point out that, unlike the multispin HP boson Fock states, the map $\eta_0$ cannot be applied to the boson Fock states for a finite collection of quantum harmonic oscillators (not even for a single oscillator). This is because the single spin HP boson Fock spaces we encounter in Nature are always finite in dimension (due to the action of the operator $\hat{\Lambda}(j,n)$, Eq. \eqref{eq:cut-off_2}) -- this is in stark contrast to the boson Fock space related to a single quantum harmonic oscillator, which is of infinite dimension.
\subsection{Eigendecomposition of Isotropic Multispin Hamiltonians}
\par We discuss here how the HP transformation and the index compression map $\eta_0$ can be deployed to solve the eigendecomposition problem for isotropic multispin Hamiltonians. We will be led to a computational scheme that fails to scale polynomially with the number of spins in some cases, but yet far better than the tout court approach. One can consider what we discuss below as the first step to systematically address the curse of dimensionality within the occupation number representation approach, with particular emphasis on the HP transformation. It is our opinion that numerical analysis of the problem alone won't get us far enough and that we should resort to the latter after we have exhausted our mathematical tricks and conveniences within a given theoretical approach. One of the advantages of the HP transformation is that it has the tendency of making it possible to find analytical solutions to seemingly intractable problems. This analytical element can be very useful when writing algorithms for multispin systems.
\par Symmetry arguments cannot be ignored in problems like the one we are about to discuss, and we will greatly make use of them. But it is important to note that symmetry arguments can get us to confidently reach certain conclusions without sharing light on \emph{how} to effectively carry on the needed calculations. This is where the HP transformation and the index compression map $\eta_0$ come to our aid. To make the symmetry arguments more intelligible we are not going to consider the general isotropic multispin Hamiltonian in the presence of a static magnetic field straightaway, but we will get there through a gradual process. And along the way, we will discuss various symmetries and analyze the spin Hamiltonians we encounter with the help of the HP transformation and $\eta_0$.
\par Without loss of generality, we have set $\hbar = 1$ in the following.
\subsubsection{The isotropic multispin Hamiltonian in the absence of an external magnetic field}
\par In the introduction section of the current paper, we saw that in the absence of external magnetic fields, the isotropic spin Hamiltonian of an arbitrary collection of spins $\mathpzc{A}=\{j_1,j_2,\ldots,j_N\}$ takes the form of $\op{H}_{spin-spin}$, Eq. \eqref{eq:H_spin-spin}, which we now rewrite as\citep{art:Corio-1960}:
				\begin{equation}
				\label{eq:H_spin-spin_2}
				\op{H}_{spin-spin} =  \sum_{i > i'} T_{i,i'} \vspin{J}_i \cdot \vspin{J}_{i'} = \sum_{i > i'} T_{i,i'} \left[\spin{J}^z_i \spin{J}^z_{i'} + \frac{1}{2} \left( \spin{J}^+_i \spin{J}^-_{i'} + \spin{J}^-_i \spin{J}^+_{i'}\right)\right] \ .
				\end{equation}			  
$H_{spin-spin}$ manifestly remains the same in all directions (rotationally invariant). Not surprisingly, it is a zero rank spherical tensor operator. Another way of proving this is by recalling the definition of spherical tensor operators. A spherical tensor operator of rank $k$ can be defined as an operator with $(2k+1)$ irreducible components which obey the following commutation relations\citep{book:Sakurai-2011}:
				\begin{subequations}
				\label{eq:def_spherical_tensors}
				\begin{align}
				\left[\spin{J}^z, \hat{T}^{(k)}_q \right] & = \hbar q \hat{T}^{(k)}_q \label{eq:def_spherical_tensors_a}\\
				\left[\spin{J}^\pm, \hat{T}^{(k)}_q \right] & = \hbar \sqrt{(k \mp q)(k \pm q +1)}  \hat{T}^{(k)}_{q\pm 1}
				\end{align}
				\end{subequations}
with $k=0, 1, 2, \ldots$ and $-k \leq q \leq +k$. (Spherical tensor operators are often defined according to how they behave under rotation\citep{book:Zare-2018}. This involves the use of Wigner matrices and we would like to avoid them for now. We therefore stick to the definition above, which is much simpler to handle.) For example, if we compare the commutation relations in Eqs. \eqref{eq:commutations_J_z_J_pm} with Eqs. \eqref{eq:def_spherical_tensors}, we note that the trio $\spin{J}^z, \spin{J}^\pm$ must be proportional to the components of a rank $k=1$ spherical tensor. Indeed, if we set:
				\begin{equation}
				\label{eq:examples_rank_1_tensors}
				\hat{T}^{(1)}_{0} = \spin{J}^z \qquad \hat{T}^{(1)}_{\pm 1} = \mp \frac{1}{\sqrt{2}} \spin{J}^{\pm} 
				\end{equation}
we see that the relations in Eq. \eqref{eq:def_spherical_tensors} are perfectly observed.
\par Going back to $\op{H}_{spin-spin}$, it can easily be verified that:
				\begin{subequations}
				\label{eq:H_spin_spin_spherical_commutations}
				\begin{align}
				\left[\spin{J}^z_{tot}, \op{H}_{spin-spin} \right] & = \hat{0} \label{eq:H_spin_spin_spherical_commutations_a}\\
				\left[\spin{J}^\pm_{tot},  \op{H}_{spin-spin} \right] & = \hat{0} \label{eq:H_spin_spin_spherical_commutations_b}
				\end{align}
				\end{subequations}
where, $\spin{J}^z_{tot} = \sum^N_{i=1} \spin{J}^z_i$ and $\spin{J}^\pm_{tot} = \sum^N_{i=1} \spin{J}^\pm_i$. In line with Eqs. \eqref{eq:def_spherical_tensors}, we observe that since $\op{H}_{spin-spin}$ is not a null operator, Eqs. \eqref{eq:H_spin_spin_spherical_commutations} hold only if $\op{H}_{spin-spin}$ is proportional to a zero rank tensor ($k=0$). From Eq. \eqref{eq:H_spin_spin_spherical_commutations_b}, we also derive that:
				\begin{subequations}
				\begin{align}
				\left[\spin{J}^x_{tot},  \op{H}_{spin-spin} \right] & = \hat{0} \label{eq:H_spin_spin_spherical_commutations_c}\\
				\left[\spin{J}^y_{tot},  \op{H}_{spin-spin} \right] & = \hat{0} \label{eq:H_spin_spin_spherical_commutations_d}
				\end{align}
				\end{subequations}
The conclusion we draw from the commutation relations stated in Eqs. \eqref{eq:H_spin_spin_spherical_commutations_a}, \eqref{eq:H_spin_spin_spherical_commutations_c} and \eqref{eq:H_spin_spin_spherical_commutations_d} is that the total spin angular momentum must be conserved along the three axes, just as one would expect from a rotationally invariant operator like $\op{H}_{spin-spin}$. The operators $\spin{J}^x_{tot}, \spin{J}^y_{tot}, \spin{J}^z_{tot}$ are thus constants of motion for an isolated spin system whose spin Hamiltonian is given by $\op{H}_{spin-spin}$. Note that any linear combination of these three operators is also a constant of motion. Thus, the total spin angular momentum vector operator $\vspin{J}_{tot}$, for example, is also a constant of motion. Not only that: any operator of the form $\hat{X}^r$, where $r=1,2,3,\ldots$ and $\hat{X}$ is any of the operators $\vspin{J}_{tot}$,  $\spin{J}^x_{tot}$, $\spin{J}^y_{tot}$, $\spin{J}^z_{tot}$, is also a constant of motion. The same applies to any linear combination of such powers of operators. However, not all pairs of operators of this vast family of constants of motions commute with each other. For example, $\spin{J}^x_{tot}$ does not commute with $\spin{J}^y_{tot}$. This is crucial because the easiest way to determine the energy spectrum of $\op{H}_{spin-spin}$, for example, is to express it as a function of a subset of mutually commuting operators which belong to this infinitely numerable family of constants of motion. Unfortunately, as far as we know at the moment, $\op{H}_{spin-spin}$ as given in Eq. \eqref{eq:H_spin-spin_2} fails to succumb to this mathematical contrivance due to the arbitrary difference between the coupling constants $\{T_{i,i'}\}$. Numerical diagonalization of $\mathscr{H}_{spin-spin}$, therefore, seems the most reasonable route to the eigenvalues and eigenvectors of $\op{H}_{spin-spin}$, especially when dealing with multispin systems.
\par As mentioned earlier, one could employ the tout court approach whereby one simply diagonalizes $\mathscr{H}_{spin-spin}$ in whole. The other approach is to first divide $\mathscr{H}_{spin-spin}$ into independent sub-units and then proceed with the diagonalization of each sub-unit. The HP transformation and the index compression map $\eta_o$ discussed in \S \ref{subsec:HP_transformation} and \S \ref{subsec:Index_comp_map}, respectively, enables us to break $\mathscr{H}_{spin-spin}$ into sub-units in a computationally efficient manner.
\par First of all, consider two arbitrary multispin states $\ket{\mathbb{n}}(=\ket{n_1, \ldots, n_N})$ and $\ket{\mathbb{n}'}(=\ket{n'_1, \ldots, n'_N}) $. From Eq. \eqref{eq:H_spin_spin_spherical_commutations_a}, we have that:
				\begin{equation}
				\label{eq:matrixel_[J_z,H_spin]}
				\matrixel{\mathbb{n}}{\left[\spin{J}^z_{tot}, \op{H}_{spin-spin}  \right]}{\mathbb{n}'}=0 \ .
				\end{equation}
Given that,
				\begin{equation}
				\label{eq:multi_J_z}
				\spin{J}^z_{tot} = \sum^N_{i=1} \left(j_i - \hat{b}^\dagger_i \hat{b}_i\right) = J_0 - \op{N} 
				\end{equation}
(where $J_0 \equiv \sum^N_{i=1} j_i$ is the total spin of the system, and $\op{N} \equiv \sum^N_{i=1} \hat{b}^\dagger_i \hat{b}_i$ the total HP bosons occupation number operator), we note that -- as one would expect -- the kets $\{\ket{\mathbb{n}}\}$ are eigenvectors of $\spin{J}^z_{tot}$:
				\begin{equation}
				\label{eq:n_eigenvect_J_z}
				\spin{J}^z_{tot} \ket{n_1, \ldots, n_N} = \spin{J}^z_{tot} \ket{\mathbb{n}} =\left(J_0- n \right) \ket{\mathbb{n}}
				\end{equation}
where $n=\sum^N_{i=1} n_i$ is the total number of HP bosons contained in $\ket{\mathbb{n}}$. In light of Eq. \eqref{eq:n_eigenvect_J_z}, Eq. \eqref{eq:matrixel_[J_z,H_spin]} reduces to the form:
				\begin{equation}
				\label{eq:N_conserved}
				\left(n-n' \right) \matrixel{\mathbb{n}}{\op{H}_{spin-spin}}{\mathbb{n'}}=0
				\end{equation}
where it becomes evident that $\matrixel{\mathbb{n}}{\op{H}_{spin-spin}}{\mathbb{n'}}=0$ when $n \neq n'$. In other words, a necessary but not sufficient condition for the matrix element $\matrixel{\mathbb{n}}{\op{H}_{spin-spin}}{\mathbb{n'}}$ to be nonzero is that the two multispin states $\ket{\mathbb{n}}$ and $\ket{\mathbb{n}'}$ are represented by the same total number of HP bosons. The total number of HP bosons is thus conserved for a spin system with Hamiltonian $\op{H}_{spin-spin}$, Eq. \eqref{eq:H_spin-spin_2}; i.e. $\left[\op{N}, \op{H}_{spin-spin}\right]=0$ (this commutation relation easily follows from Eqs.  \eqref{eq:H_spin_spin_spherical_commutations_a} and \eqref{eq:n_eigenvect_J_z}). 
\par The importance of the relation in Eq. \eqref{eq:N_conserved} even goes further than what we have concluded so far. Indeed, if $\matrixel{\mathbb{n}}{\op{H}_{spin-spin}}{\mathbb{n'}}$ is identically zero for any pair of states $\ket{\mathbb{n}}$ and $\ket{\mathbb{n}'}$ with different number of total HP bosons, it implies that in the matrix representation of $\op{H}_{spin-spin}$ according to the basis $\{\ket{\mathbb{n}}\}$, the matrix elements between states with the same total number $n$ of HP bosons constitute a block which is orthogonal to all other possible blocks characterized by a different total number of HP bosons. In other words, if we decompose the Hamiltonian $\op{H}_{spin-spin}$ in the basis $\{\ket{\mathbb{n}}\}$, all the multispin states $\ket{\mathbb{n}}$ occupied by the same total number $n$ of HP bosons compose a subspace $\mathcal{B}_n$ of the HP bosons Fock space, and subspaces characterized by different $n$ will be orthogonal to each other, i.e. $\mathcal{B}_n \perp \mathcal{B}_{n'}$ if $n \neq n'$. But since there is a one-to-one correspondence between the states $\{\ket{\mathbb{n}}\}$ and the normal multispin states $\{\ket{j_1m_1,\ldots,j_Nm_N}\}$, it means that the normal multispin Hilbert space of the system is also decomposed into orthogonal subspaces in similar fashion. In more concise mathematical terms, the subspace $\mathcal{B}_n$ is simply the set of kets defined as:
				\begin{equation}
				\label{eq:def_mathcal_B_n}
				\mathcal{B}_n := \{\ket{\mathbb{n}} \ | \ \op{N} \ket{\mathbb{n}}=n\ket{\mathbb{n}},\ \mathbb{0} \leq \mathbb{n} \leq (D_{\mathcal{H}}-1)\} 
				\end{equation}		
(recall $D_{\mathcal{H}}$ is the dimension of the Hilbert space).
\par It is only natural at this point to ask ourselves to determine the range of $n$ for a given multiset $\mathpzc{A}$ of spins. Since $n=\sum^N_{i=1} n_i$, and $0 \leq n_i \leq 2j_i$, it is clear that $0 \leq n \leq 2J_0$. Consequently, for a given collection of spins whose Hamiltonian operator is the $q=0$-th component of a rank $k$ spherical tensor (see Eq. \eqref{eq:def_spherical_tensors}) like $\op{H}_{spin-spin}$, the system's Hilbert space can always be decomposed into $(2J_0+1)$ orthogonal subspaces. This particular observation is far from new: as a matter of fact, it is well known that whenever an operator commutes with $\spin{J}^z_{tot}$, the total spin magnetic quantum number $M_z$ is conserved -- which leads to the creation of orthogonal subspaces each characterized by a particular $M_z$. Since the total spin is $J_0$, the range of $M_z$ is $-J_0 \leq M_z \leq J_0$, which means there are $(2J_0 +1)$ distinct possible values of $M_z$, hence $(2J_0+1)$ orthogonal subspaces. Indeed, as it follows from Eq. \eqref{eq:n_eigenvect_J_z}, the connection between $M_z$ and $n$ is very simple: $M_z = J_0 - n$. Nonetheless, the use of $n$ is far more convenient computationally than $M_z$ because, unlike $M_z$ whose values can be either all integers or all half-integers, the possible values of $n$ are always nonnegative integers, independent of the collection of spins at hand. This is not the only advantage of using the HP transformation, and we shall discuss others shortly.
\par The realization of the orthogonal subspaces $\{\mathcal{B}_n\}$ also implies that the matrix representation of $\op{H}_{spin-spin}$, $\mathscr{H}_{spin-spin}$, is block-diagonalized in the basis $\{\ket{\mathbb{n}}\}$. In fact, the relation
					\begin{equation}
					\label{eq:H_spin_spin_diag}
					\mathscr{H}_{spin-spin} = \bigoplus^{2J_0}_{n=0} \mathscr{B}_n = \diag \left(\mathscr{B}_0, \mathscr{B}_1, \ldots , \mathscr{B}_{2J_0} \right) = 
					  \renewcommand{\arraystretch}{1.2}
  \left(
  \begin{array}{ c c | c c | c c }
  \cline{1-2}
    \multicolumn{1}{|c}{} & &  & \mc{} &  &  \\
    \multicolumn{2}{|c|}{\raisebox{.6\normalbaselineskip}[0pt][0pt]{$\mathscr{B}_0$}} &  & \mc{} &  &  \\
    \cline{1-4}
     &  & & &  &  \\
    &  & \multicolumn{2}{c|}{\raisebox{.6\normalbaselineskip}[0pt][0pt]{$\ddots$}} &  &  \\
    \cline{3-6}
     & \mc{} &  &  & & \multicolumn{1}{c|}{} \\
     & \mc{} &  &  & \multicolumn{2}{c|}{\raisebox{.6\normalbaselineskip}[0pt][0pt]{$\mathscr{B}_{2J_0}$}} \\ \cline{5-6}
  \end{array}
  \right)
					\end{equation}
holds true. The block matrix $\mathscr{B}_n$ collects the matrix elements of $\op{H}_{spin-spin}$ between kets belonging to the subspace $\mathcal{B}_n$, i.e.
					\begin{equation}
					\label{eq:def_mathscr_B_n}
					\mathscr{B}_n := \left\lbrace \matrixel{\mathbb{n}}{\op{H}_{spin-spin}}{\mathbb{n'}} \ \Big\vert \ \forall \mathbb{n}, \mathbb{n'} \in \mathcal{B}_n \right\rbrace \ .
					\end{equation}
Given a multiset $\mathpzc{A}$ of spins, we shall indicate the dimension of the block matrix $\mathscr{B}_n$ as $\Omega_{\mathpzc{A},n}$, which is also the dimension of the orthogonal subspace $\mathcal{B}_n$. Thus, for consistency,
					\begin{equation}
					\label{eq:W_R_0_sum_Omega_n}
					D_{\mathcal{H}} = \sum^{2J_0}_{n=0} \Omega_{\mathpzc{A},n} \ .
					\end{equation}
\par The very crucial implication of Eq. \eqref{eq:H_spin_spin_diag} is that the eigendecomposition of $\mathscr{H}_{spin-spin}$ can be equally achieved by eigendecomposing each $\mathscr{B}_n$ independently, with the benefit of reducing the computational cost (in both computing time and memory). If the computational cost of the eigendecomposition of a square matrix of dimension $M$ is $\mathcal{O}(M^p)$ (in most cases $p \approx 3$), we see that the computational cost of eigendecomposing $\mathscr{H}_{spin-spin}$ tout court is $\mathcal{O}(D^p_{\mathcal{H}})$. In contrast, if we choose to determine the eigenvalues and eigenvectors of $\mathscr{H}_{spin-spin}$  by eigendecomposing the block matrices $\mathscr{B}_n$, then the computational cost turns out to be: $\mathcal{O}(\Omega^p_{\mathpzc{A},0}) + \mathcal{O}(\Omega^p_{\mathpzc{A},1}) + \ldots + \mathcal{O}(\Omega^p_{\mathpzc{A},2J_0})$. Due to the relation in Eq. \eqref{eq:W_R_0_sum_Omega_n}, it is patently obvious that for $p>1$,
					\begin{equation}
					\mathcal{O}(D^p_{\mathcal{H}}) > \mathcal{O}(\Omega^p_{\mathpzc{A},0}) + \mathcal{O}(\Omega^p_{\mathpzc{A},1}) + \ldots + \mathcal{O}(\Omega^p_{\mathpzc{A},2J_0}) 
					\end{equation}
which confirms our assertion that eigendecomposing the block matrices $\mathscr{B}_n$ costs less than the tout court eigendecomposition of $\mathscr{H}_{spin-spin}$. (In this analysis, we have tacitly assumed that the computational cost of generating the $\mathscr{H}_{spin-spin}$ matrix is the same as the total cost of generating the block matrices $\{\mathscr{B}_n\}$.)

\paragraph{Integer partitions. Dimension of the submatrices $\mathscr{B}_n$. Density and sparseness of $\mathscr{H}_{spin-spin}$}\label{par:submatrices_dimension}
\par Another significant feature of the HP transformation is that it allows us to easily determine the dimension $\Omega_{\mathpzc{A},n}$ of the various submatrices $\mathscr{B}_n$, analytically. This is a feat hardly achievable using the normal spin representation.
\par The analytical determination of the value of $\Omega_{\mathpzc{A},n}$ has been extensively covered in \citep{art:Gyamfi-2018} so we shall just limit ourselves to some key points in the following. The interested Reader may see \citep{art:Gyamfi-2018} for a more elaborate exposition of the problem.
\par Recall that $\ket{\mathbb{n}}=\ket{n_1,n_2,\ldots,n_N}$, and $\mathbb{n}=\eta_o(n_1,n_2,\ldots,n_N)$ (see Eq. \eqref{eq:eta_0}). As already seen above, if $\op{N} \ket{\mathbb{n}}=n \ket{\mathbb{n}}$, then $n= \sum^N_{i=1} n_i$ (see Eqs. \eqref{eq:multi_J_z} and \eqref{eq:n_eigenvect_J_z}). We also need to bear in mind that each $n_i$ is restricted to the range $0 \leq n_i \leq 2j_i$. Since $\Omega_{\mathpzc{A},n}$ is the number of kets $\ket{\mathbb{n}}$ which contain the same total number of HP bosons (Eqs. \eqref{eq:def_mathcal_B_n} and \eqref{eq:def_mathscr_B_n}), the problem at hand is equivalent to asking: in how many distinct ways can we distribute $n$ indistinguishable objects among $N$ sites, knowing that the $i-$th site can contain at most $2j_i$ objects? The number of ways this can be done is exactly $\Omega_{\mathpzc{A},n}$. Counting problems like this is the subject of a branch of discrete mathematics called Enumerative Combinatorics\citep{misc:Stanley-2011}, which fundamentally deals with how to count the elements of a finite set. Several ways of computing $\Omega_{\mathpzc{A},n}$ are illustrated in \citep{art:Gyamfi-2018} but we shall focus here on just one of these, namely the generating function approach. Generating functions in Enumerative Combinatorics are \emph{formal} power series (in one or multiple variables) whose coefficients are proportional to the solutions to a counting problem.
\par Consider for example the partition of integers. The partition of a positive integer $n$ is a way one can obtain $n$ through the sum of positive integers, order being irrelevant. For example, take the integer $4$, since:
				\begin{equation}
				\begin{split}
				4 & = 4 \\
				  & = 3 + 1 \\
				  & = 2 + 2 \\
				  & = 2 + 1 + 1 \\
				  & = 1 + 1 + 1 + 1 
				\end{split}
				\end{equation}
the sums on the RHS are all partitions of $4$. Each summand in a given partition is referred to as a \emph{part}. Thus, the partitions $3+1$ and $2+2$ have both two parts, while the partition $2+1+1$ has three parts. Say $p(n)$ the total number of partitions of the integer $n$. $p(n)$ is known as the \emph{partition function}. We see that $p(n=4)=5$, for example, since the integer $4$ can be partitioned in five different ways as illustrated above. We may then pose the following problem: given an arbitrary (positive) integer $n$, in how many ways can we partition it? Or, in other words, what is the value of its $p(n)$? Variant forms of this problem have been proposed throughout centuries. According to known records, it seems Gottfried Wilhelm von Leibniz (1646-1716) was the first to pose a variant of this problem in a letter to Johann Bernoulli (1667-1748)\citep{book:Dickson-2005}. Leibniz was interested in how many ways an integer could be partitioned into two, three, etc., parts. In Leibniz's problem, we clearly see there is a constrain on the number of parts. This falls under what we call today \emph{restricted partitions}. For example, there is only one way to partition $4$ into three parts, and that is $2+1+1$. In a letter to the great Leonhard Euler, one Naudé wanted to know how many ways the integer 50 can be obtained from the sum of seven parts which are unequal to each other\citep{misc:Euler_Archive-2019, misc:Euler_Bell-2007, inbook:Andrews-2013}. Here, the restriction is both on the nature of the parts (i.e. each part cannot appear more than once) and the total number of parts. It is not far fetched to assume that Naudé's letter is what led the great Euler to his memorable \emph{Observationes analyticae variae de combinationibus} [\emph{Various analytical observations about combinations}] presented to the St. Petersburg Academy in 1741 but published in 1751\citep{misc:Euler_Archive-2019}. This is where Euler introduced for the first time the concept of generating function (though he did not coin the name) in the theory of partitions (and by extension, number theory). For example, he showed that $p(n)$ is the coefficient of $q^n$ when $\prod^{\infty}_{i=1} \frac{1}{1-q^i}$ is expanded in powers of $q$. Namely\citep{misc:Euler_Bell-2007, inbook:Andrews-2013},
				\begin{equation}
				\label{eq:gen_func_p(n)}
				\begin{split}
				\prod^{\infty}_{i=1} \frac{1}{1-q^i} & = \sum^\infty_{n=0} p(n) q^n \\
				& = 1+ q + 2q^2 + 3q^3 + 5q^4 + 7q^5 + 11q^6 + 15q^7 + 22 q^8 + \ldots
				\end{split}
				\end{equation}
The function $\prod^{\infty}_{i=1} \frac{1}{1-q^i}$ is thus said to be the \emph{generating function} for $p(n)$. Recall at the beginning of this section we defined generating functions as being \emph{formal} power series. This is so because their variables have no intrinsic meaning. In the case of Eq. \eqref{eq:gen_func_p(n)}, $q$ is the variable but though its exponents and coefficients have definite interpretations in relation to our counting problem, $q$ is devoid of any. Perhaps, the most famous generating function in Number theory and Combinatorics is the one reported in Eq. \eqref{eq:gen_func_p(n)}. In response to Naudé's problem, Euler showed that if we denote with $\tilde{p}_k(n)$ the number of ways $n$ can be partitioned into $k$ mutually unequal parts, then\citep{misc:Euler_Archive-2019, misc:Euler_Bell-2007}:
				\begin{equation}
				\label{eq:gen_func_p_tilde_k_n}
				q^{k(k+1)/2} \prod^{k}_{i=1} \frac{1}{1-q^i} = \sum^{\infty}_{n=0} \tilde{p}_k(n) q^n \ .
				\end{equation}
Thus, $q^{k(k+1)/2} \prod^{k}_{i=1} \frac{1}{1-q^i}$ is the \emph{generating function} for $\tilde{p}_k(n)$. To solve Naudé's original problem, we set $n=50$ and $k=7$; so from Eq. \eqref{eq:gen_func_p_tilde_k_n} it follows that $\tilde{p}_7(50)=522$, which is obtained by expanding  $q^{28} \prod^{7}_{i=1} \frac{1}{1-q^i}$ and taking the coefficient of the term $q^{50}$. Hence, there are $522$ ways one can write the integer $50$ as the sum of exactly seven mutually unequal positive integers. Euler's original paper, now translated into English by Jordan Bell\citep{misc:Euler_Bell-2007}, is an excellent and easy-to-read introduction to generating functions and it is accessible to anyone with a high school knowledge of algebra. We strongly recommend it to Readers who are new to the concept.
\par Instead of determining the solution to counting problems by means of generating functions, we may seek explicit formulae for the desired quantities. For instance, we may ask ourselves if there is an explicit formula for $p(n)$. It is a problem which quite a number of generations of mathematicians had to wrestle with. The first significant breakthrough came in Godfrey H. Hardy (1877-1947) and Srinivasa Ramanujan's (1887-1920) celebrated 1918 paper entitled \emph{Asymptotic formulae in combinatory analysis}\citep{art:Hardy_Ramanujan-1918}. In that paper, Hardy and Ramanujan presented an asymptotic series for $p(n)$ but whose main defect was that it failed to converge. A couple of decades had to pass before Hans Rademacher (1892-1969) derived an improved version of Hardy and Ramanujan's result which, finally, had no convergence problem\citep{art:Rademacher-1938}.
\par The Reader might be wondering why we have dedicated so many lines to integer partitions. The reason is very simple: the determination of the value of $\Omega_{\mathpzc{A},n}$ may be reinterpreted as an integer partition problem. This is how the close relation between multispin dynamics, on one hand, and Number theory and Enumerative Combinatorics (specifically in this case, the theory of partitions) elegantly emerges from the HP transformation.  Indeed, if, as noted earlier, $\Omega_{\mathpzc{A},n}$ is the number of ways the integer $n$ can be obtained through the sum $n=\sum^N_{i=1} n_i$, where $0 \leq n_i \leq 2j_i$, it is clear that:
\begin{quote}
$\Omega_{\mathpzc{A},n}$ is the number of partitions of the integer $n$ into $N$ parts, with the $i-$th part restricted to the range $0 \leq n_i \leq 2j_i$.  
\end{quote}  
Certainly, just like $\tilde{p}_k(n)$, $\Omega_{\mathpzc{A},n}$ is restricted to a fixed number of parts. But unlike $p(n)$ and $\tilde{p}_k(n)$, $\Omega_{\mathpzc{A},n}$ has the following properties:
\begin{enumerate}
\item the minimum value of a part in any of its partitions is $0$, not $1$;
\item the partitions of interest here are all \emph{ordered} (so for example, $2+1+1$ must be considered different from $1+2+1$); this is because each part $n_i$ refers to a distinct spin.
\end{enumerate}
Consider for example the deuterated hydroxymethyl radical (\ce{^{.}CH2OD}), which corresponds to the multiset of spins $\mathpzc{A}=\{j_1,j_2,j_3,j_4\}=\{\frac{1}{2},\frac{1}{2},\frac{1}{2},1\}$ (see \S \ref{subsec:Index_comp_map}). Therefore, according to the conventions discussed in \S \ref{subsec:Index_comp_map}, we know that since $j_1=j_2=j_3=\frac{1}{2}$ and $0 \leq n_i \leq 2j_i$,  then $0 \leq n_1,n_2,n_3 \leq 1$, while $0 \leq n_4 \leq 2$. $\Omega_{\mathpzc{A},n}$ is then the number of ordered partitions of $n$ into exactly $N=4$ parts, such that the first three parts are at most $1$ and the fourth part is at most $2$ (with $0$ being an admissible value of a part). We also need to bear in mind that $\Omega_{\mathpzc{A},n}$ is also equivalent to the number of states $\ket{\mathbb{n}}$ which contain exactly $n$ HP bosons. For example, with $n=2$, we find from table \ref{tab:CH2OD_table_1} that the states $\ket{\mathbb{n}}$ which contain in total two HP bosons are:
				\begin{align*}
				\ket{\mathbb{2}} & = \ket{0,0,0,2} &  \ket{\mathbb{4}}& = \ket{0,0,1,1}  & \ket{\mathbb{7}} & = \ket{0,1,0,1}\\
				\ket{\mathbb{9}} & = \ket{0,1,1,0} &  \ket{\mathbb{13}}& = \ket{1,0,0,1}  & \ket{\mathbb{15}} & = \ket{1,0,1,0}\\
				&   &  \ket{\mathbb{18}}& = \ket{1,1,0,0}  &  & 
				\end{align*}
Thus, for the radical 	\ce{^{.}CH2OD}, $\Omega_{\mathpzc{A},2}	= 7 $. That is, there are seven ways of partitioning the integer $2$ into four parts, with the restriction that the first three parts cannot exceed the value of 1 and the fourth part cannot be greater than 2. Thus, the subspace $\mathcal{B}_{n=2}$ for \ce{^{.}CH2OD} is of dimension $\Omega_{\mathpzc{A},2}	= 7$. In the particular case of \ce{^{.}CH2OD}, $n=2$ corresponds to the total spin magnetic number $M_z = J_0 - n = \frac{5}{2} -2 = +\frac{1}{2}$. Note that these results actually still hold for any multispin system with three spin-$\frac{1}{2}$ and one spin-$1$.
\par How can we determine the value of $\Omega_{\mathpzc{A},n}$ without having to explicitly write down all the kets $\ket{\mathbb{n}}$ (and how the HP bosons are disposed in them)? This problem has been solved and discussed in \citep{art:Gyamfi-2018}. We discuss here only the generating function for $\Omega_{\mathpzc{A},n}$ without going into much details. The interested Reader may see \citep{art:Gyamfi-2018} for further discussions, explicit formulae for $\Omega_{\mathpzc{A},n}$ and proofs.
\par Before we present the generating function for $\Omega_{\mathpzc{A},n}$ it is advisable we briefly see what is meant by the \emph{q-analogue} of a nonnegative integer $n$. The $q-$analogue of the integer $n$, indicated as $[n]_q$, is defined as\citep{art:Gyamfi-2018}:
				\begin{equation}
				[n]_q := \frac{1-q^n}{1-q} = 1 + q + q^2  + \ldots + q^{n-1} \ .
				\end{equation}
Note that $\lim_{q \to 1} [n]_q = n$.
\par It can be shown that the generating function for $\Omega_{\mathpzc{A},n}$, $G_{\mathpzc{A},\Omega}(q)$, is\citep{art:Gyamfi-2018}:
				\begin{equation}
				\label{eq:generating_func_Omega_a}
				G_{\mathpzc{A},\Omega}(q) = \prod_{\alpha} \left(\left[2j_\alpha + 1 \right]_q \right)^{N_\alpha} 
				\end{equation}
where the index $\alpha$ runs over distinct values of the spin multiset $\mathpzc{A}$ and $N_\alpha$ is the multiplicity of the $\alpha-$th distinct element in $\mathpzc{A}$. Hence,
				\begin{equation}
				\label{eq:generating_func_Omega_b}
				\prod_{\alpha} \left(\left[2j_\alpha + 1 \right]_q \right)^{N_\alpha} = \sum^{2J_0}_{n=0} \Omega_{\mathpzc{A},n} q^n \ .
				\end{equation}
It is worth noting that if we take the limit $q \to 1$ of Eq. \eqref{eq:generating_func_Omega_b} we get Eq. \eqref{eq:W_R_0_sum_Omega_n}.
To illustrate the use of Eq. \eqref{eq:generating_func_Omega_b}, let us consider once again \ce{^{.}CH2OD}. We know that for this radical, $\mathpzc{A}=\{\frac{1}{2},\frac{1}{2},\frac{1}{2},1\}$. $\mathpzc{A}$ has thus only two distinct elements: $j_{\alpha=1}=\frac{1}{2}$ and $j_{\alpha=2}=1$. The multiplicity of $j_{\alpha=1}$ is $3$ while that of $j_{\alpha=2}$ is $1$. Therefore, $N_{\alpha=1}=3$ and $N_{\alpha=2}=1$. Moreover, $\left[2j_{\alpha=1} + 1 \right]_q= \left[2 \right]_q= (1+q)$ and $\left[2j_{\alpha=2} + 1 \right]_q= \left[3 \right]_q= (1+q + q^2)$. Thus, in the case of the radical \ce{^{.}CH2OD}, Eq. \eqref{eq:generating_func_Omega_b} becomes:
				\begin{equation}
				(1+q)^3 (1+q+q^2) = \sum^{5}_{n=0} \Omega_{\mathpzc{A},n} q^n \ .
				\end{equation}
Since,
				\begin{equation}
				\label{eq:example_1aaa}
				(1+q)^3 (1+q+q^2) = 1 + 4 q + 7 q^2 + 7 q^3 + 4 q^4 + q^5 
				\end{equation}
we conclude that:
				\begin{align}
				\label{eq:example_1bb}
				\Omega_{\mathpzc{A},0} =\Omega_{\mathpzc{A},5}=1 &  & \Omega_{\mathpzc{A},1} =\Omega_{\mathpzc{A},4}=4 &  & \Omega_{\mathpzc{A},2} =\Omega_{\mathpzc{A},3}=7
				\end{align}
which is in agreement with the previous value we found for $\Omega_{\mathpzc{A},2}$. In regards to \ce{^{.}CH2OD}, its subspaces $\mathcal{B}_n$, their respective dimension and basis elements are reported in table \ref{tab:CH2OD_table_2}. While the correspondence between the first two columns will always remain the same, the basis kets $\ket{\mathbb{n}}$ spanning each subspace depends on the order chosen when labelling the spins. The kets in table \ref{tab:CH2OD_table_2} are the same kets in table \ref{tab:CH2OD_table_1}, therefore they correspond to the same ordered multiset of spins $\mathpzc{A}=\{j_1,j_2,j_3,j_4\}=\{\frac{1}{2},\frac{1}{2},\frac{1}{2},1\}$.
\begin{table}[htb]
\centering
\begin{tabular}{|c|c|c|}
\hline 
Subspace $\mathcal{B}_n$ & Dimension of subspace, $\Omega_{\mathpzc{A},n}$ & Basis elements, $\ket{\mathbb{n}}$ \\
\hline \hline
$\mathcal{B}_0$ & 1 &$\ket{\mathbb{0}}$\\
$\mathcal{B}_1$ & 4 & $\ket{\mathbb{1}}$, $\ket{\mathbb{3}}$, $\ket{\mathbb{6}}$, $\ket{\mathbb{12}}$\\
$\mathcal{B}_2$ & 7 & $\ket{\mathbb{2}}$, $\ket{\mathbb{4}}$, $\ket{\mathbb{7}}$, $\ket{\mathbb{9}}$, $\ket{\mathbb{13}}$, $\ket{\mathbb{15}}$, $\ket{\mathbb{18}}$\\
$\mathcal{B}_3$ & 7 & $\ket{\mathbb{5}}$, $\ket{\mathbb{8}}$, $\ket{\mathbb{10}}$, $\ket{\mathbb{14}}$, $\ket{\mathbb{16}}$, $\ket{\mathbb{19}}$, $\ket{\mathbb{21}}$\\
$\mathcal{B}_4$ & 4 & $\ket{\mathbb{11}}$, $\ket{\mathbb{17}}$, $\ket{\mathbb{20}}$, $\ket{\mathbb{22}}$\\
$\mathcal{B}_5$ & 1 & $\ket{\mathbb{23}}$\\
\hline
\end{tabular}
\caption{Orthogonal subspaces $\mathcal{B}_n$, their respective dimension and basis kets for the radical \ce{^{.}CH2OD} in the case whereby the system's spin Hamiltonian is proportional to the $q=0-$th component of a spherical tensor of rank $k$ -- $\op{H}_{spin-spin}$, Eq. \eqref{eq:H_spin-spin_2}, being an example.}
\label{tab:CH2OD_table_2}
\end{table}
\par What immediately catches our attention looking at the RHS of Eq. \eqref{eq:example_1aaa} (or the second column of table \ref{tab:CH2OD_table_2}) is the "palindromic" distribution of $\Omega_{\mathpzc{A},n}$'s values. This is not an exception but the rule. Indeed, one can prove that the generating function $G_{\mathpzc{A},\Omega}(q)$ is a \emph{reciprocal} polynomial. Namely\citep{art:Gyamfi-2018}:
				\begin{equation}
				\label{eq:reciprocal_identity}
				\Omega_{\mathpzc{A},n} =\Omega_{\mathpzc{A},2J_0-n} \ .
				\end{equation}
A polynomial $P(x)=a_0 + a_1 x + a_2 x^2 + \ldots + a_s x^s$ with real coefficients is said to be reciprocal (or \emph{palindromic}) if its coefficients are such that $a_i = a_{s-i}$ for all $i = 0,1,\ldots, s$\citep{book:Andrews-1976}. An equally equivalent definition is that $P(x)$ is reciprocal if $x^s P\left(\frac{1}{x}\right)=P(x)$\citep{book:Andrews-1976}.  
The implication of \eqref{eq:reciprocal_identity} is that the subspaces $\mathcal{B}_n$ and $\mathcal{B}_{2J_0-n}$ have the same dimension. We shall give a simple proof of Eq. \eqref{eq:reciprocal_identity} later when we discuss time-reversal symmetry.  For now, we also point out that $G_{\mathpzc{A},\Omega}(q)$ is also \emph{unimodal}\citep{book:Andrews-1976}, namely, there exists an integer $n'$ such that:
				\begin{equation}
				\Omega_{\mathpzc{A},0} \leq \Omega_{\mathpzc{A},1} \leq \Omega_{\mathpzc{A},2} \leq \ldots \leq \Omega_{\mathpzc{A},n'} \geq \Omega_{\mathpzc{A},n'+1} \geq \Omega_{\mathpzc{A},n'+2} \geq \ldots \geq \Omega_{\mathpzc{A},2J_0} \ .
				\end{equation}
This $n'$ is found to be $n'=\lfloor J_0\rfloor$\citep{art:Gyamfi-2018}, where $\lfloor \bullet \rfloor$ is the floor function. Clearly, $\Omega_{\mathpzc{A},\lfloor J_0\rfloor}$ is the maximum value the $\{\Omega_{\mathpzc{A},n}\}$ may assume. Note that $\Omega_{\mathpzc{A},\lfloor J_0\rfloor}$ may not be the only $\Omega_{\mathpzc{A},n}$ with this maximum value (afterall, $G_{\mathpzc{A},\Omega}(q)$ is reciprocal). For example, in the case of \ce{^{.}CH2OD}, whose $\mathpzc{A}=\{\frac{1}{2}^3, 1\}$, $J_0=5/2$; hence, $\lfloor J_0\rfloor = \lfloor \frac{5}{2} \rfloor = 2$. Therefore, we expect $\Omega_{\mathpzc{A},2}$ to have the maximum value any $\Omega_{\mathpzc{A},n}$ here can possibly have. In fact, that is the case but $\Omega_{\mathpzc{A},2}$ is not the only $\Omega_{\mathpzc{A},n}$ with this maximum value since $\Omega_{\mathpzc{A},3}$ also has the same value as $\Omega_{\mathpzc{A},2}$ (Eq. \eqref{eq:example_1bb}). For more on the properties of $G_{\mathpzc{A},\Omega}(q)$ see \citep{art:Gyamfi-2018}.
\par In the limit case whereby all the spins are spin-$\frac{1}{2}$, i.e. the spin system is univariate (see \S \ref{sec:Notations}) with $j=\frac{1}{2}$, $\mathpzc{A}=\left\lbrace \frac{1}{2}^N\right\rbrace$, the generating function $G_{\mathpzc{A},\Omega}(q)$ takes the simple form:
				\begin{equation}
				\label{eq:gen_Omega_half_a}
				\begin{split}
				G_{\mathpzc{A},\Omega}(q) & = (1+q)^N\\
				& = \sum^{N}_{n=0} \Omega_{\mathpzc{A},n} q^n
				\end{split}
				\end{equation}
Consequently, the dimension of the subspace $\mathcal{B}_n$, $\Omega_{\mathpzc{A},n}=\Omega_{\left\lbrace \frac{1}{2}^N\right\rbrace,n} $, in this limit case is given by a binomial coefficient:
				\begin{equation}
				\label{eq:Omega_N_spin_half}
				\Omega_{\left\lbrace \frac{1}{2}^N\right\rbrace,n} = \binom{N}{n} \ .
				\end{equation}
Let us consider for instance a spin system which consists of $N=10$ spin-$\frac{1}{2}$s, i.e. $\mathpzc{A}=\left\lbrace \frac{1}{2}^{10}\right\rbrace$. We are dealing here with a univariate spin system. If the spin Hamiltonian of the system commutes with $\spin{J}^z_{tot}$ (so the Hamiltonian is proportional to a spherical tensor of rank $k$ but $q=0$, see Eq. \eqref{eq:def_spherical_tensors}), the tout court approach will have us diagonalize a matrix of dimension $2^{10}=1024$. But with the creation of the subspaces $\mathcal{B}_n$, we know that
				\begin{equation}
				\begin{split}
				G_{\mathpzc{A},\Omega}(q) & = (1+q)^{10} \\
				& = 1 + 10q + 45q^2 + 120q^3 + 210q^4 + 252q^5 + 210q^6 + 120q^7 + 45q^8 + 10q^9 + q^{10}
				\end{split}
				\end{equation}
Thus, instead of eigendecomposing a matrix of dimension $1024$ to determine the eigenvectors and eigenvalues of the spin Hamiltonian, we can equally obtain the same results by eigendecomposing smaller matrices of dimension $1,10,45,120,210$ and $252$.
\par The distribution of the values of $\Omega_{\mathpzc{A},n}$ does not only inform us about the dimension of the subspaces $\mathcal{B}_n$, but they also help us to quantify how sparse the matrix $\mathscr{H}_{spin-spin}$ is. Let us define the density $\zeta(A)$ of an arbitrary matrix $A$ as the ratio between the number of its nonzero elements and the dimension of $A$. Analogously, we define the sparseness $\chi(A)$ of the matrix $A$ as the fraction of the elements of the latter which are identically zero. Naturally, for any given matrix $A$:
				\begin{equation}
				\zeta(A) + \chi(A) = 1 \ .
				\end{equation}
It is easy to prove that $\zeta(\mathscr{H}_{spin-spin})$ is subject to the tight upper bound: 
				\begin{equation}
				\label{eq:density_upper_bound}
				\zeta(\mathscr{H}_{spin-spin}) \leq \frac{\sum^{2J_0}_{n=0} \Omega^2_{\mathpzc{A},n}}{D^2_{\mathcal{H}}} 
				\end{equation}
independent of the specific values of the coupling constants $T_{i,i'}$. If we go back to the radical \ce{^{.}CH2OD}, for example, $\zeta(\mathscr{H}_{spin-spin}) \leq \frac{132}{24^2} \sim 0.23$. This means no matter what the values of the constants $T_{i,i'}$ are, the matrix representation of the Hamiltonian $\op{H}_{spin-spin}$ cannot have more than $23\%$ of its elements being nonzero. This is not only true for \ce{^{.}CH2OD}, but also for all multiset of spins $\mathpzc{A}=\left\lbrace \frac{1}{2}^3,1 \right\rbrace$. Indeed, Eq. \eqref{eq:density_upper_bound} holds for any arbitrary multiset of spins whose spin Hamiltonian is proportional to the zero-th component of a spherical tensor of rank $k$, where $k=0,1,2,\ldots$
\par For a univariate spin system of $N$ spin-$\frac{1}{2}$, i.e. $\mathpzc{A}=\left\lbrace \frac{1}{2}^N \right\rbrace$, it follows from Eqs. \eqref{eq:Omega_N_spin_half} and \eqref{eq:density_upper_bound} that:
				\begin{equation}
				\label{eq:density_N_spin_half_0}
				\zeta(\mathscr{H}_{spin-spin}) \leq \frac{\sum^{N}_{n=0} \binom{N}{n}^2}{2^{2N}} =  \frac{\binom{2N}{N}}{4^{N}} 
				\end{equation}
from which one derives that for very large $N$,
				\begin{equation}
				\label{eq:density_N_spin_half}
				\zeta(\mathscr{H}_{spin-spin}) \lesssim \frac{1}{\sqrt{\pi N}} \ .
				\end{equation}		
Eq. \eqref{eq:density_N_spin_half} shows in unambiguous terms that for a system like $\mathpzc{A}=\left\lbrace \frac{1}{2}^N \right\rbrace$, $\zeta(\mathscr{H}_{spin-spin})$ tends to zero as $N$ becomes very large. For instance, if we consider the spin system $\mathpzc{A}=\left\lbrace \frac{1}{2}^{1000} \right\rbrace$, i.e. a collection of $1000$ spins, all of spin-$\frac{1}{2}$, then $\zeta(\mathscr{H}_{spin-spin}) \lesssim \frac{1}{\sqrt{\pi 1000}} \approx 0.018$; which means the elements of $\mathscr{H}_{spin-spin}$ which are identically zero will never drop below $98\%$ of the total entries. 
\par Interestingly, we note that if we begin with a spin system $\mathpzc{A}=\left\lbrace \frac{1}{2}^N \right\rbrace$, and substitute one of the spins with a particle whose spin quantum number is greater than $1/2$, the new $\mathscr{H}_{spin-spin}$ is less denser than the original. We can thus imagine creating any collection of $N$ spins from the multiset $\mathpzc{A}=\left\lbrace \frac{1}{2}^N \right\rbrace$ by substitution. Given that anytime we substitute a spin-$1/2$ with a greater spin the density $\zeta(\mathscr{H}_{spin-spin})$ reduces, it implies that for any imaginable multiset $\mathpzc{A} \neq \left\lbrace \frac{1}{2}^N \right\rbrace$ of spins, 
				\begin{equation}
				\zeta(\mathscr{H}_{spin-spin}) <  \frac{\binom{2N}{N}}{4^{N}} 
				\end{equation}
and in the limit of a large number of spins, 
				\begin{equation}
				\zeta(\mathscr{H}_{spin-spin}) <  N^{-\delta} \ ,  \qquad \delta=\frac{1}{2}(1+ \log_N \pi) \ .
				\end{equation} 
\paragraph{Eigenenergies of a system of equivalent spins in the absence of an external field.}\label{par:eigenenergies_equivalent_spins}				
\par We illustrate here another powerful application of the HP transformation. In the limit case whereby the multiset $\mathpzc{A}=\{j_1, j_2 , \ldots, j_N\}$ consists of solely $N$ interacting \emph{equivalent} spins (\S \ref{sec:Notations}), $\op{H}_{spin-spin}$ (Eq. \eqref{eq:H_spin-spin_2}) reduces to the form:
				\begin{equation}
				\label{eq:H_spin-spin_3}
				\op{H}_{spin-spin} = T \sum_{i > i'}  \vspin{J}_i \cdot \vspin{J}_{i'} = T \sum_{i > i'} \left[\spin{J}^z_i \spin{J}^z_{i'} + \frac{1}{2} \left( \spin{J}^+_i \spin{J}^-_{i'} + \spin{J}^-_i \spin{J}^+_{i'}\right)\right] \ .
				\end{equation}
Since,
				\begin{equation}
				\vspin{J}^2_{tot} = \sum_i \vspin{J}^2_i + 2 \sum_{i>i'} \vspin{J}_i \cdot \vspin{J}_{i'} \ ,
				\end{equation}	
$\op{H}_{spin-spin}$ in Eq. \eqref{eq:H_spin-spin_3} may be written as:
				\begin{equation}
				\op{H}_{spin-spin} = \frac{T}{2} \left( \vspin{J}^2_{tot} - \sum_i \vspin{J}^2_i \right) \ .
				\end{equation}
Given that $\vspin{J}_{tot}$ commutes with $\vspin{J}_i$, we can easily determine the eigenvalues of $\op{H}_{spin-spin}$	if we know those of $\vspin{J}^2_{tot}$ and $\vspin{J}^2_i$.	Fortunately, the expression for the eigenvalues of this set of commuting operators is well known. From Eq. \eqref{eq:J^2}, for example, we have that:
				\begin{equation}
				\label{eq:H_spin-spin-4}
				\op{H}_{spin-spin} = \frac{T}{2} \left[ J_{tot}\left(J_{tot} + 1 \right) - \sum_i j_i(j_i + 1) \right] \hat{\mathbb{I}} \ .
				\end{equation}		
As simple as Eq. \eqref{eq:H_spin-spin-4} may seem, the actual computation of the eigenenergies is not an easy task for a generic $\mathpzc{A}$ with $N>2$. The difficulty here lies in computing the various total spin angular momentum $J_{tot}$ and their multiplicities. But the $J_{tot}$ and their respective multiplicities are the Clebsch-Gordan series, which is the multiset of all the possible total (spin) angular momenta one can get by coupling the $N$ elements of $\mathpzc{A}$. For example, if $N=2$ (in which case Eqs. \eqref{eq:H_spin-spin_3} and \eqref{eq:H_spin-spin-4} hold irrespective of whether the two spins are equivalent or not), we well know that $|j_1-j_2| \leq J_{tot} \leq j_1+j_2$.  Certainly, when $N>2$ one could determine the Clebsch-Gordan series by coupling $j_1$ and $j_2$, and then couple the resulting angular momenta with $j_3$, followed by the coupling of the new set of resulting angular momenta with $j_4$, and one repeats the scheme till one gets to $j_N$. Needless to say, this is truly cumbersome. It is understandable that if one has an easy and computationally efficient method of computing the Clebsch-Gordan series for a generic collection of spins, one also reduces dramatically the computational cost of determining the eigenvalues of $\op{H}_{spin-spin}$ in Eq. \eqref{eq:H_spin-spin_3}. Here again, the HP transformation is invaluable.
\par The problem of determining analytically the Clebsch-Gordan series for an arbitrary collection of spins $\mathpzc{A}$ has been solved in \citep{art:Gyamfi-2018}. We will therefore limit ourselves here to citing some salient results from \citep{art:Gyamfi-2018} without providing any proof.
\par Consider an arbitrary collection of spins $\mathpzc{A}=\{j_1, j_2, \ldots, j_N\}$. The addition of the spin angular momentum of the elements of $\mathpzc{A}$ will result in a collection (or multiset) of total spin angular momenta $\{J_{tot}\}$. Let us indicate the \emph{distinct} elements of $\{J_{tot}\}$ as $J_0, J_1, J_2, \ldots, J_m$, with $J_0$ being the maximum, $J_1$ the second highest and so forth. Consecutive distinct $J_{tot}$ differ by $1$, hence:
				\begin{equation}
				\label{eq:J_kappa}
				J_\kappa = J_0 - \kappa \ , \ \mbox{where }\kappa=0,1,2,\ldots,m \ .
				\end{equation}
Clearly, $J_0=j_1 + j_2 + \ldots + j_N$. The first big challenge we encounter here is computing the value of $J_m$. It can be proved that\citep{art:Gyamfi-2018}:
				\begin{equation}
				\label{eq:J_m}
				J_{m}  = \upsilon_{\mathpzc{A}} \cdot H(\upsilon_{\mathpzc{A}}) + \left(1- H(\upsilon_{\mathpzc{A}})\right)\cdot \frac{(2J_0 \hspace{-0.3cm}\mod 2 )}{2}		
				\end{equation}
where,
				\begin{equation}
				\label{eq:upislon_A}
				\upsilon_{\mathpzc{A}} := 2\cdot \max \mathpzc{A} - J_0 
				\end{equation}
and where $H\left( x \right)$ is the Heaviside step function, defined here to be
				\begin{equation}
				\label{eq:Heaviside}
				H\left( x \right) := \begin{cases}
				0 , \mbox{ if }x < 0\\
				1 , \mbox{ if }x \geq 0 \ .
				\end{cases}
				\end{equation}	
Say $N_J$ the total number of distinct elements in $\{J_{tot}\}$. Evidently, $N_J = J_0 - J_m + 1$. At this point, we may represent the multiset $\{J_{tot}\}$ as: $\{J_{tot}\} = \left\lbrace J^{\lambda_0}_0, J^{\lambda_1}_1, \ldots , J^{\lambda_m}_m\right\rbrace$, where $\lambda_\kappa$ is the multiplicity of $J_\kappa$. 
\par Having determined all the distinct elements of $\{J_{tot}\}$ thanks to Eqs. \eqref{eq:J_kappa} and \eqref{eq:J_m}, we are half-way through in getting to the Clebsch-Gordan series. All we need to do now is to determine the multiplicities $\{\lambda_\kappa\}$. Once again, we can easily solve the problem by making use of the HP transformation. One can show that the generating function for $\lambda_\kappa$, $G_{\mathpzc{A},\lambda}(q)$, is related to $G_{\mathpzc{A},\Omega}(q)$ -- Eq. \eqref{eq:generating_func_Omega_a} -- through the relation\citep{art:Gyamfi-2018}:
					\begin{equation}
					\label{eq:generating_lambda}
					G_{\mathpzc{A},\lambda}(q) = (1-q) G_{\mathpzc{A},\Omega}(q) = \sum^{2J_0+1}_{\kappa=0} \lambda_\kappa q^\kappa \ .
					\end{equation}
The polynomial $G_{\mathpzc{A},\lambda}(q)$ is \emph{antipalindromic} since,
					\begin{equation}
					\lambda_\kappa = - \lambda_{2J_0+1-\kappa} \ .
					\end{equation}
Moreover, it can be easily proved that:
					\begin{equation}
					\label{eq:sum_rule_lambda}
					\sum^m_{\kappa=0} \lambda_\kappa = \Omega_{\mathpzc{A},m} 
					\end{equation}
i.e. the cardinality of the Clebsch-Gordan series is exactly $\Omega_{\mathpzc{A},m}$.
\par In our current quest to determine the eigenvalues of the Hamiltonian $\op{H}_{spin-spin}$ of Eq. \eqref{eq:H_spin-spin_3}, we see that since the multiset $\mathpzc{A}$ is fixed, the eigenvalues of $\op{H}_{spin-spin}$ can be distinguished on the basis of $J_{tot}$. We may thus indicate these eigenenergies as $E_{J_{tot}}$, or $E_\kappa$ on the basis of Eq. \eqref{eq:J_kappa}. We shall employ the latter in the following. It then follows from Eqs. \eqref{eq:H_spin-spin-4} and \eqref{eq:J_kappa} that:
					\begin{equation}
					\label{eq:E_kappa}
					E_\kappa = E_0 - \frac{T}{2} \kappa \left( 2J_0 + 1 - \kappa \right) 
					\end{equation}	
where $\kappa=0,1,2,\ldots,m=(J_0-J_m)$, and
					\begin{equation}
					E_0 := \frac{T}{2} \left( J_0^2  - \sum_i j^2_i\right) \ .
					\end{equation}		
The following observations on the system readily follows from Eq. \eqref{eq:E_kappa}:
\begin{enumerate}
\item if $T>0$, then $E_m$ is the ground state $(J_{tot}=J_m)$ energy and $E_0$ is the energy of the highest excited state;
\item if $T<0$, then $E_0$ is the energy of the ground state $(J_{tot}=J_0)$, and $E_m$ is the energy of the highest excited state;
\item $E_{\kappa +1} - E_\kappa = - T J_\kappa \ $. This means that the energy of multispin states with consecutive $J_{tot}$ at the lower end of $\{J_{tot}\}$ (i.e. as $J_\kappa$ approaches $J_m$) are relatively less spaced compared to their counterparts at the higher end. 
\end{enumerate}	
Furthermore, given that the degeneracy of $J_\kappa$ is $(2J_\kappa + 1)$, the total degeneracy of the energy level $E_\kappa$ will thus be given by the product:  $\lambda_\kappa (2J_\kappa + 1)$. The sum total of the degeneracy of the energy levels must return the dimension of the system's spin Hilbert space, $D_{\mathcal{H}}$. This leads us to the relation:
					\begin{equation}
					\label{eq:sum_rule_W_lambda}
					D_{\mathcal{H}} = \sum^m_{\kappa=0} \lambda_\kappa (2J_\kappa + 1)
					\end{equation}			
which is a simple sum rule one can deploy to spot flaws in the calculations. Other sum rules can be derived from Eq. \eqref{eq:sum_rule_W_lambda} by combining it with Eqs. \eqref{eq:J_kappa} and \eqref{eq:sum_rule_lambda}.
\par To illustrate the usefulness and potential of the relations and techniques discussed above, let us consider the multiset of spins $\mathpzc{A}=\left\lbrace \frac{1}{2}^7, 1^3\right\rbrace$. Assuming $\mathpzc{A}$ is a multiset of equivalent spins whose Hamiltonian is given by \eqref{eq:H_spin-spin-4}, we ask: what are the energy levels of the system and their respective degeneracy? As simple as this problem may appear, it is rather difficult -- if not computationally time consuming -- to solve using the conventional spin representation. The dimension of the spin Hilbert space alone is $D_{\mathcal{H}}=3456$, though less than $0.96\%$ of the matrix elements of the spin Hamiltonian is nonzero. It would be a waste of resources to construct such a matrix and then eigendecompose it to find the energy levels. It is even more computationally challenging to directly employ Eq. \eqref{eq:H_spin-spin-4} to compute the energy levels because -- to the best of our knowledge -- there is not, hitherto, a generally valid efficient algorithm to compute the Clebsch-Gordan series for arbitrary multispin systems. In the framework of the HP transformation, such a problem can be effortlessly solved without the need to eigendecompose $\mathscr{H}_{spin-spin}$.
\par Naturally, the eigenenergies $\{E_\kappa\}$ of the system is given by Eq. \eqref{eq:E_kappa}. Concerning the distinct values of the multiset $\{J_{tot}\}$, we easily determine its maximum to be $J_0=\frac{13}{2}$. From Eq. \eqref{eq:J_m}, the minimum $J_{tot}$ is found to be $J_m = \frac{1}{2}$. Thus, there are $N_J = J_0-J_m+1=7$ distinct values of $J_{tot}$, and $m=J_0-J_m=6$. The distinct $J_{tot}$ are: $\{J_0,J_1,J_2,J_3,J_4,J_5,J_6\}=\left\lbrace \frac{13}{2},\frac{11}{2},\frac{9}{2},\frac{7}{2},\frac{5}{2}, \frac{3}{2}, \frac{1}{2}\right\rbrace$. We now determine the multiplicity of each $J_{tot}$ by employing the generating function $G_{\mathpzc{A},\lambda}(q)$, Eq. \eqref{eq:generating_lambda}. First of all, from Eq. \eqref{eq:generating_func_Omega_a}, we have that $G_{\mathpzc{A},\Omega}(q)=(1+q)^7(1+q+q^2)^3$. Therefore,
					\begin{equation}
					\begin{split}
					G_{\mathpzc{A},\lambda}(q) & = (1-q)(1+q)^7(1+q+q^2)^3 \\
					& = 1 + 9q + 38q^2 + 99q^3 + 174q^4 + 207q^5 + 145q^6 +\ldots - q^{14}\\
					& = \sum^{14}_{\kappa=0} \lambda_\kappa q^\kappa
					\end{split}
					\end{equation}	
We therefore conclude that the Clebsch-Gordan series for $\mathpzc{A}=\left\lbrace \frac{1}{2}^7, 1^3\right\rbrace$ is given by the multiset $\{J_{tot}\}=\left\lbrace  \frac{13}{2},\frac{11}{2}^9,\frac{9}{2}^{38},\frac{7}{2}^{99},\frac{5}{2}^{174}, \frac{3}{2}^{207}, \frac{1}{2}^{145}\right\rbrace$. With these values, we can now easily compute all the eigenenergies $E_\kappa$ of the system from Eq. \eqref{eq:E_kappa}. The results are reported in table \ref{tab:7_half_3_one}, where we have also calculated the degeneracy of each energy level. As expected on the basis of Eq. \eqref{eq:sum_rule_W_lambda}, the sum of these degeneracies gives exactly $3456$ -- which we recall is the dimension of the spin Hilbert space of the system. 		 		%
\begin{table}[htb]
\centering
\begin{tabular}{|C{1.5cm}|C{1.5cm}|C{1.5cm}|C{2.0cm}|c|}
\hline 
$\kappa $ & $J_\kappa $ & $\lambda_\kappa$ & $E_\kappa - E_0$ & Degeneracy of $E_\kappa$  \\
\hline \hline
$0$ & $\frac{13}{2}$ &  $1$ &  $0$ &  $14$  \\
$1$ & $\frac{11}{2}$ & $9$  &  $-\frac{13}{2}T$  & $108$   \\
$2$ & $\frac{9}{2}$  & $38$  & $-12T$  &  $380$  \\
$3$ & $\frac{7}{2}$  & $99$ &  $-\frac{33}{2}T$ &  $792$  \\
$4$ & $\frac{5}{2}$  & $174$ &  $-20T$ & $1044$   \\
$5$ & $\frac{3}{2}$  & $207$  & $-\frac{45}{2}T$  & $828$   \\
$6$ & $\frac{1}{2}$  &  $145$ &  $-24 T$ &   $290$ \\
\hline
\multicolumn{4}{c|}{} &   $\sum = 3456$ \\
\cline{5-5}
\end{tabular}
\caption{Energy levels and their respective degeneracy for a system of equivalent spins composed of seven spin-$1/2$ and three spin-$1$, $\mathpzc{A}=\left\lbrace \frac{1}{2}^7, 1^3\right\rbrace$, whose Hamiltonian is given by Eq. \eqref{eq:H_spin-spin_3}.}
\label{tab:7_half_3_one}
\end{table}		
\par It is worth considering the same problem but this time with $\mathpzc{A}=\left\lbrace \frac{1}{2}^N\right\rbrace$. Surely, $J_0=N/2$, and from Eq. \eqref{eq:J_m}, we derive that,
					\[
					J_m = \begin{cases}
					\frac{(2J_0 \hspace{-0.1cm}\mod 2 )}{2} & \mbox{if }N>2 \\
					0 & \mbox{if } N = 2 \ .
					\end{cases}
					\]	
For $N>2$, we observe that $J_m=0$ if $N$ is even, while $J_m=1/2$ if $N$ is odd. Thus, the number of distinct $J_{tot}$ to expect is $N_J=\lceil \frac{N-1}{2}\rceil + 1$, and $m=\lceil \frac{N-1}{2}\rceil$, where $\lceil \bullet \rceil$ is the ceiling function. Therefore, the distinct $J_{tot}$ we get from the Clebsch-Gordan series for $\mathpzc{A}=\left\lbrace \frac{1}{2}^N\right\rbrace$ are: $\{J_0, J_1, \ldots, J_{\lceil \frac{N-1}{2}\rceil}\}=\{N/2, N/2-1, \ldots, N/2-\lceil \frac{N-1}{2}\rceil\}$. Here, the multiplicity $\lambda_\kappa$ of $J_\kappa$ obeys the simple relation:
					\begin{equation}
					\label{eq:lambda_N_half_a}
					\lambda_\kappa = \binom{N}{\kappa} - \binom{N}{\kappa-1} = \frac{N+1-2\kappa}{N+1-\kappa}\binom{N}{\kappa} \ , \quad \kappa = 0,1,2, \ldots,m \ .
					\end{equation}
Eq. \eqref{eq:lambda_N_half_a} readily follows from Eqs. \eqref{eq:gen_Omega_half_a}, \eqref{eq:Omega_N_spin_half} and \eqref{eq:generating_lambda}.

\paragraph{Time-reversal symmetry. Creation of submatrices $\mathscr{B}_n$.}\label{par:Time-reversal}
To the best of our knowledge, in the general context of Eq. \eqref{eq:H_spin-spin_2}, it is not in general possible to easily determine the eigenvalues of $\op{H}_{spin-spin}$ like we just did for an arbitrary collection of equivalent spins in the last section. In general, the eigendecomposition of $\op{H}_{spin-spin}$ requires, first of all, creating the submatrices $\{\mathscr{B}_n\}$ and then eigendecomposing each separately, Eq. \eqref{eq:H_spin_spin_diag}. To create the submatrix $\mathscr{B}_n$ in the HP representation, we need a general formula for $\op{H}_{spin-spin}$'s matrix elements in terms of the HP bosons occupation numbers. Indeed, given any two basis kets $\ket{\mathbb{n}'}=\ket{n'_1, n'_2, \ldots, n'_N}$ and $\ket{\mathbb{n}}=\ket{n_1, n_2, \ldots, n_N}$, we derive from Eq. \eqref{eq:H_spin-spin_2} that:
				\begin{multline}
				\label{eq:H_spin-spin_matrix_el}
				\matrixel{\mathbb{n}'}{\op{H}_{spin-spin}}{\mathbb{n}}  = \matrixel{n'_1, n'_2, \ldots, n'_N}{\op{H}_{spin-spin}}{n_1, n_2, \ldots, n_N}\\
				 = \delta_{\mathbb{n}', \mathbb{n}} \sum_{i>k} T_{i,k} (j_i-n_i)(j_k-n_k)\\
				 + \frac{1}{2}  \sum_{i>k} T_{i,k} \sqrt{\left(n_i + \frac{1 \pm 1}{2} \right)\left(2j_i-n_i+\frac{1 \mp 1}{2} \right)\left(2j_k-n_k+ \frac{1 \pm 1}{2} \right)\left(n_k + \frac{1 \mp 1}{2} \right)} \\
				 \times \ \Delta_{i,k} \ \delta_{n'_i,n_i \pm 1} \ \delta_{n'_k,n_k \mp 1}
				\end{multline}
where,
				\begin{equation}
				\Delta_{i,k} := \prod_{l \neq i,k} \braket{n'_l}{n_l} \ .
				\end{equation}
In deriving Eq. \eqref{eq:H_spin-spin_matrix_el}, we have made use of Eqs. \eqref{eq:J_+_in_b}, \eqref{eq:J_-_in_b} and \eqref{eq:b_dagger_b}. Eq. \eqref{eq:H_spin-spin_matrix_el} confirms once again that $\matrixel{\mathbb{n}'}{\op{H}_{spin-spin}}{\mathbb{n}}$ is identically zero when $\ket{\mathbb{n}'}$ and $\ket{\mathbb{n}}$ differ in the total number of HP bosons they contain.					 
\par It is worth noting that the creation of the submatrices $\mathscr{B}_n$ can be greatly simplified and done in an efficient manner if we make use of the time-reversal symmetry (or $T-$symmetry).
\par In the field of magnetic resonance (and in quantum chemistry, in general), we are accustomed to space symmetries like rotational and translational invariances, but we seldomly speak of time-reversal symmetry. One of the reasons for this, we suppose, can be attributed to the fact that unlike other symmetry operations like rotations and translations which are represented by unitary operators in quantum mechanics, $T-$ symmetry is represented by an \emph{antiunitary} operator (see below) -- whose properties depart considerably from the known unitary operators. In certain areas of quantum chemistry like vibrational spectroscopy, $T-$symmetry can be of no enlightening use, but in magnetic resonance it is often of vital importance as we shall shortly see. For more on $T-$symmetry, interested Readers may see (in this order): \citep{book:Zee-2016, book:Sakurai-2011, book:Sachs-1987}. For an in-depth introduction embedded in some interesting philosophical discussions, see \citep{art:Bryan-2016}.
\par Let $\hat{\Theta}$ be the time-reversal operator. That is,
				\begin{equation}
				\hat{\Theta} \ t \ \hat{\Theta}^{-1} = -t
				\end{equation}
where $t$ is the real parameter which indicates time. Like any operator representing a symmetry transformation, $\hat{\Theta}  \hat{\Theta}^{-1} = \hat{\Theta}^{-1}  \hat{\Theta}= \hat{\mathbb{1}}$. As already remarked above, unlike space rotation or translation operators which are unitary, the time-reversal operator $\hat{\Theta}$ is antiunitary. The major difference between these two types of symmetry operators is how they operate on complex scalars: unitary operators leave complex scalars intact, while antiunitary operators change complex scalars into their corresponding complex conjugate. For example,  $\hat{\Theta} e^{i\alpha} \hat{\Theta}^{-1}=e^{-i\alpha^*}$, where $\alpha$ is a complex number whose complex conjugate is $\alpha^*$.	
\par There are many interesting properties of the time-reversal operator, but for our purposes, it suffices to know that all angular momentum operators (orbital and spin) are odd under the operation of time-reversal. Namely\citep{book:Sakurai-2011,book:Sachs-1987,art:Bryan-2016,book:Wigner-1959},
				\begin{equation}
				\label{eq:time-reversal_J}
				\hat{\Theta} \vspin{J} \hat{\Theta}^{-1} = -\vspin{J} \ .
				\end{equation} 	
This simple relation has many profound consequences which will unfold before us shortly. To begin, it follows from Eq. \eqref{eq:time-reversal_J} that 	$\hat{\Theta} \spin{J}^z \hat{\Theta}^{-1} = -\spin{J}^z$. Combining this with Eq. \eqref{eq:J_z_in_b_b}, we find that:
				\begin{equation}
				\label{eq:time_reversal_no_op}
				\hat{\Theta} \ \hat{b}^\dagger \hat{b} \ \hat{\Theta}^{-1} = 2j - \hat{b}^\dagger \hat{b}\ .
				\end{equation}
Till now, we have always seen spin kets of a spin-$j$ as being represented by a number of HP bosons occupying a certain vacuum space which can accommodate at most $2j$ bosons. Another way of seeing it is to imagine this vacuum state as comprised of $2j$ holes (called HP holes), where each hole can be filled with no more than one HP boson at a time. In this picture, the sum of the number of HP bosons and holes is always $2j$. Therefore, if  $\hat{b}^\dagger \hat{b}$ counts the number of HP bosons, then $(2j-\hat{b}^\dagger \hat{b})$ counts the number of holes. In other words, $(2j-\hat{b}^\dagger \hat{b})$ is the number operator for the holes. Going back to Eq. \eqref{eq:time_reversal_no_op}, the effect of the time-reversal operator now becomes perspicuous: it transforms the HP occupation number operator into a hole number operator. Put in another way, $\Theta$ instantly converts HP bosons into holes, and vice versa.
\par Before we proceed, let us see how $\hat{\Theta}$ transforms the single spin ket $\ket{n}$ and the multispin ket $\ket{\mathbb{n}}$. First of all, we note that Eq. \eqref{eq:time_reversal_no_op} may be rewritten as:
				\begin{equation}
				\label{eq:time_reversal_no_op_b}
				\left[\ \hat{\Theta}, \ \hat{b}^\dagger \hat{b} \ \right]_+ = 2j\hat{\Theta}
				\end{equation}
where $\left[\hat{A},\hat{B}\right]_+(:= \hat{A}\hat{B} + \hat{B}\hat{A})$ denotes the anticommutation between $\hat{A}$ and $\hat{B}$. Say $\ket{n}$ a state ket of a particle of spin-$j$ according to the HP representation, where, as usual, $n$ represents the number of HP bosons. We define the state $\Theta \ket{n} \equiv \ket{\overline{n}}$ as the \emph{time-reversed} or \emph{hole complement of}  $\ket{n}$. The state $\ket{\overline{n}}$ must necessarily be an admissible spin state, else the symmetry operation enacted by $\Theta$ would not be reversible. The reversibility of the time-reversal symmetry operation is guaranteed by the fact that $\hat{\Theta}  \hat{\Theta}^{-1} = \hat{\Theta}^{-1}  \hat{\Theta}= \hat{\mathbb{I}}$.  After multiplying Eq. \eqref{eq:time_reversal_no_op_b} from the left by $\ket{n}$, followed by an appropriate rearrangement of the terms, we get:
				\begin{equation}
				\label{eq:n_bar}
				\hat{b}^\dagger \hat{b} \ket{\overline{n}} = (2j-n) \ket{\overline{n}} 
				\end{equation}
where we have made use of Eq. \eqref{eq:b_occupation_no}. Comparing Eq. \eqref{eq:n_bar} with \eqref{eq:b_occupation_no}, it becomes immediately clear that:
				\begin{equation}
				\hat{\Theta} \ket{n} \equiv \ket{\overline{n}} = \ket{2j-n} 
				\end{equation}
which is, once again, in line with the interpretation that $\hat{\Theta}$ converts HP bosons into holes, and holes into HP bosons. The dual ket relative to $\hat{\Theta} \ket{n}$ is $ \bra{n}\hat{\Theta}^{-1} \equiv \bra{\overline{n}}$.
\par Let us now turn our attention to how $\hat{\Theta}$ operates on a multispin ket $\ket{\mathbb{n}}$. Consider the spin multiset $\mathpzc{A}=\{j_1, j_2, \ldots, j_N\}$. From Eqs. \eqref{eq:multi_J_z} and \eqref{eq:time_reversal_no_op}, it follows that:
				\begin{equation}
				\label{eq:time_reversal_no_op_c}
				\left[\ \hat{\Theta}, \ \op{N} \ \right]_+ = 2J_0\hat{\Theta} 
				\end{equation}
which is the analogous of Eq. \eqref{eq:time_reversal_no_op_b} for multispin systems. Say $\hat{\Theta} \ket{\mathbb{n}} \equiv \ket{\overline{\mathbb{n}}}$ the hole complement of $\ket{\mathbb{n}}(=\ket{n_1,n_2,\ldots,n_N})$. We remind the Reader that the HP bosons occupation numbers $n_i$ in $\ket{n_1,n_2,\ldots,n_N}$ are related to the integer $\ket{\mathbb{n}}$ through the index compression map $\eta_0$, Eq. \eqref{eq:eta_0}. If we now multiply Eq. \eqref{eq:time_reversal_no_op_c} from the right by $\ket{\mathbb{n}}$, we end up with the relation:
				\begin{equation}
				\label{eq:n_bar_multispin}
				\op{N} \ket{\overline{\mathbb{n}}} = (2J_0-n) \ket{\overline{\mathbb{n}}} 
				\end{equation}
which implies that if $\ket{\mathbb{n}}$ belongs to the subspace $\mathcal{B}_n$, then its hole complement $\ket{\overline{\mathbb{n}}}$ is an element of the subspace $\mathcal{B}_{2J_0-n}$. Indeed, since there is a one-to-one correspondence between $\ket{\mathbb{n}}$ and $\ket{\overline{\mathbb{n}}}$, we also conclude that the hole complements of the basis kets in $\mathcal{B}_n$ constitute the basis kets of the subspace $\mathcal{B}_{2J_0-n}$. In other words, the two subspaces $\mathcal{B}_n$ and $\mathcal{B}_{2J_0-n}$ are related by time-reversal symmetry, and this holds independent of the nature of the spin Hamiltonian. The one-to-one correspondence we just mentioned, implies that the dimension of the subspaces $\mathcal{B}_n$ and $\mathcal{B}_{2J_0-n}$ must coincide, and indeed, that is something we already know from Eq. \eqref{eq:reciprocal_identity}.
\par Note that,
				\begin{equation}
				\begin{split}
				\hat{\Theta} \ket{\mathbb{n}}  = \hat{\Theta} \ket{n_1,n_2,\ldots,n_N} & = \ket{\overline{n}_1,\overline{n}_2,\ldots,\overline{n}_N} \\
				& = \ket{2j_1-n_1,2j_2-n_2,\ldots,2j_N-n_N} \ .
				\end{split}
				\end{equation}
Thus,
				\begin{equation}
				\ket{\overline{\mathbb{n}}} = \ket{2j_1-n_1,2j_2-n_2,\ldots,2j_N-n_N} \ .
				\end{equation}
Hence, by virtue of the index compression map $\eta_0$, it follows from Eq. \eqref{eq:eta_0} that:
				\begin{equation}
				\overline{\mathbb{n}} = \sum^N_{i=1} \mathcal{W}_{R,i} \cdot (2j_i-n_i) = \sum^N_{i=1} \mathcal{W}_{R,i} \cdot (2j_i) - \mathbb{n} \ .
				\end{equation}
The term $\sum^N_{i=1} \mathcal{W}_{R,i} \cdot (2j_i)$ corresponds to the multispin state whereby all the single HP vacuum states are filled with the maximum number of HP bosons they can accommodate. But we know the index compression map $\eta_0$ always assigns to this state the integer $(D_{\mathcal{H}}-1)$. Thus,
				\begin{equation}
				\overline{\mathbb{n}} = D_{\mathcal{H}}-1 - \mathbb{n} 
				\end{equation}				 
which leads us to conclude that:
				\begin{equation}
				\label{eq:n_bar_n_multispin}
				\hat{\Theta} \ket{\mathbb{n}} \equiv \ket{\overline{\mathbb{n}}} = \ket{D_{\mathcal{H}}-1 - \mathbb{n}} 
				\end{equation}
where we recall once again that $D_{\mathcal{H}}$ is the dimension of the multispin Hilbert space. If we take the  \ce{^{.}CH2OD} radical, for example, $D_{\mathcal{H}}=24$, so $\ket{\overline{\mathbb{n}}} = \ket{\mathbb{23}-\mathbb{n}}$. This means that, for example, the multispin kets $\ket{\mathbb{9}}$ and $\ket{\mathbb{14}} (=\ket{\bar{\mathbb{9}}})$ are related by time-reversal symmetry. In fact, from table \ref{tab:CH2OD_table_1}, one observes that if one converts all the HP bosons in $\ket{\mathbb{9}}$ into holes, and vice-versa, one obtains $\ket{\mathbb{14}}$. It is worth noting that for a given collection of spins $\mathpzc{A}$, Eq. \eqref{eq:n_bar_n_multispin} is always valid, independent of how we choose to number-label the spins in $\mathpzc{A}$. The corresponding dual ket of $\hat{\Theta}\ket{\mathbb{n}}$ is $\bra{\mathbb{n}}\hat{\Theta}^{-1} \equiv \bra{\overline{\mathbb{n}}}=\bra{D_{\mathcal{H}}-1 - \mathbb{n}}$.
\par We are now ready to see what the time-reversal symmetry reveals about the multispin system. To understand the importance of time-reversal symmetry here, it is worth considering how it transforms the multispin Hamiltonian $\op{H}_{spin-spin}$. From Eq. \eqref{eq:H_spin-spin_2} and \eqref{eq:time-reversal_J}, it follows that:
					\begin{equation}
					\label{eq:Theta_H_spin_spin_commutation}
					\hat{\Theta} \op{H}_{spin-spin} \hat{\Theta}^{-1} = \op{H}_{spin-spin}
					\end{equation}
(note that the coupling constants $T_{i,i'}$ are all real) which means that $\op{H}_{spin-spin}$ is invariant under time-reversal. This is exactly what we should expect since $\op{H}_{spin-spin}$ describes the Hamiltonian of an isolated system, and we know that for such systems the homogeneity of time applies (i.e. the energy -- or, in general, the physics -- of the system remains the same under any time translation), so $\op{H}_{spin-spin}$ must certainly commute with $\hat{\Theta}$. Consider now the matrix element between the bra $\bra{\mathbb{n}'}$ and ket $\ket{\mathbb{n}}$ according to Eq. \eqref{eq:Theta_H_spin_spin_commutation}:
					\begin{subequations}
					\begin{align}
					\matrixel{\mathbb{n}'}{\hat{\Theta}^{-1} \op{H}_{spin-spin} \hat{\Theta}}{\mathbb{n}} & = \matrixel{\mathbb{n}'}{\op{H}_{spin-spin}}{\mathbb{n}}\\
					\matrixel{\overline{\mathbb{n}}'}{\op{H}_{spin-spin}}{\overline{\mathbb{n}}} & = \matrixel{\mathbb{n}'}{\op{H}_{spin-spin}}{\mathbb{n}} \ . \label{eq:matrixel_H_s-s_time_reversed}
					\end{align}
					\end{subequations}
Eq. \eqref{eq:matrixel_H_s-s_time_reversed} is of vital importance: it tells us that if we know the matrix element $\matrixel{\mathbb{n}'}{\op{H}_{spin-spin}}{\mathbb{n}}$ between the kets $\ket{\mathbb{n}}$ and $\ket{\mathbb{n}'}$ which are elements of the subspace $\mathcal{B}_n$, then we also know the corresponding matrix element between their hole complements in the subspace $\mathcal{B}_{2J_0-n}$. This observation reduces significantly the computational cost of creating the submatrices $\{\mathscr{B}_n\}$: once we know $\mathscr{B}_n$ we can easily create $\mathscr{B}_{2J_0-n}$. Another interpretation of Eq. \eqref{eq:matrixel_H_s-s_time_reversed} is that if we convert all HP bosons into HP holes, and vice versa, the physics remain the same. This is a property we may call \emph{particle-hole transformation invariance}.
\par Eq. \eqref{eq:Theta_H_spin_spin_commutation} has an even more profound implication. Since $\op{H}_{spin-spin}$ is decomposable into subspaces according to Eq. \eqref{eq:H_spin_spin_diag}, each eigenvector of $\op{H}_{spin-spin}$ will also belong to only one of these subspaces. Say $\ket{\mathscr{E}^{(n)}_\mu}$ an eigenvector of $\op{H}_{spin-spin}$ but which belongs to the subspace $\mathcal{B}_n$:
				\begin{equation}
				\op{H}_{spin-spin} \ket{\mathscr{E}^{(n)}_\mu} = \mathscr{E}^{(n)}_\mu \ket{\mathscr{E}^{(n)}_\mu} 
				\end{equation}
where $\mathscr{E}^{(n)}_\mu$ is the eigenvalue of $\op{H}_{spin-spin}$ relative to $\ket{\mathscr{E}^{(n)}_\mu}$. The index $\mu$ numbers the eigenvectors of $\op{H}_{spin-spin}$ in $\mathcal{B}_n$, so $\mu=1,2,\ldots, \Omega_{\mathpzc{A},n}$ -- where $\Omega_{\mathpzc{A},n}$ is the dimension of $\mathcal{B}_n$, \S \ref{par:submatrices_dimension}. As usual, the time-reversed state of $\ket{\mathscr{E}^{(n)}_\mu}$, $\ket{\mathscr{E}^{(\overline{n})}_\mu}$, is obtained by operating $\hat{\Theta}$ on $\ket{\mathscr{E}^{(n)}_\mu}$, i.e. $\ket{\mathscr{E}^{(\overline{n})}_\mu} = \hat{\Theta} \ket{\mathscr{E}^{(n)}_\mu}$. Given that $\ket{\mathscr{E}^{(\overline{n})}_\mu}=\ket{\mathscr{E}^{(2J_0-n)}_\mu}$, $\ket{\mathscr{E}^{(\overline{n})}_\mu}$ necessarily belongs to the subspace $\mathcal{B}_{2J_0-n}$. Since $\hat{\Theta}$ commutes with $\op{H}_{spin-spin}$ according to Eq. \eqref{eq:Theta_H_spin_spin_commutation}, it follows that:
				\begin{subequations}
				\begin{align}
				\left[\op{H}_{spin-spin}, \hat{\Theta} \right] \ket{\mathscr{E}^{(n)}_\mu} & = 0 \\
				\left(\mathscr{E}^{(\overline{n})}_\mu - \mathscr{E}^{(n)}_\mu \right) \ket{\mathscr{E}^{(\overline{n})}_\mu} & =0 \label{eq:time-reversed_eigenvalues}
				\end{align}
				\end{subequations}				 
from which we deduce that $\mathscr{E}^{(\overline{n})}_\mu = \mathscr{E}^{(n)}_\mu$. The implication of Eq.\eqref{eq:time-reversed_eigenvalues} is this: two different eigenvectors of $\op{H}_{spin-spin}$ related by time-reversal symmetry also share the same eigenvalue. A far more reaching conclusion is that once we are able to create and eigendecompose the submatrice $\mathscr{B}_n$ of $\mathscr{H}_{spin-spin}$, Eq.\eqref{eq:H_spin_spin_diag}, we do not need to create its hole complement $\mathscr{B}_{2J_0-n}$ and eigendecompose it de novo, because the eigenvalues of the two submatrices coincide and their eigenvectors are related through $\hat{\Theta}$. Finding the hole complement of an eigenvector can be easily done thanks to Eq. \eqref{eq:n_bar_n_multispin}. For example, let the spin system whose Hamiltonian is given by $\op{H}_{spin-spin}$ in Eq. \eqref{eq:H_spin-spin_2} be the \ce{^{.}CH2OD} radical. From table \ref{tab:CH2OD_table_1}, we know that a generic normalized eigenvector $\ket{\mathscr{E}^{(1)}_\mu}$ of the subspace $\mathcal{B}_1$ is given by the linear combination:
				\begin{equation}
				\ket{\mathscr{E}^{(n=1)}_\mu} = c_{\mu,1} \ket{\mathbb{1}} + c_{\mu,2} \ket{\mathbb{3}} + c_{\mu,3} \ket{\mathbb{6}} + c_{\mu,4} \ket{\mathbb{12}}
				\end{equation}
where the $c_{\mu,i}$ are real coefficients, and $\sum_i c_{\mu,i}^2 =1$. The corresponding hole complement of $\ket{\mathscr{E}^{(n=1)}_\mu}$, $\ket{\mathscr{E}^{(\overline{n}=4)}_\mu}$, is an eigenvector of $\mathcal{B}_{4}$, and
				\begin{equation}
				\begin{split}
				 \ket{\mathscr{E}^{(\overline{n}=4)}_\mu} & = \hat{\Theta}\ket{\mathscr{E}^{(n=1)}_\mu}   \\
				& = c_{\mu,1} \hat{\Theta} \ket{\mathbb{1}} + c_{\mu,2} \hat{\Theta}\ket{\mathbb{3}} + c_{\mu,3} \hat{\Theta}\ket{\mathbb{6}} + c_{\mu,4} \hat{\Theta}\ket{\mathbb{12}} \\
				& = c_{\mu,1} \ket{\mathbb{22}} + c_{\mu,2} \ket{\mathbb{20}} + c_{\mu,3} \ket{\mathbb{17}} + c_{\mu,4} \ket{\mathbb{11}}
				\end{split}
				\end{equation}
where, in getting to the last step, we applied Eq. \eqref{eq:n_bar_n_multispin}.
\par It is worth mentioning that there are instances whereby a subspace $\mathcal{B}_{n}$ coincides with its time-reversed counterpart. Recall that each subspace $\mathcal{B}_{n}$ is characterized by the total number $n$ of HP bosons each of its basis kets contains. So, if each basis ket of the hole complement of $\mathcal{B}_n$, i.e. $\mathcal{B}_{2J_0-n}$, contains $(2J_0-n)$ HP bosons, then  $\mathcal{B}_n = \mathcal{B}_{2J_0-n}$ when both subspaces are characterized by the same number of HP bosons. That is, when $n=2J_0-n$, which happens only when $n=J_0$. But $n$ can assume the value of $J_0$ only when $J_0$ is an integer. This means that for multispin systems whose total spin quantum number $J_0$ is an integer, the subspace $\mathcal{B}_{n=J_0}$ is its own hole complement. While the eigenvalues of all the other subspaces are atleast double degenerate due to Eq. \eqref{eq:time-reversed_eigenvalues}, the same cannot be said of $\mathcal{B}_{n=J_0}$. Conversely, we also conclude that for multispin systems whose total spin quantum number $J_0$ is a half-integer, if the system's Hamiltonian is invariant under time-reversal symmetry, then the energy eigenvalues of the system are all atleast double degenerate. This conclusion is perfectly inline with Kramer's degeneracy theorem\citep{book:Sakurai-2011,book:Wigner-1959}.

\subsubsection{The isotropic multispin Hamiltonian in the presence of an external static magnetic field}\label{par:external_field}
\par The theoretical machinery developed above are still useful and relevant when we subject the previously isolated multispin system to a static magnetic field $\mathbf{B}_o$. The new spin Hamiltonian of the system, $\op{H}_o$, is given by Eq. \eqref{eq:H_o}. If we take the direction of the external magnetic field $\mathbf{B}_o$ to be the axis of quantization $\gv{e}_z$, so that $\mathbf{B}_o=B_o \gv{e}_z$, then Eq. \eqref{eq:H_o} becomes:
				\begin{equation}
				\label{eq:H_o_b}
				\op{H}_o = \op{Z} + \op{H}_{spin-spin} 
				\end{equation}
where $\op{H}_{spin-spin}$ is still given by Eq. \eqref{eq:H_spin-spin_2}, and
				\begin{equation}
				\label{eq:Zeeman}
				\op{Z} := -B_o \sum_{i} \gamma_i  \spin{J}^z_i \ .
				\end{equation}
We know from Eq. \eqref{eq:examples_rank_1_tensors} that $\spin{J}^z$ is the zeroth component of a rank $k=1$ spherical tensor, so $\op{Z}$ is also the zeroth component of a rank $k=1$ tensor. We have also already seen that $\op{H}_{spin-spin}$ is a zero rank spherical tensor. Thus, $\op{H}_o$ is the sum of the zeroth components of spherical tensors of different ranks. But we know from Eq. \eqref{eq:def_spherical_tensors_a} that $\spin{J}^z_{tot}$ commutes with the zeroth component $(q=0)$ of \emph{any} spherical tensor operator. It therefore follows from these considerations that:
				\begin{equation}
				\left[\spin{J}^z_{tot}, \op{H}_o \right] = 	\hat{0}
				\end{equation}
which is in complete analogy to Eq. \eqref{eq:H_spin_spin_spherical_commutations_a}. This is significant because it means that all the results we derived above for the isolated multispin system on the basis of Eq. \eqref{eq:H_spin_spin_spherical_commutations_a} also apply here. For example, $\op{H}_o$, like $\op{H}_{spin-spin}$, conserves the total number of HP bosons. So, $\matrixel{\mathbb{n}'}{\op{H}_o}{\mathbb{n}} = 0$, if the multispin kets $\ket{\mathbb{n}'}$ and $\ket{\mathbb{n}}$ do not contain the same total number of HP bosons. The conservation of the total number of HP bosons also implies that, just like in the case of $\op{H}_{spin-spin}$, $\op{H}_o$ subdivides the system's Hilbert space into $(2J_0+1)$ subspaces: $\mathcal{B}_0, \mathcal{B}_1, \ldots , \mathcal{B}_{2J_0}$. Hence, the matrix representation of $\op{H}_o$, $\mathscr{H}_o$, can be written in the block diagonalized form:
					\begin{equation}
					\label{eq:H_o_diag}
					\mathscr{H}_{o} = \bigoplus^{2J_0}_{n=0} \mathscr{B}_n = \diag \left(\mathscr{B}_0, \mathscr{B}_1, \ldots , \mathscr{B}_{2J_0} \right) = 
					  \renewcommand{\arraystretch}{1.2}
  \left(
  \begin{array}{ c c | c c | c c }
  \cline{1-2}
    \multicolumn{1}{|c}{} & &  & \mc{} &  &  \\
    \multicolumn{2}{|c|}{\raisebox{.6\normalbaselineskip}[0pt][0pt]{$\mathscr{B}_0$}} &  & \mc{} &  &  \\
    \cline{1-4}
     &  & & &  &  \\
    &  & \multicolumn{2}{c|}{\raisebox{.6\normalbaselineskip}[0pt][0pt]{$\ddots$}} &  &  \\
    \cline{3-6}
     & \mc{} &  &  & & \multicolumn{1}{c|}{} \\
     & \mc{} &  &  & \multicolumn{2}{c|}{\raisebox{.6\normalbaselineskip}[0pt][0pt]{$\mathscr{B}_{2J_0}$}} \\ \cline{5-6}
  \end{array}
  \right)
					\end{equation}	
just as we saw for $\mathscr{H}_{spin-spin}$ at Eq. \eqref{eq:H_spin_spin_diag}, where the definition for $\mathcal{B}_n$ -- Eq. \eqref{eq:def_mathcal_B_n} -- still applies; and $\mathscr{B}_n$, in analogy to Eq. \eqref{eq:def_mathscr_B_n}, is now defined as:
					\begin{equation}
					\label{eq:def_mathscr_B_n_b}
					\mathscr{B}_n := \left\lbrace \matrixel{\mathbb{n}}{\op{H}_o}{\mathbb{n'}} \ \Big\vert \ \forall \mathbb{n}, \mathbb{n'} \in \mathcal{B}_n \right\rbrace \ .
					\end{equation}
\par The generating function for the dimensions $\{\Omega_{\mathpzc{A},n}\}$ of the subspaces $\{\mathcal{B}_n\}$ is, therefore, still given by $G_{\mathpzc{A},\Omega}(q)$, Eq. \eqref{eq:generating_func_Omega_a}. 
\par The expression for the generic matrix element of $\op{H}_o$ between $\ket{\mathbb{n}'}=\ket{n'_1, n'_2, \ldots, n'_N}$ and $\ket{\mathbb{n}}=\ket{n'_1, n'_2, \ldots, n'_N}$, follows directly from Eq. \eqref{eq:H_spin-spin_matrix_el} and \eqref{eq:Zeeman}: 
				\begin{multline}
				\label{eq:H_o_matrix_el}
				\matrixel{\mathbb{n}'}{\op{H}_{o}}{\mathbb{n}}  = \matrixel{n'_1, n'_2, \ldots, n'_N}{\op{H}_{o}}{n_1, n_2, \ldots, n_N}\\
				 = \delta_{\mathbb{n}', \mathbb{n}} \left[-B_o \sum_{i} \gamma_i  (j_i-n_i) + \sum_{i>k} T_{i,k} (j_i-n_i)(j_k-n_k)\right]\\
				 + \frac{1}{2}  \sum_{i>k} T_{i,k} \sqrt{\left(n_i + \frac{1 \pm 1}{2} \right)\left(2j_i-n_i+\frac{1 \mp 1}{2} \right)\left(2j_k-n_k+ \frac{1 \pm 1}{2} \right)\left(n_k + \frac{1 \mp 1}{2} \right)} \\
				 \times \ \Delta_{i,k} \ \delta_{n'_i,n_i \pm 1} \ \delta_{n'_k,n_k \mp 1}
				\end{multline}
where we can once again observe that $\op{H}_o$ conserves the total number of HP bosons. The submatrices $\mathscr{B}_n$ can be easily created by employing Eq. \eqref{eq:H_o_matrix_el}. 
\par Here too, the subspaces $\mathcal{B}_n$ and $\mathcal{B}_{2J_0-n}$ are related by time-reversal symmetry. However, the striking difference between the isolated $\op{H}_{spin-spin}$ and $\op{H}_o$ is how they are transformed under the time-reversal operator $\hat{\Theta}$. Unlike $\op{H}_{spin-spin}$, $\op{H}_o$ is not time-reversal symmetric. In fact, from Eqs. \eqref{eq:time-reversal_J} and \eqref{eq:Theta_H_spin_spin_commutation}, we derive that:	
					\begin{equation}
					\label{eq:Theta_H_o_commutation}
					\hat{\Theta} \op{H}_{o} \hat{\Theta}^{-1} = \op{H}_o- 2\op{Z} \ .
					\end{equation}
It must be emphasized that the time-reversal operator $\hat{\Theta}$ we have employed so many times above, including Eq. \eqref{eq:Theta_H_o_commutation}, is not a universal time-reversal operator but a local one restricted to the multispin system. Had $\hat{\Theta}$ been the time-reversal operator acting on the whole Universe, then $\op{H}_o$ would certainly commute with $\hat{\Theta}$ since the Universe is an isolated system. Similar results also follow if we consider the multispin system together with the external magnetic field $\mathbf{B}_o$ as one isolated system, and $\hat{\Theta}$ their combined time-reversal operator. By restricting $\hat{\Theta}$ only to the multispin system, we have inadvertently partitioned the system into two parts: a focus system (the spins) and its environment (the external magnetic field), which are both odd (i.e. they change sign) respect to time-reversal operation. This bipartite partitioning therefore leads to a broken $T-$symmetry in $\op{H}_o$.  
\par Taking the matrix element of Eq. \eqref{eq:Theta_H_o_commutation} between the two generic multispin states $\ket{\mathbb{n}'}$ and $\ket{\mathbb{n}}$, we obtain:
					\begin{equation}
					\label{eq:matrixel_H_o_time-reversal}
					\matrixel{\overline{\mathbb{n}}'}{\op{H}_o}{\overline{\mathbb{n}}} = \matrixel{\mathbb{n}'}{\op{H}_o}{\mathbb{n}} - \delta_{\mathbb{n}',\mathbb{n}} \ 2 \mathscr{Z}_{\mathbb{n},\mathbb{n}}
					\end{equation}
where, $\mathscr{Z}_{\mathbb{n},\mathbb{n}} := \matrixel{\mathbb{n}}{\op{Z}}{\mathbb{n}}$. If $\ket{\mathbb{n}'}$ and $\ket{\mathbb{n}}$ belong to the subspace $\mathcal{B}_n$, then $\ket{\overline{\mathbb{n}}'}$ and $\ket{\overline{\mathbb{n}}}$ are elements of $\mathcal{B}_{2J_0-n}$. Thus, Eq. \eqref{eq:matrixel_H_o_time-reversal} provides an easy way to generate the submatrix $\mathscr{B}_{2J_0-n}$ when $\mathscr{B}_n$ is known.
\par An even more important consequence of Eq. \eqref{eq:Theta_H_o_commutation} is yet to be unveiled. It has to do with the relationship between the eigenvalues and eigenvectors of $\mathcal{B}_n$ and those of $\mathcal{B}_{2J_o-n}$. We assume $\mathcal{B}_n$ is not its own hole complement. Say $\ket{E^{(n)}_\mu}$ an eigenvector of $\op{H}_o$ in the subspace $\mathcal{B}_n$:
					\begin{equation}
					\op{H}_o \ket{E^{(n)}_\mu} = E^{(n)}_\mu \ket{E^{(n)}_\mu} \ , \qquad \ \mu = 1, 2, \ldots, \Omega_{\mathpzc{A},n}
					\end{equation}
where $E^{(n)}_\mu$ is the eigenvalue of $\ket{E^{(n)}_\mu}$ according to $\op{H}_o$. Surely, the hole complement of $\ket{E^{(n)}_\mu}$, that is:
					\begin{equation}
					\label{eq:H_o_eigenvector_time_inverse}
					\hat{\Theta} \ket{E^{(n)}_\mu}\equiv \ket{E^{(\overline{n})}_\mu}=\ket{E^{(2J_0-n)}_\mu} 
					\end{equation}
is also an eigenvector of $\op{H}_o$, but belongs to the subspace $\mathcal{B}_{2J_0-n}$:
					\begin{equation}
					\op{H}_o \ket{E^{(\overline{n})}_\mu} = E^{(\overline{n})}_\mu \ket{E^{(\overline{n})}_\mu} \ , \qquad \ \mu = 1, 2, \ldots, \Omega_{\mathpzc{A},n}
					\end{equation}		
where we recall that $\Omega_{\mathpzc{A},n}=\Omega_{\mathpzc{A},2J_0-n}$ according to Eq. \eqref{eq:reciprocal_identity}.	Now, from Eq. \eqref{eq:Theta_H_o_commutation} we easily derive the commutation relation between $\op{H}_o$ and $\hat{\Theta}$:			
					\begin{equation}
					\label{eq:Theta_H_o_commutation_c}
					\left[\ \op{H}_o \ , \ \hat{\Theta} \ \right] = 2 \op{Z} \hat{\Theta} \ .
					\end{equation}
If we now take the matrix element of Eq. \eqref{eq:Theta_H_o_commutation_c} between the eigenvector $\ket{E^{(n)}_\mu}$ and its hole complement $\ket{E^{(\overline{n})}_\mu}$, we get:
					\begin{subequations}
					\begin{align}
					\matrixel{E^{(\overline{n})}_\mu}{\left[\ \op{H}_o \ , \ \hat{\Theta} \ \right]}{E^{(n)}_\mu} & = 2\matrixel{E^{(\overline{n})}_\mu}{\op{Z} \hat{\Theta}}{E^{(n)}_\mu} \\
					E^{(\overline{n})}_\mu - E^{(n)}_\mu & = 2\matrixel{E^{(\overline{n})}_\mu}{\op{Z}}{E^{(\overline{n})}_\mu}
					\end{align}
					\end{subequations}
from which follows that:
					\begin{equation}
					\label{eq:H_o_eigenvalues_time-reversal_relation}
					E^{(\overline{n})}_\mu = E^{(n)}_\mu - 2 \matrixel{E^{(n)}_\mu}{\op{Z}}{E^{(n)}_\mu} \ .
					\end{equation}
Eq. \eqref{eq:H_o_eigenvalues_time-reversal_relation} asserts that we can easily compute the eigenvalue of the eigenvector $\ket{E^{(\overline{n})}_\mu}$ from its hole complement $\ket{E^{(n)}_\mu}$ if we already know the latter and its eigenvalue. The significance of Eq. \eqref{eq:H_o_eigenvalues_time-reversal_relation} cannot be stressed enough. For example, suppose we have a univariate spin system of $10$ spin-$\frac{1}{2}$, i.e. $\mathpzc{A}=\left\lbrace \frac{1}{2}^{10}\right\rbrace$, whose Hamiltonian in the presence of a static magnetic field is given by $\op{H}_o$, Eq. \eqref{eq:H_o_b}. The total spin quantum number for the system is $J_0=5$, so, in principle, we can determine the eigenvectors and eigenvalues of  $\op{H}_o$ by eigendecomposing the submatrices $\mathscr{B}_0, \mathscr{B}_1, \mathscr{B}_2, \ldots, \mathscr{B}_{10}$, Eq.\eqref{eq:H_o_diag}. But by virtue of Eqs. \eqref{eq:H_o_eigenvector_time_inverse} and \eqref{eq:H_o_eigenvalues_time-reversal_relation}, there is no need to create and eigendecompose all the eleven submatrices: we only need $\mathscr{B}_0, \mathscr{B}_1, \mathscr{B}_2, \mathscr{B}_3,\mathscr{B}_4, \mathscr{B}_5$ to completely determine the eigenvalues and eigenvectors of $\op{H}_o$. This is an enormous simplification.
\par Most multispin systems of chemical interest present one or more groups of equivalent spins. We can further reduce the computational cost of diagonalizing $\op{H}_o$ if we exploit the presence of these groups in the system. The multiset $\mathpzc{A}$ of spins can be viewed as the \emph{sum} of multisets:
				\begin{equation}
				\mathpzc{A} = \biguplus^{\phi(\mathpzc{A})}_g \mathpzc{A}_g
				\end{equation}
(for example, $\{1^2,2,5\} \uplus \{1^3,2^2,3,5\}=\{1^5,2^3,3,5^2\}$) where $\phi(\mathpzc{A})$ is the number of groups of equivalent spins present in $\mathpzc{A}$; the index $g$ runs over the groups of equivalent spins, and $\mathpzc{A}_g$ is the spin multiset for the $g-$th group of equivalent spins. Each $\mathpzc{A}_g$ is then transformed into its Clebsch-Gordan series: $\mathpzc{A}_g \mapsto \widetilde{\mathpzc{A}}_g$ (see \S \ref{par:eigenenergies_equivalent_spins}). Since the elements of the Clebsch-Gordan series $\widetilde{\mathpzc{A}}_g$ are independent of each other (and irreducible), each element of the Cartesian product: 
				\begin{equation}
				\bigtimes^{\phi(\mathpzc{A})}_g \widetilde{\mathpzc{A}}_g = \widetilde{\mathpzc{A}}_1 \times \widetilde{\mathpzc{A}}_2 \times \ldots \times \widetilde{\mathpzc{A}}_{\phi(\mathpzc{A})} 
				\end{equation}
constitutes an independent multiset of spins subject to the same \emph{form} of spin Hamiltonian and can be diagonalized independently. Take for example the \ce{^{.}CH2OD} radical. If the experimental conditions are such that the two hydrogen nuclei can be considered as equivalent spins, then the radical consists of three groups of equivalent spins: 1) the unpaired electron, $\mathpzc{A}_1=\left\lbrace \frac{1}{2}\right\rbrace$; 2) the two hydrogen nuclei, $\mathpzc{A}_2=\left\lbrace \frac{1}{2}^2\right\rbrace $;  and 3) the deuterium nucleus, $\mathpzc{A}_3=\{1\}$. $\mathpzc{A}_1$ and $\mathpzc{A}_3$ have only one element each so they coincide with their Clebsch-Gordan multisets $\widetilde{\mathpzc{A}}_g$. On the other hand, we immediately have that $\widetilde{\mathpzc{A}}_2=\left\lbrace 0,1\right\rbrace$. Hence,
				\begin{equation}
				\begin{split}
				\widetilde{\mathpzc{A}}_1 \times \widetilde{\mathpzc{A}}_2 \times \widetilde{\mathpzc{A}}_3 & = \left\lbrace \frac{1}{2} \right\rbrace \times \left\lbrace 0,1 \right\rbrace \times \left\lbrace 1 \right\rbrace  = \left\lbrace  \left\lbrace \frac{1}{2},0,1 \right\rbrace ,  \left\lbrace \frac{1}{2},1^2 \right\rbrace\right\rbrace \ .
				\end{split}
				\end{equation}
What this means is that when the two hydrogen nuclei in \ce{^{.}CH2OD} are considered equivalent, the radical can be viewed as the sum of two smaller multispin systems, independent of each other: the first system consists of a spin-$1/2$ (the unpaired electron) and a spin-$1$ particle (the deuterium nucleus), while the second system consists of a spin-$1/2$ (the same electron) and two spin-$1$ particles (the deuterium nucleus and the triplet state of the two hydrogen nuclei). The Hamiltonian of both subsystems is still of the form given in Eq. \eqref{eq:H_o_b}. The Hilbert space of both subsystems can be block diagonalized as we saw above. For the first subsystem, the dimension of the block matrices are coefficients of the polynomial:
				\begin{equation}
				(1+q)(1+q+q^2)= 1+2q+2q^2+q^3 \ .
				\end{equation}		
And the dimension of the block matrices for the second subsystem are the coefficients of the following polynomial:
				\begin{equation}
				(1+q)(1+q+q^2)^2= 1+3q+5q^2+5q^3 + 3q^4 + q^5 \ .
				\end{equation}		
For each subsystem, all that has been discussed above in this section still applies. To recapitulate, we see that instead of diagonalizing a matrix of dimension $4$ and one of dimension $7$ to find the eigenvalues and eigenvectors of $\op{H}_o$ for the radical \ce{^{.}CH2OD} as we saw previously, we only need to diagonalize matrices of dimension $2,3$ and $5$ when the two hydrogen nuclei are equivalent. When applied to large spin systems with groups of equivalent spins, this approach reduces the computational cost of diagonalizing the multispin Hamiltonian significantly.  
\par If we take the naphthalene anion for example, we are dealing with a univariate system of $9$ spin-1/2 particles, i.e. $\mathpzc{A}=\left\lbrace \frac{1}{2}^9\right\rbrace$. The dimension of the Hilbert space is, therefore, $512$. Suppose we ignore the presence of equivalent spins in the system. Then, to engeindecompose the systems $\op{H}_o$, we will have to effectively diagonalize four matrices whose dimensions are $9,36,84$ and $126$, according to the HP transformation scheme discussed above. If the experimental conditions are sufficiently favorable, the system can be thought of as consisting of three groups of equivalent spins: 1) the electron,  $\mathpzc{A}_1=\left\lbrace \frac{1}{2}\right\rbrace$; 2) a collection four hydrogen nuclei,  $\mathpzc{A}_2=\left\lbrace \frac{1}{2}^4\right\rbrace$; and 3) another collection of four hydrogen nuclei,  $\mathpzc{A}_3=\left\lbrace \frac{1}{2}^4\right\rbrace$. Then, $\widetilde{\mathpzc{A}}_2 = \{2,1^3,0^2\}$, and $\widetilde{\mathpzc{A}}_3 = \{2,1^3,0^2\}$. Thus,
				\begin{multline}
				\widetilde{\mathpzc{A}}_1 \times \widetilde{\mathpzc{A}}_2 \times \widetilde{\mathpzc{A}}_3 = \left\lbrace \frac{1}{2}\right\rbrace \times \{2,1^3,0^2\} \times \{2,1^3,0^2\} \\
				 = \left\lbrace \left\lbrace \frac{1}{2},2,2 \right\rbrace,  \left\lbrace \frac{1}{2},2,1 \right\rbrace^3, \left\lbrace \frac{1}{2},2,0 \right\rbrace^2  , \left\lbrace \frac{1}{2},1,2 \right\rbrace^3, \left\lbrace \frac{1}{2},1,1 \right\rbrace^9, \right.  \\  
				 \left. \left\lbrace \frac{1}{2},1,0\right\rbrace^6, \left\lbrace \frac{1}{2},0,2 \right\rbrace^2, \left\lbrace \frac{1}{2},0,1 \right\rbrace^6, \left\lbrace \frac{1}{2},0,0 \right\rbrace^4 \right\rbrace \ .
				\end{multline}
So the eigendecomposition of the naphthalene anion's Hamiltonian $\op{H}_o$ can be done by considering a series of smaller but independent multisets: i) one multispin system comprised of a spin-$1/2$ and two spin-$2$ particles, i.e. $\{1/2,2,2\}$; ii) three multispin systems each of the type $\{1/2,2,1\}$, etc. For each of these subsystems, we can apply the HP transformation and create the block matrices. In the case of the naphthalene anion, the largest block matrix we shall encounter comes from the subsystem $\{1/2,2,2\}$. For this particular subsystem, the dimension of the block matrices follows from the generating function:
				\begin{equation}
				(1+q)(1+q+q^2+q^3+q^4)^2 = 1+3q+5q^2+7q^3+9q^4+9q^5+7q^6+5q^7+3q^8+q^9\ .
				\end{equation}
Hence, if we take into account the two groups of equivalent spins present in the naphthalene anion, the largest matrix we will ever have to diagonalize is of dimension $9$, which is remarkable. 
\par To fully exploit the presence of groups of equivalent spins in the multispin system, we need the Clebsch-Gordan coefficients. This constitutes the primary computational challenge in the method just illustrated. One can thus incorporate optimized subroutines for the calculation of Clebsch-Gordan coefficients into ones multispin algorithm. The somehow comforting observation we can make here is that, in most of the multispin systems of interest, groups of equivalent spins hardly exceed six in number -- which means a subroutine which can handle the Clebsch-Gordan coefficients for the addition of up to six angular momenta suffices for most routine computations. 		
\subsubsection{Eigendecomposition of Liouvillians}
\par In the normal settings of quantum mechanics, as we have done above, one works in the so-called Hilbert space, where observables like energy, magnetization, etc. are operators and the quantum states are vectors. In this way of doing quantum mechanics, the system's energy are obtained as the expectation values of its Hamiltonian. These energy values are, in principle, defined up to a given constant. But in spectroscopic experiments, the properties we do observe or measure are essentially related -- not to the absolute energies -- but to the difference between these eigenenergies. This fact is nicely conveyed in the resonance conditions of said experiments. An alternative way of doing quantum mechanics whereby such energy differences naturally come up as the expectation values of some operator is thus more suitable to spectroscopy. Such an alternative is accomplished when we formulate quantum mechanics in the so-called Liouville space\citep{inbook:Jeener-1982}. In addition, the Liouville space formulation makes it relatively easier to treat relaxation processes, compared to the Hilbert space\citep{inbook:Jeener-1982}. It is therefore understandable why the Liouville space formalism is popular in various fields of spectroscopy, including magnetic resonance. The catch, however, is that given a Hilbert space of finite dimension $D_\mathcal{H}$, the dimension of its corresponding Liouville space is $D^2_\mathcal{H}$ -- which roughly translates into even higher computational costs when one works in Liouville space. 
\par As it is well-known, operators in Hilbert space, including the density matrix, become vectors (\emph{supervectors}) in Liouville space. Likewise, operations which transformed one operator into another in Hilbert space become operators (\emph{superoperators}) in Liouville space. 
\par Recall that the Hamiltonian is known to be the generator of the dynamics of the density matrix in Hilbert space. When the density matrix becomes a supervector in Liouville space, its dynamics are generated by the superoperator called the \emph{Liouvillian}. Just as we eigendecompose the Hamiltonian $\op{H}$ in the Hilbert space to get the eigenenergies of the system, the eigendecomposition of its corresponding Liouvillian $\sop{L}$ returns all possible pairwise differences between the eigenenergies. The relation between a given Hamiltonian $\op{H}$ and its Liouvillian $\sop{L}$ is:
					\begin{equation}
					\label{eq:def_Liouvillian}
					\sop{L} = \op{H} \otimes \hat{\mathbb{I}} -  \hat{\mathbb{I}} \otimes \op{H}^* 
					\end{equation}
where "$\otimes$" denotes the operation of vector space tensor direct product; $\hat{\mathbb{I}}$ is the identity operator defined on the same Hilbert space as $\op{H}$, and $\op{H}^*$ is the complex conjugate of $\op{H}$. We once again observe from Eq. \eqref{eq:def_Liouvillian} that the dimension of the linear space where $\sop{L}$ operates, i.e. the Liouville space, is of dimension $(D_{\mathcal{H}})^2$. For example, if we have a multispin system $\mathpzc{A}=\left\lbrace \frac{1}{2}^{10}\right\rbrace$, whose Hamiltonian is given by $\op{H}_o$ in Eq. \eqref{eq:H_o}, the simulation of the multispin system's magnetic spectra would require eigendecomposing a matrix of dimension $2^{10}=1024$ according to the tout court approach if we work in Hilbert space. If we choose to work in the Liouville space, the same tout court approach will have us eigendecompose the matrix representation of the Liouvillian, which is of dimension $2^{20}=1,048,576$. 
\par This apparent inconvenience in computational costs one has to grapple with when working in the Liouville space can be easily overcome. Indeed, if $\mathfrak{U}$ is the matrix (\emph{supermatrix}) which diagonalizes $\mathfrak{L}$, i.e.
					\begin{equation}
					\mathfrak{U} \ \mathfrak{L}\ \mathfrak{U}^{-1} = \mathfrak{D} 
					\end{equation}
where $\mathfrak{D}$ is the diagonal supermatrix of the eigenvalues of $\mathfrak{L}$, and $\mathscr{U}$ is the eigenvector matrix of $\mathscr{H}$:
					\begin{equation}
					\mathscr{U} \mathscr{H} \mathscr{U}^{-1} = \mathscr{D} 
					\end{equation}
where $\mathscr{D}$ is the diagonal matrix containing the eigenvalues of $\mathscr{H}$, then it can be proved that:
					\begin{equation}
					\label{eq:Liouville_eigenvectors_in_Hilbert}
					\mathfrak{U} = \mathscr{U} \otimes \left( \mathscr{U}^{-1}\right)^T 
					\end{equation}
and
					\begin{equation}
					\label{eq:Liouville_eigenvalues_in_Hilbert}
					\mathfrak{D} = \mathscr{D}\otimes \mathbb{I} - \mathbb{I} \otimes \mathscr{D} 
					\end{equation}
where "$\otimes$" in Eqs. \eqref{eq:Liouville_eigenvectors_in_Hilbert} and \eqref{eq:Liouville_eigenvalues_in_Hilbert} indicates the operation of Kronecker (or matrix direct) product, and $\mathbb{I}$ is the identity matrix of the same dimension as $\mathscr{D}$. Eq. \eqref{eq:Liouville_eigenvectors_in_Hilbert} is of great significance because it enables us to compute the eigenvector supermatrix $\mathfrak{U}$ directly from $\mathscr{U}$ by means of a simple matrix direct product, without having to diagonalize the Liouvillian $\mathfrak{L}$ de novo. Note that when the Hamiltonian is real, like in the case of the multispin Hamiltonian $\op{H}_o$ seen above, $\mathscr{U}$ is an orthogonal matrix, and so Eq. \eqref{eq:Liouville_eigenvectors_in_Hilbert} reduces to the form:
					\begin{equation}
					\label{eq:Liouville_eigenvectors_in_Hilbert_b}
					\mathfrak{U} = \mathscr{U} \otimes \mathscr{U} \ . 
					\end{equation}
Combining the relations given in Eq. \eqref{eq:Liouville_eigenvectors_in_Hilbert} and \eqref{eq:Liouville_eigenvalues_in_Hilbert} with the HP transformation and related techniques discussed in previous sections could be very useful in reducing the computational cost of simulating the magnetic resonance spectra of multispin systems described by isotropic Hamiltonians like $\op{H}_o$ in Liouville space. For example, going back to our previous example with the multiset $\mathpzc{A}=\left\lbrace \frac{1}{2}^{10}\right\rbrace$, if we want to work in Liouville space, instead of eigendecomposing a square matrix of dimension $2^{20}$, we can obtain the same eigen-supervectors and -supervalues through Eq. \eqref{eq:Liouville_eigenvectors_in_Hilbert_b} and \eqref{eq:Liouville_eigenvalues_in_Hilbert}, respectively, by eigendecomposing only the submatrices $\mathscr{B}_0,\mathscr{B}_1 , \mathscr{B}_2, \mathscr{B}_3, \mathscr{B}_4, \mathscr{B}_5$ which are of dimension $1,10, 45, 120, 210$ and $252$, respectively.  The computational cost can be further drastically reduced if we take into consideration the presence of groups of equivalent spins as explained in \S \ref{par:external_field}.

\section{On Schwinger bosons}\label{sec:Schwinger_bosons}
\par We conclude this paper with a brief introduction to Schwinger bosons, which is another kind of spin representation commonly used in condensed matter physics. Schwinger bosons are closely related to the Holstein-Primakoff bosons, but the two representations are suitable for certain specific applications. The HP bosons are particularly useful in transforming quantum mechanical problems into counting problems. The Schwinger bosons are extremely powerful tools when we want to describe the behavior of spin states under unitary transformations like rotations. The literature provides a plethora of applications of the Schwinger bosons, mainly in quantum magnetic studies. This paper won't be complete without this brief introduction to Schwinger bosons. However, the introduction we provide below departs greatly from what one may find elsewhere. Besides the simplicity of our exposition below, we stress on the close link between Schwinger bosons and HP bosons -- which is rarely done in such details in the literature. Readers may see \cite{book:Schwinger-1952, book:Auerbach-1998} for the mainstream exposition, interpretation and applications of Schwinger bosons.  
\par We begin our introduction to Schwinger bosons by noting that the presence of the square root of operators in the HP transformation, Eqs. \eqref{eq:J_+_in_b} and \eqref{eq:J_-_in_b}, makes its use in the study of important problems like the rotation of spin states very inconvenient. The Schwinger transformation, from which derives the Schwinger bosons, provides an alternative way to represent the spin operators $\spin{J}^{\pm}$ and $\spin{J}^z$ in a very simple way without trace of any square roots. Interestingly, as we show below, the Schwinger representation follows directly from the HP transformation.
\par When we introduced the time-reversal operator in \S \ref{par:Time-reversal}, Eq. \eqref{eq:time_reversal_no_op}, we saw that when it acts on a spin state represented by HP bosons, it transforms the HP bosons present into holes and the holes into HP bosons, contemporarily. We also saw that the sum of the number of holes and HP bosons in any given basis state $\ket{n}$ of a spin-$j$ particle is always $2j$. And that, while $\hat{b}^\dagger \hat{b}$ counts the number of HP bosons, the operator $(2j-\hat{b}^\dagger \hat{b})$ counts the number of holes. The transition from HP transformation to the Schwinger transformation relies on one simple trick: treat the holes and HP bosons as two set of independent particles that can be created and annihilated separately (with the caveat that the sum of their occupation numbers remain constant). If we do so, then we need to assign an occupation number operator to the holes. Let us indicate this occupation number operator as $\hat{a}^\dagger \hat{a}$. Thus,
				\begin{equation}
				\label{eq:conservation_aa_bb}
				\hat{a}^\dagger \hat{a} = 2j-\hat{b}^\dagger \hat{b} \ .
				\end{equation}
Clearly, $\hat{a}^\dagger$ and $\hat{a}$ are the creation and annihilation operators for the holes, respectively; and they obey the same commutation rules as $\hat{b}^\dagger$ and $\hat{b}$. Eq. \eqref{eq:time_reversal_no_op} may, therefore, be rewritten as:
				\begin{equation}
				\label{eq:time_reversal_no_op_holes}
				\hat{\Theta} \ \hat{b}^\dagger \hat{b} \ \hat{\Theta}^{-1} = \hat{a}^\dagger \hat{a} \ .
				\end{equation}
Since $\{\hat{b},\hat{b}^\dagger\}$	effect only the HP bosons and $\{\hat{a},\hat{a}^\dagger\}$ act on only the holes, the two set of operators commute with each other.	
For the sake of clarity, let $\hat{n}_b \equiv \hat{b}^\dagger \hat{b}$ and 
$\hat{n}_a \equiv \hat{a}^\dagger \hat{a}$; thus, the nonnegative integers $n_b$ and $n_a$ will indicate the number of HP bosons and holes, respectively. We are therefore representing spin states with two types of particles, i.e. HP bosons and (HP) holes:
				\begin{align*}
				\mbox{HP representation} &  \hspace{1.2cm} \mapsto    & \mbox{Schwinger representation}\\
				\ket{n} & \hspace{1.2cm}\mapsto \hspace{-1.2cm} & \ket{n_b,n_a}
				\end{align*}
where, obviously, $n=n_b$. Hence, in the Schwinger representation we need two occupation numbers to indicate a single spin eigenvector of $\spin{J}^z$. Moreover, given that the two sets of operators $\{\hat{b},\hat{b}^\dagger\}$ and $\{\hat{a},\hat{a}^\dagger\}$ operate on different particles, it follows that:
				\begin{align}
				\hat{b} \ket{n_b,n_a} & = \sqrt{n_b} \ket{n_b-1,n_a}&     \hat{b}^\dagger \ket{n_b,n_a} & = \sqrt{n_b+1} \ket{n_b+1,n_a} \label{eq:Schwinger_HP_bosons_effects}\\
			\hat{a} \ket{n_b,n_a} & = \sqrt{n_a} \ket{n_b,n_a-1}&        \hat{a}^\dagger \ket{n_b,n_a} & = \sqrt{n_a+1} \ket{n_b,n_a+1} \label{eq:Schwinger_HP_holes_effects} 
				\end{align}
where the nonnegative integers $n_a$ and $n_b$ are subject to the contraint: $n_a+n_b = 2j$, Eq. \eqref{eq:conservation_aa_bb}. 
\par In regards to the operator $\spin{J}^z$, we note that if we combine Eq. \eqref{eq:J_z_in_b_b} with Eq. \eqref{eq:conservation_aa_bb}, we obtain the following expression for $\spin{J}^z$ in function of $\hat{n}_b$ and $\hat{n}_a$:
				\begin{equation}
				\label{eq:Schwinger_J_z}
				\spin{J}^z = \frac{1}{2} \left( \hat{a}^\dagger \hat{a} -\hat{b}^\dagger \hat{b}\right) = \frac{1}{2} \left( \hat{n}_a -\hat{n}_b \right) \ .
				\end{equation}
(recall we have set $\hbar=1$). It is clear from Eq. \eqref{eq:Schwinger_J_z} that the usual spin magnetic number $m$ in $\ket{j,m}$ relates to $n_a$ and $n_b$ through the expression:
				\begin{equation}
				\label{eq:aa_bb_condition_1}
				m=\frac{1}{2} (n_a - n_b) \ .
				\end{equation}				 
Moreover, from Eq. \eqref{eq:conservation_aa_bb}, we also have the condition:
				\begin{equation}
				\label{eq:aa_bb_condition_2}
				j = \frac{1}{2} (n_a + n_b)
				\end{equation}
Eqs. \eqref{eq:aa_bb_condition_1} and \eqref{eq:aa_bb_condition_2}, together, constitute the conditions the integers $n_a$ and $n_b$ must satisfy in order to represent the usual spin state $\ket{j,m}$.
\par What about the Schwinger representation for $\spin{J}^\pm$? To begin, recall the operator $\spin{J}^+=\sqrt{2j-\hat{b}^\dagger \hat{b}} \ \hat{b}$ in the HP representation, Eq. \eqref{eq:J_+_in_b}. From this expression, we see that the operator $\hat{b}$ first  reduces the number of HP bosons by $1$ while the operator $\sqrt{2j-\hat{b}^\dagger \hat{b}}$ leaves the resulting state unchanged in the number of HP bosons. The net effect of $\spin{J}^+$ is, therefore, to reduce $n_b$ by $1$. But, given that $n_a+n_b=2j$, Eq. \eqref{eq:aa_bb_condition_2}, it follows that if $\spin{J}^+$ has the effect of reducing $n_b$ by $1$, then it must also have the effect of increasing the number of holes by the same quantity, i.e. $n_a \mapsto n_a + 1$. The operator which increases $n_a$ by $1$ is $\hat{a}^\dagger$. Therefore, $\spin{J}^+$ must be proportional to $\hat{a}^\dagger \hat{b}$:
				\begin{equation}
				\spin{J}^+ = c \ \hat{a}^\dagger \ \hat{b} = c \ \hat{b}\ \hat{a}^\dagger 
				\end{equation}				
where $c$ is the proportionality constant. If we let $\spin{J}^+$ operate on $\ket{n_b,n_a}$, we find that:
				\begin{subequations}
				\begin{align}
				\spin{J}^+ \ket{n_b,n_a} & = c \ \hat{a}^\dagger \ \hat{b} \ket{n_b,n_a} \\
				& = c \sqrt{(n_a+1)}\sqrt{n_b} \ \ket{n_b-1,n_a+1}\\
				& = c \sqrt{(2j-n_b+1)}\sqrt{n_b} \ \ket{n_b-1,n_a+1} \label{eq:Schwinger_J_+}
				\end{align}
				\end{subequations}	
where we have made use of Eqs. \eqref{eq:Schwinger_HP_bosons_effects} and \eqref{eq:Schwinger_HP_holes_effects}. If we compare Eq. \eqref{eq:Schwinger_J_+} with Eq. \eqref{eq:J_+_in_b}, we immediately reach the conclusion that $c=1$ (bear in mind that $n=n_b$, and as stated above, we have set $\hbar=1$ throughout this section). Thus,
				\begin{equation}
				\label{eq:Schwinger_J_+_2}
				\spin{J}^+ = \hat{a}^\dagger \ \hat{b} = \hat{b}\ \hat{a}^\dagger \ .
				\end{equation}	
Hence, it also follows that:
				\begin{subequations}
				\begin{align}
				\spin{J}^- & = \left(\spin{J}^+ \right)^\dagger \nonumber \\
				           & = \hat{a} \ \hat{b}^\dagger = \hat{b}^\dagger \ \hat{a} \ . \label{eq:Schwinger_J_-}
				\end{align}
				\end{subequations}	
Eqs. \eqref{eq:Schwinger_J_z}, \eqref{eq:Schwinger_J_+_2} and \eqref{eq:Schwinger_J_-} constitute the Schwinger transformation. Unlike the HP transformation, we see that the Schwinger transformation is linear in the operators $\hat{a}^\dagger, \hat{a}, \hat{b}^\dagger, \hat{b}$.					
\par In \S \ref{subsec:HP_transformation}, we saw that the vacuum state $\ket{n=0}$ in the HP representation of the $\spin{J}^z$ eigenvectors of spin-$j$ corresponds to the state $\ket{j,j}$ in the normal $\ket{j,m}-$representation. The Schwinger representation also has its vacuum state, namely $\ket{n_b=0,n_a=0}=\ket{0,0}$. This vacuum state, unlike the one we encountered in the HP representation, does not correspond to any specific spin state, and its energy is undefined. However, one interesting characteristic of the Schwinger vacuum is that it is the same for all spins. This is in net contrast to the HP vacuum. \par Just as generic spin states in the HP representation can be created from the HP vacuum state by filling the latter with a number of HP bosons, generic spin states in the Schwinger representation can be easily created from the Schwinger vacuum $\ket{0,0}$ by creating a number of holes and HP bosons. Thus,
				\begin{equation}
				\label{eq:Schwinger_spin_state}
				\ket{j,m} \mapsto \ket{n_b,n_a} = \frac{\left(\hat{a}^\dagger \right)^{n_a}}{\sqrt{n_a!}}\frac{\left(\hat{b}^\dagger \right)^{n_b}}{\sqrt{n_b!}} \ket{0,0}
				\end{equation}
where the factor $\frac{1}{\sqrt{n_a! n_b!}}$ is necessary to keep the state on the RHS normalized like $\ket{n_b,n_a}$. If we express $n_a$ and $n_b$ in Eq. \eqref{eq:Schwinger_spin_state} in terms of $j$ and $m$, using Eqs. \eqref{eq:aa_bb_condition_1} and \eqref{eq:aa_bb_condition_2}, it turns out that:
				\begin{equation}
				\label{eq:Schwinger_spin_state_2}
				\ket{j,m} \mapsto \ket{j-m,j+m} = \frac{\left(\hat{a}^\dagger \right)^{j+m}}{\sqrt{(j+m)!}}\frac{\left(\hat{b}^\dagger \right)^{j-m}}{\sqrt{(j-m)!}} \ket{0,0}\ .
				\end{equation}	
In the Schwinger representation, the HP bosons and HP holes are collectively called \emph{Schwinger bosons}.
\par Among the many useful applications of the Schwinger bosons is the relatively straightforward ease with which they allow the derivation of the famous Majorana formula\citep{art:Schwinger-1977}, which states the general expression for the transition probability between two generic spin states of an arbitrary spin in the presence of: 1) a static magnetic field along a specified direction in space (chosen as the quantization axis), and 2) a rf field perpendicular to the static field. Other applications of the Schwinger bosons include the derivation of analytical expressions for the matrix elements of the rotation operator. In particular, they can be used to easily derive a close expression for the elements of the Wigner $d-$matrix\citep{book:Sakurai-2011, book:Schwinger-1952}. Closed expressions for the Clebsch-Gordan coefficients for the addition of two, three and four angular momenta can also obtained using the Schwinger bosons\citep{book:Schwinger-1952}.
\section{Final Remarks}
\par The Second Peloponnesian War, which began in 431 BCE between the Leagues led by the two rival ancient Greek cities of Athens and Sparta, marked for the former -- a city which, prior to the war, was living perhaps its most remarkable period (the "age of Pericles") of fortunes and great achievements in the arts, culture, and philosophy -- a regrettable departure from its very recent glorious past. As if the war wasn't enough a tribulation, Fate bestowed on the city of Athens and neighboring city-states, around 430 BCE, the great plague which claimed the lives of about a quarter of its population, including its "first citizen" -- the great Pericles (495-429 BCE). Idiosyncratic of the time, the people believed the plague was a punishment from the gods. Legend has it that  a delegation was, therefore, dispatched to the oracle of Apollo at Delos to seek guidance on how to stave off the plague, to which they were instructed to double the size of the altar of Apollo and make sacrifice on it. The problem, however, was that the altar in question was a cube and the instruction was understood -- in line with the traditions of ancient Greek geometry -- by all as exactly duplicating the volume of the altar by means of only two instruments: compass and unmarked straightedge. The people went on to diligently double each side of the altar, only to find that the volume of the new cubical altar was not twofold but eightfold the original. Many more attempts followed but the problem stubbornly persisted, but thankfully, the plague eventually worn out with time.
\par Stated in modern terms, the problem was to solve the equation $x^3=2$ with compass and unmarked straightedge. This problem, which came to be known as the \emph{Delian problem}, bedeviled for centuries the great minds of Antiquity. As Boyer puts it, "Not until some 2000 years later was it recognized that the oracle had sardonically proposed an unsolvable problem"\citep{art:Boyer-1949}. The proof that $x^3=2$ cannot be solved with compass and unmarked straightedge was to be found in analytic geometry, a new mathematical discipline which was essentially born out of the insights of two Frenchmen in the early half of the seventeenth century: René Descartes (1596-1650) and Pierre de Fermat (1601-1665), with the fecund ideas of Nicole Oresme (1323?-1382) and François Viète (1540-1603) -- both French as well -- playing the role of important precursors. Before the advent of this new discipline, geometry had been primarily only known for about 2000 years in the form of synthetic geometry (that is a form of geometry which relies only on axioms and construction methods using compass and straightedge) since the days of Thales of Miletus (ca. 624-548 BCE). But, adopting the algebraic notations introduced by Viéte and improving them himself, Descartes in his \emph{La Géométrie}, first published in 1637, went on to systematically study the relationship between algebra and geometry. This allowed him to translate geometrical problems into algebraic equations and infer some important properties of the geometrical figures he was studying. With such a powerful method in his possession, Descartes even went on to solve the general $n-$lines Pappus problem whose solution had even escaped "The Great Geometer" of Antiquity, Appollonius of Perga (ca. 262 - ca. 190 BCE). 
\par The history of mathematics runs abundant with episodes like the above, and we could make a long list of them. If any lesson is to be learnt from these, it cannot but be the fact that seemingly intractable mathematical problems can be hacked with just the right set of tools in hand. The process, almost invariably, involves translating the original problem into an equivalent problem of a different field. In this sense, the HP transformation does for multispin magnetic resonance what algebra, so elegantly, did for geometry. And the results presented in this paper, \S \ref{sec:HP_trans_and_multispin}, are a clear indication that the HP transformation has all the resemblance of a much needed tool in multispin magnetic resonance. The majority of these results, for example when we considered in \S \ref{par:eigenenergies_equivalent_spins} the energy levels and their respective degeneracy in a system of equivalent spins, are hardly imaginable with the usual spin representations. One important drawback of the results presented above, though, is that since the dimension of the block matrices $\mathscr{B}_n$ does not scale polynomially with the number of spins $N$, the simulation of the magnetic resonance spectra of multispins based on the eigendecomposition of these block matrices is also not going to scale polynomially with $N$, unless one takes advantage of the eventual presence of groups of equivalent spins (with cardinality $\geq 2$) in the system, as we saw in \S \ref{par:external_field}. In fact, if we ignore the presence of these equivalent groups, the scaling is exponential in the number of spins $N$, but still far better than the exponentially scaling one has with the tout court approach. If we consider a multiset $\mathpzc{A}=\{\frac{1}{2}^N\}$ with Hamiltonian given by $\op{H}_o$, Eq. \eqref{eq:H_o_b}, for example, we see that the computational cost of diagonalizing $\mathscr{H}_o$ according to the tout court approach scales $\mathcal{O}\left(2^{Np}\right)$ if the computational cost (according to the algorithm we are using) of diagonalizing a matrix of dimension $M$ is $\mathcal{O}(M^p)$, where $p$ is a positive real number. With the HP transformation, we will have to diagonalize the matrices $\mathscr{B}_0,\mathscr{B}_1,\ldots,\mathscr{B}_{N/2}$ (without loss of generality, we assume $N$ is even) to determine the eigenvalues and eigenfunctions of $\op{H}_o$. Knowing that the dimension of $\mathscr{B}_n$ in this case is given by the binomial coefficient $\binom{N}{n}$, Eq. \eqref{eq:Omega_N_spin_half}, it then follows that the computational cost of diagonalizing $\op{H}_o$ within the HP transformation framework is:
			\begin{equation}
			\mathcal{O} \left(\sum^{N/2}_{n=0} \binom{N}{n}^p \right) \stackrel{N \gg 1}{\sim} \mathcal{O} \left(\frac{2^{Np}}{2\sqrt{p}} \left(\frac{2}{\pi N} \right)^{(p-1)/2} \right) \ .
			\end{equation}
Therefore, the complexity of diagonalizing $\op{H}_o$ with HP transformation scheme is reduced by a factor proportional to $N^{(p-1)/2}$ with respect to the tout court approach. With normal diagonalization algorithms $p\approx 3$, so the reduction factor scales linearly with $N$. This is certainly an improvement but not enough when the number of spins exceeds about 15 (and if we don't exploit the presence of groups of equivalent spins). The argument just outlined goes on to show how the diagonalization approach to the simulation of magnetic resonance spectra -- despite being exact --  is a less favorable one when dealing with a large spin system which lacks groups of equivalent spins with cardinality $\geq 2$. Simulation protocols based on the HP transformation and a density matrix propagation method like the state-space restricted (SSR)\citep{art:Kuprov-2007,art:Kuprov-2008} method are sure to give a better performance. 
\par Density matrix propagation methods for the simulation of multispin magnetic resonance spectra are heavily dependent on the ability to easily index the multispin states at various stages of the algorithm. This is where the HP transformation and the index compression map $\eta_o$ become even more vital. The analytical elements these two mathematical tools bring to a propagation method can also be advantageous and may help simplify some heavy computations.
\par The multispin Hamiltonians considered in this paper are all given by the sum of operators which are proportional to the $q=0$ component of some spherical tensors. Given that the leading term in the spin Hamiltonian of a significant number of systems studied in magnetic resonance (especially in high-resolution problems) are also proportional to the $q=0$ component of some spherical tensor, the results presented here are very useful in perturbative calculations on such systems.
\par We have also shown how with the Holstein-Primakoff bosons, we may easily translate certain problems in multispin magnetic resonance into counting problems. This connection may be an indication of a profound and no less consequential relationship between magnetic resonance and discrete mathematics; a relationship, which, if ultimately fully understood, might help free -- to a significant degree -- multispin magnetic resonance from the spell of the curse that is of dimensionality. On this note, it is only reasonable that we envisage a future where the full power and glory of discrete mathematics are unequivocally brought to bear in the study of magnetic resonance. Here, computer algebra -- now almost ubiquitous in modern computing languages and which is finding its way also into magnetic resonance \citep{art:Anand-2007, art:Filip-2010} -- is going to play a central role in the simulation of multispin magnetic resonance spectra. And the recurrent use of generating functions like $G_{\mathpzc{A},\Omega}(q)$ and $G_{\mathpzc{A},\lambda}(q)$ in this paper strongly suggests the pivotal role polynomial algebra is also going to assume in magnetic resonance simulations.   
\par It is reported that when the Delian problem was put before Plato (ca. 427 - ca. 347), he interpreted it as a command to the people to study geometry. By the same token, this paper is a call to integrate discrete mathematics into the theory of magnetic resonance. We hope the discussions and results presented here may kindle at once the curiosity of the Reader to further investigate the relation between magnetic resonance, second quantized spin representations like the HP and Schwinger bosons, and discrete mathematics. The current paper only minimally scratches the surface of the subject. As Feynman once said, "There's plenty of room at the bottom".
\section*{Acknowledgments}
The author expresses his gratitude to Prof. Henrik Koch and Prof. Antonino Polimeno for stimulating discussions.

\bibliography{QMME_1_ArXiv_biblio}

\end{document}